%% file: main.tex
\documentclass[acmsmall,screen]{acmart}

\input{packages_and_commands}
\AtBeginDocument{%
  \providecommand\BibTeX{{%
    \normalfont B\kern-0.5em{\scshape i\kern-0.25em b}\kern-0.8em\TeX}}}

\setcopyright{acmcopyright}
\copyrightyear{2021}
\acmYear{2021}
\acmDOI{xxx}

\acmJournal{TIIS}
\acmVolume{xx}
\acmNumber{xx}
\acmArticle{ACM Article No}
\acmMonth{1}

\begin{document}

\title[\ours]{\ours: Interactive Query-Oriented Text Analytics for Comprehensive Investigation of Historical News Archives}

\author{Abram Handler}
\email{abram.handler@colorado.edu}
\affiliation{%
  \institution{Department of Information Science, University of Colorado, Boulder}
  \streetaddress{1045 18th Street}
  \city{Boulder}
  \state{CO}
  \country{USA}
  \postcode{80309}
}

\author{Narges Mahyar}
\author{Brendan O'Connor}
\affiliation{%
  \institution{Manning College of Information and Computer Sciences, University of Massachusetts, Amherst}
  \streetaddress{140 Governors Dr.}
  \city{Amherst}
  \state{MA}
  \country{USA}
  \postcode{01003}
}

\renewcommand{\shortauthors}{Handler, Mahyar and O'Connor.}

\input{samples/abstract}

\ccsdesc[500]{Human-centered computing~Interactive systems and tools}

\keywords{digital humanities, history, interactive text analytics, user interfaces}

\maketitle

\section{Introduction}\label{s:intro}
\input{samples/introv2}

\section{Related work}\label{s:related}
\input{samples/related_work}

\section{Needfinding study: defining practices and requirements}\label{s:needs}

\input{samples/user_needs}

\section{System}\label{s:system}
\input{samples/system}

\section{Expert interview study procedure}\label{s:usabilitystudy}
\input{samples/expert_interview_procedure}

\section{Expert interview study results}\label{s:qualresults}
\input{samples/expert_interview_results}

\section{Field study}\label{s:fieldstudy}
\input{samples/field_study}

\section{A quantitative comparison with \Baselongname~tools}\label{s:crowdstudy}
\input{samples/crowd_study}

\section{Discussion}\label{s:discussion}
\input{samples/discussion}

\section{Limitations and future work}\label{s:limits_and_future}
\input{samples/limitations_and_future}

\section{Conclusion}\label{s:conclusion_cc}
\input{samples/conclusion}

\begin{acks}
We thank Mahmood Jasim for detailed feedback throughout our research process.  We also thank Su Lin Blodgett, Javier Burroni, Katherine A. Keith and Kalpesh Krishna for helpful discussions, suggestions and edits. Finally, we thank Alyx Burns, Lucy Cousins, Prachi Modi and Ali Sarvghad for offering feedback on the user interface.
\end{acks}

\bibliographystyle{ACM-Reference-Format}
\bibliography{sample-base,new-cites}


\appendix
\input{appendix_content}

\end{document}
\endinput


\begin{figure}[h]
\fbox{\includegraphics[width=1.0\textwidth]{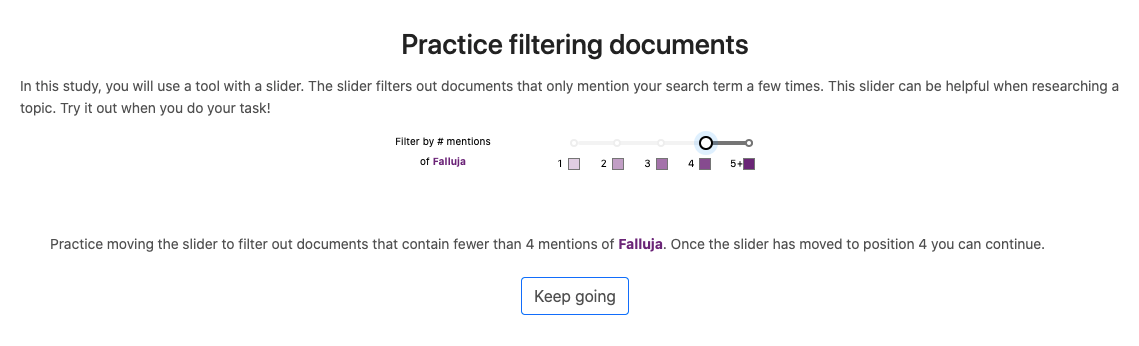}}
\caption{In the our crowd study, participants used this screen to practice using the \ours~slider}
\end{figure}

\begin{figure}[h]
\fbox{\includegraphics[width=1.0\textwidth]{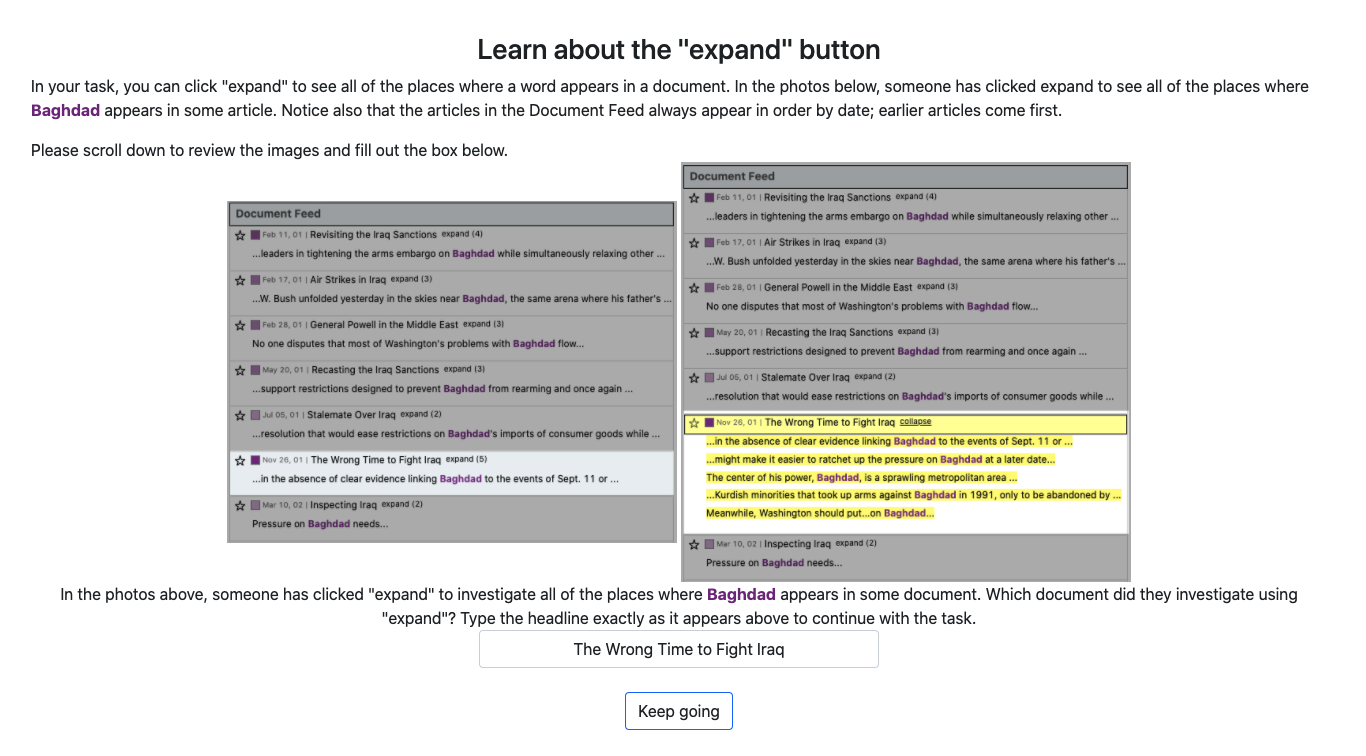}}
\caption{In the our crowd study, participants used this screen to practice using the \ours~expand button}
\end{figure}

\begin{figure}[h]
\fbox{\includegraphics[width=1.0\textwidth]{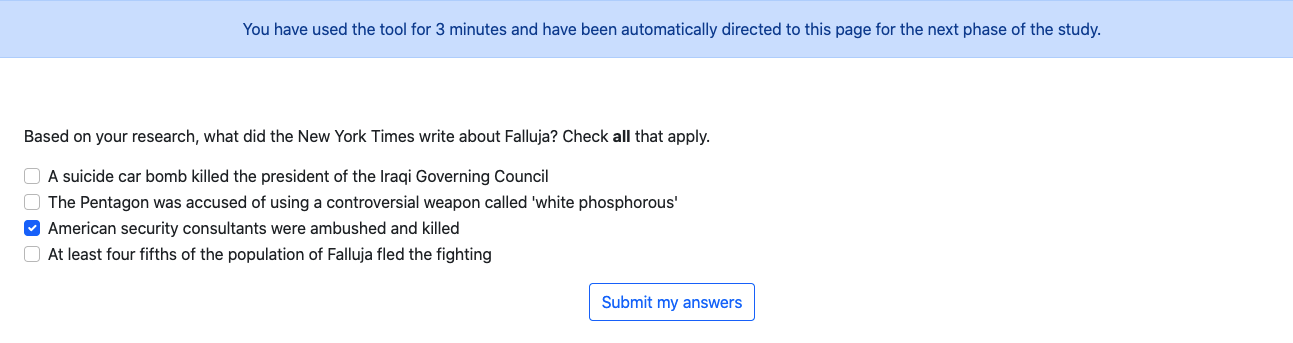}}
\caption{In the our crowd study, participants used this screen to enter answers for the pretest. The only correct answer is shown with a check mark.}
\end{figure}

%% file: packages_and_commands.tex
\newcount\Comments  
\newcommand{\deltaCQengaged}{0.519} 
\newcommand{\deltaCQall}{$0.399$}

\newcommand{\cohensengaged}{0.495} 
\newcommand{\cohensALL}{0.352}  
\newcommand{\nengaged}{62}

\newcommand{\cqNperfectscore}{8}
\newcommand{\irNperfectscore}{4}
\newcommand{\pAll}{0.030} 
\newcommand{\pAttentive}{0.029} 

\newcommand{\totalNmin}{40.3} 
\newcommand{\totalN}{121} 

\usepackage{color}

\usepackage{tikz}
\definecolor{CCPurple}{rgb}{.405, .144, .450} 
\definecolor{darkgreen}{rgb}{.106, .471, .216}
\definecolor{gold}{RGB}{240,205,17}

\newcommand{\Q}{$Q$}
\newcommand{\mentions}{$\mathcal{M}(\Q)$} 
\newcommand{\specificmention}{$i \in \mathcal{M}(Q)$} 

\newcommand{\archive}{$\mathcal{A}$} 
\newcommand{\mentionincontext}{$\mathcal{C}(i)$}
\newcommand{\simplifiedsentence}{ $\mathcal{C}_{s^{\prime}}(i)$}

\newcommand{\kibitz}[2]{\ifnum\Comments=1\textcolor{#1}{#2}\fi}
\newcommand{\ione}{I1}
\newcommand{\itwo}{I2}
\newcommand{\ithree}{I3}
\newcommand{\ifour}{I4}
\newcommand{\ifive}{I5}
\newcommand{\rhone}{H1}
\newcommand{\rhtwo}{H2}
\newcommand{\ours}{\textsc{ClioQuery}}

\newcommand{\burdensome}{mention gathering}

\newcommand{\titleoffset}{-3cm}

\newcommand{\roverview}{R1}
\newcommand{\rcomprehensive}{R2}
\newcommand{\rcontext}{R3}
\newcommand{\rnoconfound}{R4}

\newcommand{\Baselongname}{keyword document search}
\newcommand{\BaselongnameCap}{Keyword document search}

\newcommand{\leftwidth}{1.9cm}
\newcommand{\familypicwidth}{3.9cm}
\newcommand{\familypicwidthPlus}{5.5cm}

\usepackage{multirow}
\usepackage{makecell}
\usepackage{subcaption}
\usepackage{bm}
\usepackage{enumitem}
\usepackage[normalem]{ulem} 

\usepackage{cleveref} 

%% file: samples/abstract.tex
\begin{abstract}
Historians and archivists often find and analyze the occurrences of query words in newspaper archives, to help answer fundamental questions about society.
But much work in text analytics focuses on helping people investigate other textual units, such as events, clusters, ranked documents, entity relationships, or thematic hierarchies.
Informed by a study into the needs of historians and archivists, we thus propose \ours, a text analytics system uniquely organized around the analysis of query words in context.
\ours~applies text simplification techniques from natural language processing to help historians quickly and comprehensively gather and analyze all occurrences of a query word across an archive. 
It also pairs these new NLP methods with more traditional features like linked views and in-text highlighting to help engender trust in summarization techniques.
We evaluate \ours~with two separate user studies, in which historians explain how \ours's novel text simplification features can help facilitate historical research. 
We also evaluate with a separate quantitative comparison study, which shows that \ours~helps crowdworkers find and remember historical information.
Such results suggest possible new directions for text analytics in other query-oriented settings.
\end{abstract}

%% file: samples/introv2.tex
Newspaper archives are fundamental resources for historians, librarians, and social scientists \cite{Chassanoff,allen} because they offer a detailed primary source record of how social processes evolve across time \cite{pierson2004politics}.
For instance, social researchers have used news archives to examine vital questions such as why the United States abolished slavery \cite{foner1995free} and how different jurisdictions slowed the spread of the 1918 flu \cite{jama}.
While historians are known to use archives in different ways (e.g., sequential browsing \cite{allen}), prior work reports that historians often look for ``specific keywords'' \cite[p. 2]{allen} in newspaper corpora.
For instance, scholars in history and social science journals describe tracking down and reviewing occurrences of words like
``William Benbow'' \cite{Putnam},
{``Frances Maule''} \cite{FrancesMaule},
{``watermelon'' }\cite{watermelon},
{``Japanese beetles''} \cite{japanesebeetles},
{``refugee''} \cite{katrinarefugee},
{``Loving''} \cite{Loving}, and
{``race suicide''} \cite{racesuicide}
in news archives
to help answer questions about society.
In this work, we describe such search terms (e.g., ``William Benbow'') as \textbf{queries} and we describe each exact occurrence of a query in an archive as a query \textbf{mention}.
(Section \ref{s:limits_and_future} discusses possible improvements to exact string matching.)
We then describe the task of locating query mentions as \textbf{mention gathering}, and the closely-related task of reviewing and drawing conclusions from query mentions within the context of surrounding text as \textbf{mention analysis}.\footnote{
Using terminology from prior work \cite{pirolli2005sensemaking}, it is possible to interpret mention gathering as a kind of information foraging and mention analysis as a kind of sensemaking.}
We formally define these terms and tasks in Section \ref{s:needs_formal_problem}.

Much prior work in interactive text analytics (Section \ref{s:related}) does not focus on helping people investigate mentions of a query word in context.
Instead, prior systems (designed for different use cases) focus on the analysis of other latent and observable textual units, such as topics \cite{tiara}, events \cite{eventriver}, document metadata \cite{pivotpaths}, clusters \cite{starspire}, interrelated entities \cite{Gorg2013JigsawReflections}, or thematic hierarchies \cite{overview}. Using terminology from Chuang et al.\ \cite{chuangheer}, who articulate best practices for text analytics, because such systems do not use query words in context as their central ``unit of analysis,'' they offer ``visual encodings,'' ``modeling decisions,'' and interactions which are are poorly ``aligned'' to the ``tasks, expectations, and background knowledge'' of historians and archivists. 

This misalignment means that prior text analytics systems have concrete downsides for mention gathering and analysis. For example, some prior systems focused on helping people analyze high-level text units such as temporal trends in word use (e.g., ThemeRiver \cite{ThemeRiver}) do not show query words in underlying text. Similarly, other systems offer only indirect and incomplete access to query words in context, via extraneous mediating abstractions. We detail these limitations in Section \ref{s:related_comparison}.

hes
\input{figures/main_figure}

However, in practice, social researchers such as Shinozuka \cite{japanesebeetles} and others \cite{Putnam, FrancesMaule, Loving, watermelon, katrinarefugee, racesuicide} do not report using specialized text analytics systems.
Instead, these experts describe using traditional {\Baselongname~engines} like ProQuest \cite{proquest} to analyze query words in corpora. 
(We use \textit{corpus} and \textit{archive} interchangeably; we assume the corpus is an archive.)
Traditional \Baselongname~tools return relevance-ranked document lists in response to a free-text query.
Because they are widely used in historical practice \cite{allen, Putnam, Chassanoff, FrancesMaule},
we propose they are baselines for mention gathering and analysis (Section \ref{s:baseline}). 

Yet \Baselongname~tools also have limitations for finding and analyzing query words in context.
First, because almost all words are very rare (a well-known property of text \cite{Zipf49}), any given query will very likely appear only a small number of times within a document.
This means that people reviewing a ranked document list will have to examine many passages within documents that do not directly mention their query term.
While search within document features (e.g.\ \texttt{control + F} in Chrome \cite{chromeF}) can certainly help, gathering and analyzing query words in context using a \Baselongname~system still requires opening each article in its own window or tab,\footnote{Showing a single document in a single window or tab is a common interface pattern. It is employed, for instance, in the Overview \cite{overview} and Jigsaw \cite{ubigsaw} document viewers, and in traditional search user interfaces, which often link to individual documents from a main search engine results page \cite[Section 6.3]{croft2010search}.} locating mentions within the article, reading passages which mention the query, and integrating information from such passages with existing knowledge, before moving on to the next document in the corpus.
This means that people must context switch across stories as they perform a multi-step process to gather and analyze query mentions in context, keeping track of information from one document as they jump to the next (see Figure \ref{f:field_study_loop}).
Navigating between documents is thought to impose cognitive costs in \Baselongname~tools \cite{orientationWhite,fluidLinks}, and context switching across views is thought to impose cognitive costs in visual analytics systems \cite{Baldonado}.

\input{tables/user_overview}

Noting the importance of historical investigation and the limitations of existing tools (for mention gathering and analysis), we propose the \ours\footnote{Clio is a common prefix (e.g., ClioVis \cite{ClioVis}), implying a connection with history.}~text analytics system, to help historians in their work investigating query words in an archive.
Unlike prior tools, which focus on the analysis of other textual units like topics or hierarchies (Section \ref{s:related_comparison}), \ours~is designed and built to help people analyze query words in context, reflecting the needs and practices of historians and archivists.
Creating tools around the ``tasks, expectations and background knowledge" of end users is believed to be a best practice in text analytics \cite{chuangheer}.

We both worked with and studied historians and archivists to prototype \ours.
This process revealed intertwined technical and design requirements for our system (Section \ref{s:needs}).
First, working with historians revealed the importance of comprehensive review in historical research (Section \ref{s:needs_comprehensive}). 
Thus, \ours~includes a novel Document Feed feature, which uses natural language processing (NLP) techniques to show a comprehensive query-focused summary of every single mention of a user's query term across a corpus.
Similarly, because we found that historians require transparency and contextual information to interpret evidence, \ours's~novel Document Feed is presented alongside a more traditional linked full-text Document Viewer, which is designed to quickly and transparently show text from the summary within the context of full-length documents.
Finally, because because temporal analysis is crucial to historians, \ours~also includes an interactive Time Series View to provide an overview of a query through time.
Together, through these and other features (Section \ref{s:system}), \ours~offers a text analytics system organized around the analysis of query mentions in context. 

In total, our work offers the following:
\begin{itemize}
\item \textbf{A synthesis of extensive prior research in text analytics} (Section \ref{s:related}). 
In reviewing prior work on interactive analysis of text across time, we found that many efforts from the NLP, HCI, and Visualization communities focus on offering overviews of corpus contents.
Such overviews might help users formulate queries, but are not designed for the query-oriented tasks of mention gathering and analysis (when the user already knows what to search for).
\item \textbf{An investigation into user needs and requirements} (Section \ref{s:needs}). 
To build our tool, we translated prior research on historians' information-seeking behavior into concrete guidelines for system design. 
We also validated and contextualized prior work by conducting five needfinding interviews with historians and archivists, while gathering feedback on early prototypes.
This process revealed a need for transparency, trustworthiness, context, and comprehensiveness in archival tools, which might 
inform future work on historical search \cite{histdiv,expedition} and text summarization \cite{das2007survey,nenkova2012survey}.
\item \textbf{The \ours~system} (Section \ref{s:system}). \ours~is an open-source, text analytics system designed to help historians find and analyze query mentions in context. The system~combines novel query-focused summarization methods with more traditional features like linked views, in-text highlighting, and time series plots to help experts quickly, comprehensively, and transparently find and review query mentions across an archive. Code for the system is available on GitHub: \texttt{https://github.com/AbeHandler/ClioQuery}.
\item \textbf{An evaluation of specific \ours~features} (Sections \ref{s:usabilitystudy} and \ref{s:qualresults}). 
To test the utility and usability of \ours, we conducted an expert interview study with five social researchers, who used the system to answer a historical question from news archives.
After methodically coding qualitative feedback, we learned that many experts found \ours's skimmable, query-focused summaries useful because they condensed documents to facilitate quick review of query mentions.
We also learned that linking summary text with underlying source documents using in-text highlighting was essential because it offered necessary context for interpreting summary output.
\item \textbf{An evaluation of \ours~in the wild} (Section \ref{s:fieldstudy}). To test \ours~in a realistic setting, we deployed the system in a field study with two historians, who used \ours~to answer questions from their own research.
In comparing experiences with \ours~to prior experiences with \Baselongname~systems, one historian explained how \ours~reduced their reading burden, and another explained how text summarization features facilitated rapid mention gathering and analysis.
\item \textbf{A quantitative comparison with \Baselongname~tools} (Section \ref{s:crowdstudy}). We conducted a quantitative crowd study to directly compare \ours~with baseline \Baselongname~search systems. In the study, we observed that participants who used \ours~to complete a reading comprehension task modeled on a real historical research question correctly answered significantly more reading comprehension questions than participants who completed the same task using a \Baselongname~system.
\end{itemize}

\noindent We conclude by discussing our findings (Section \ref{s:discussion}), reviewing limitations and future work (Section \ref{s:limits_and_future}), and describing possible applications of features and ideas from \ours~in other query-oriented settings, beyond historical research (Section \ref{s:conclusion_cc}).

\input{figures/field_study_nyt_compare}

%% file: figures/main_figure.tex
\begin{figure}[t!]
\centering
\frame{\includegraphics[width=1.0\textwidth]{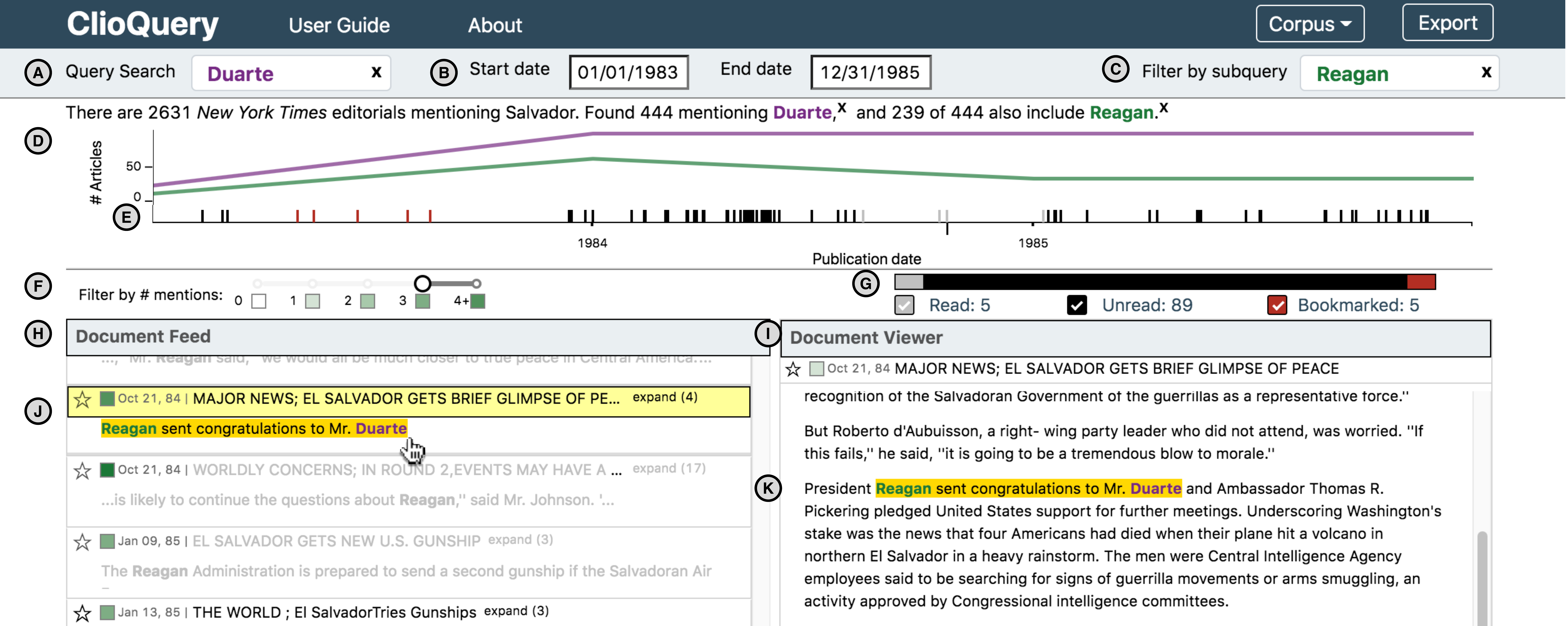}}
\caption[The \ours~interface]{
\ours, an interactive text analytics system for helping historians investigate queries in news archives.
Features (letters A to K) include: (D) a Time Series View showing the frequency of a user's query through time, (H) a linked Document Feed showing a skimmable query-oriented summary of every mention of the query in the corpus, and (I) a linked Document Viewer showing a selected news story, with text from the query-oriented summary highlighted in yellow. 
Section \ref{s:system} describes the full system, explaining each feature.\vspace{.05cm}
}
\label{f:system_cc}
\end{figure}

%% file: tables/user_overview.tex
{
\begin{table}[t!]
\centering
\begin{tabular}{@{}l | cc@{}} \toprule
                       User study   & {Num. participants} & {Total hours} \\ \hline
Needfinding study to guide system design (Sec.\ \ref{s:needs})  & 5                             & 4.5                  \\ 
Expert interview study to evaluate \ours~features (Sec.\ \ref{s:usabilitystudy}) & 5                             & 5                    \\ 
Field study to test \ours~in the wild (Sec.\ \ref{s:fieldstudy}) & 2                             & 5    \\                
Quantitative comparison study~(Sec.\ \ref{s:crowdstudy}) &             \totalN                 & \totalNmin    \\ \bottomrule                 
\end{tabular}
\caption[A summary of three separate user studies with expert historians and archivists]{This work presents four separate user studies with historians and archivists. Section \ref{s:needs} describes institutional approval. Tables in the Appendix describe the backgrounds of participants in greater detail.}\label{t:all_user_studies}
\end{table}
}

%% file: figures/field_study_nyt_compare.tex
\begin{figure}[!ht]
\begin{subfigure}{1\textwidth}
  \centering
\includegraphics[width=.71\linewidth]{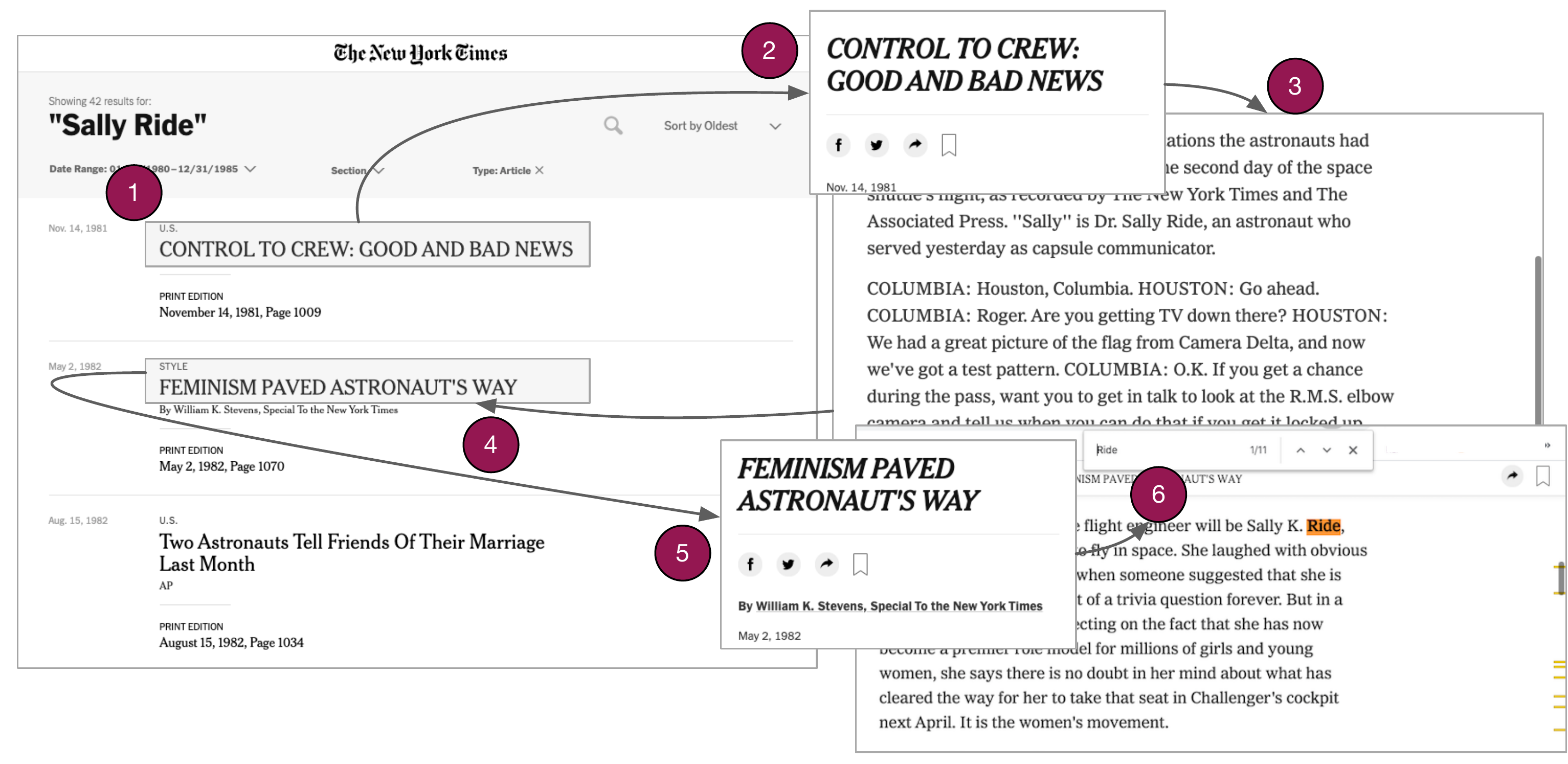}
  \subcaption{
  A user investigates Sally Ride by performing {\burdensome~and analysis} using the \textit{New York Times} web archive \cite{nytwebsite}, a baseline keyword document search interface (Section \ref{s:baseline}). They first (1) click the top headline on the search engine results page (left) in order to (2) {open} a document in a new tab (shown on the right) and then (3) scroll down to {locate} mentions of ``Ride'' in the linked news story. The user {reads} and analyzes these mentions and then (4) context switches to the second document by clicking the second headline on the results page. This (5) opens a new story in a new tab. For this second document, they (6) use a search in document feature \cite{chromeF} (i.e.\ \texttt{Control+F}) to help {locate} mentions of ``Ride'' within the story.
}\label{f:field_study_1} \vspace{.25cm}
\end{subfigure}
\begin{subfigure}{1\textwidth}
  \centering
  \includegraphics[width=.683\linewidth]{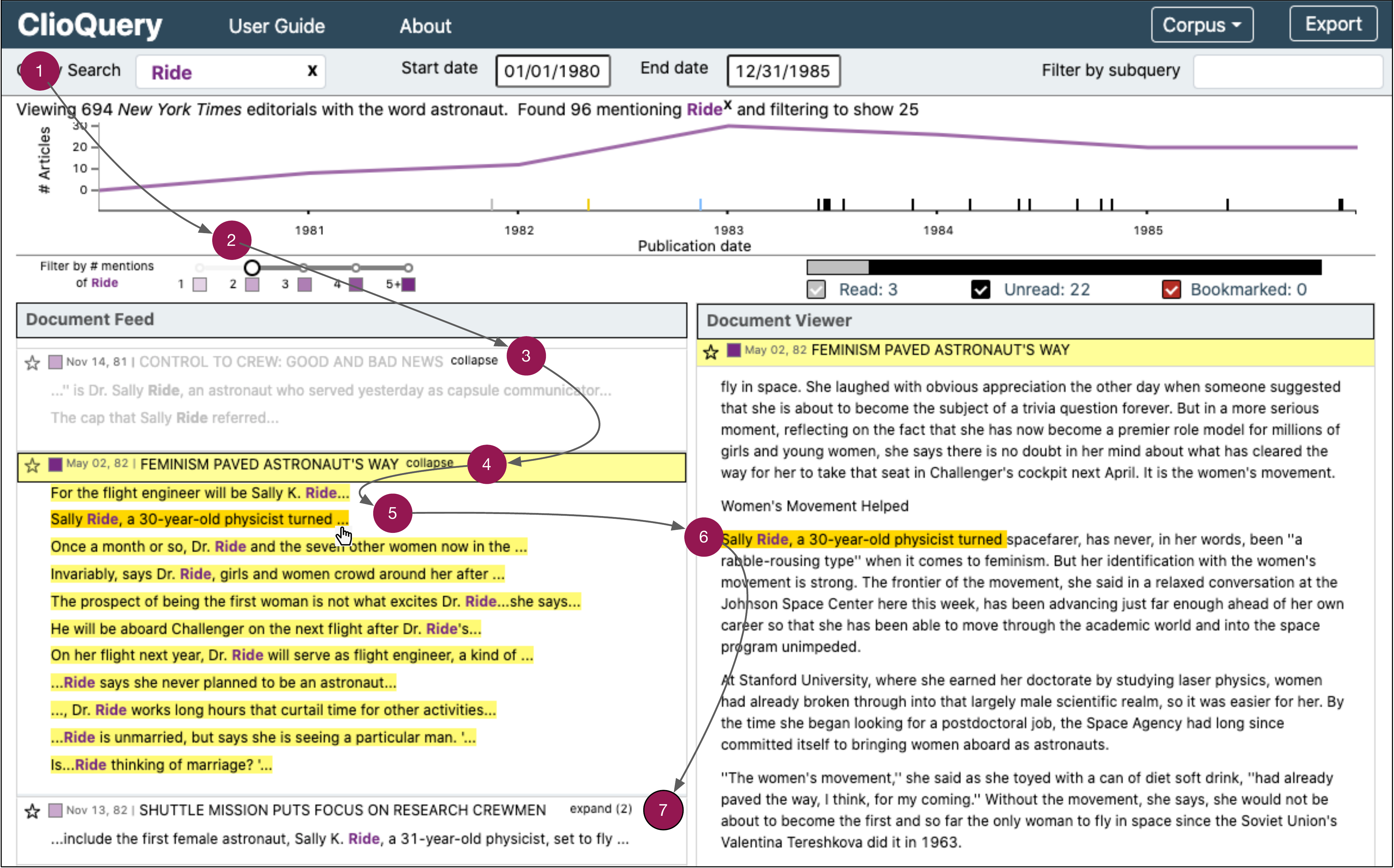}  
 \subcaption{
 A user (1) searches for ``Ride'' using \ours~and (2) sets the filter-by-count slider to limit results to stories with at least two mentions of ``Ride''. The user then clicks the expand/collapse button on two news stories (3 and 4) to review all mentions of Ride from each story in the Document Feed. They then (5) click one shortened sentence mentioning ``Ride'' (6) to read it within the context of the full original document in the linked Document Viewer, with help from automatic in-text highlighting. The user then prepares to (7) click expand to review additional mentions of Ride in the next story.}\label{f:field_study_2}
\end{subfigure}
\caption[A workflow with \ours~and a \Baselongname~tool]{
Reviewing mentions of U.S.\ astronaut Sally Ride in~\textit{The New York Times}, using \ours~(bottom) and a \Baselongname~tool (top). 
\input{figures/field_study_extra}
}\label{f:field_study_loop}
\end{figure}

%% file: figures/field_study_extra.tex
This particular example comes from our field study (Section \ref{s:fieldstudy}), where one historian commented on the advantages of the \ours~interface over a baseline \Baselongname~system.
\textit{ ``What can I do here [with \ours] that I can't do there [with \textit{New York Times} search]?"} she said. 
\textit{``It's exploring this left-hand Document Feed.''}
Where \ours~facilitates quick and comprehensive review of all query mentions, the \Baselongname~tool requires the user to read unnecessary passages and context switch across documents.

%% file: samples/related_work.tex
Historians sometimes gather and analyze mentions of specific query words in archives (Section \ref{s:intro}).
However, much prior work from the HCI, Visualization, and NLP communities focuses on helping people gain high-level overviews of large bodies of text.
We review this overview-oriented literature in Section \ref{s:related_work_overview}.
In Section \ref{s:related_work_search}, we also review another literature on search-based systems, which focus on retrieving text from a corpus in response to a user query.
This search-based approach seems better suited to mention gathering and analysis, as search-based systems can help historians find and review query mentions in a corpus.
Much evidence (Section \ref{s:intro} and Table \ref{t:baselines}) also suggests that search-based systems are central to contemporary historical practice.

In presenting prior work, we emphasize common user interface design patterns \cite{Tidwell}, shared among multiple prior systems.
Interface design patterns are ``concrete bundles of components'' \cite{DesigningInterfaces} that help a user achieve some task. 
We say that \textit{overview design patterns} (Section \ref{s:related_work_overview}) help people survey the contents of archives, and that \textit{search design patterns} (Section \ref{s:related_work_search}) help people query for specific selections from a body of text.
Table \ref{t:baselines} offers a summary of major design patterns from prior work.
Some individual systems (e.g., Expedition \cite{expedition}) may implement both overview and search patterns. Finally, we conclude this Section by discussing \ours~within the context of prior work (Section \ref{s:related_comparison})

\subsection{Overview design patterns}\label{s:related_work_overview}

\subsubsection{\textbf{Word clustering}}\label{s:word_clustering_family}

Because people often can not review every document in a large corpus, many prior text analytics tools such as Termite \cite{termite}, TIARA \cite{tiara}, Overview \cite{overview}, RoseRiver \cite{HierarchicalTopics}, TextFlow \cite{textflow}, Serendip \cite{serendip}, HierarchicalTopics \cite{HierarchicalTopics_Dou}, and ConVisIT \cite{tiisclusterthree} try to suggest overall themes in a body of text by identifying and displaying groups of thematically-related words in a user interface.
We describe this approach as the word clustering design pattern.

Many systems which implement the word clustering pattern are based on prior work from NLP, information retrieval, and text mining, focused on identifying and representing patterns of co-occurring words using methods such as topic models \cite{blei2003latent} and word embeddings \cite{word2vec}.\footnote{The system Themail \cite{themail} clusters words by time, instead of by co-occurrence statistics.
Because this system shows lists of related words (related by time period), we say the system implements word clustering.
Similarly, VisGets shows clusters (of document tags) defined by a user's selection in the interface \cite{visgets}, which we consider to be a form of clustering.}
Researchers in HCI and Visualization extend this work by considering how to present such patterns in a graphical interface;
some systems show changes in cluster patterns across time \cite{tiara,HierarchicalTopics,textflow} (e.g., Figure \ref{f:wordclustering_family}), others do not show time-based topics \cite{termite,overview}. 
Because automatic clusters may not match human mental models of a corpus, one line of work investigates human-in-the-loop techniques, which allow people to modify word clusters through interactions with a GUI \cite{Interactive_Topic_Modeling, tiisclusterone, tiisclustertwo, tiisclusterthree, architext, topiclens, starspire}.

Word clustering has a clear role in historical research.
In query-oriented settings, clustering methods may help people formulate queries they had not considered \cite{Underwood}. 
Moreover, specialized and computationally-oriented digital humanists \cite{poetics_issue} and historians \cite{programminghistorianldatutorial} have used word clusters from topic models for corpus analysis.
Nevertheless, successful application of topic modeling requires specialized knowledge and extensive interpretive effort \cite{Baumer,schmidt2012words}, making this method less accessible to a broader audience of historians.  
Additionally, many historians approach archives looking for mentions of what Allen and Sieczkiewicz describe as ``specific keywords'' \cite{allen} rather than looking to explore word cluster overviews from a topic model API.
Because we design for historians investigating known query terms (Section \ref{s:intro}), we do not employ the word clustering pattern in the \ours~interface.

\subsubsection{\textbf{Textual and visual summaries}}\label{s:textual_summary_family}

Rather than showing lists of related words to offer a corpus overview, a large body of work on text summarization from NLP \cite{das2007survey} instead attempts to create short paragraphs which convey the most ``important'' information in a corpus, by selecting a collection of sentences or sentence fragments from input documents to form an output summary.
(This is sometimes described as extractive summarization \cite{das2007survey} because the output text is extracted from input text.) 
User-facing systems such as Newsblaster \cite{NewsblasterMain} and NSTM \cite{bambrick-etal-2020-nstm} apply this research by showing such textual summaries in a graphical interface.
We say that such tools implement the textual summary design pattern (Figure \ref{f:textual_summary_family}). 
Other closely related work from text visualization considers how to present summary text in specialized visual layouts such as Document Cards \cite{DocumentCards}, Phrase Nets \cite{phrasenet}, or Word Trees \cite{wordtree}.
We say that these interfaces offer structured visual summaries, as they place summary text within some structured visual format (e.g., a directed graph \cite{phrasenet}).

Like word clusters, both traditional text summaries and structured visual summaries do not seem to help with mention gathering and analysis. A user can't turn to these forms of summaries to find and review query mentions because ``important'' sentences selected for inclusion in summary output may or may not contain a given query word.
Moreover, traditional approaches typically do not explain \textit{how} ``important'' information is chosen, which may be important in the history domain (Section \ref{s:discussion_NLP}).

However, two ideas from the text summarization literature may help historians perform mention gathering and analysis.
First, work in query-focused summarization tries to identify the most salient information in a corpus, based on a user's query \cite{nenkova2012survey}.
Historians might use such query-focused summaries to review keywords in text. 
Query-focused summaries which define \textit{all} query mentions as important enough to warrant inclusion in summary output may be especially helpful (see Section \ref{s:needs_comprehensive}). 
Second, work on sentence compression \cite{Knight2000StatisticsBasedS,filippova-strube-2008-dependency,filippova2015sentence} tries to shorten individual sentences by removing words, usually for the purpose of including more (shortened) sentences in a fixed-length summary. 
These methods, or closely-related sentence fusion techniques \cite{barzilay-mckeown-2005-sentence}, might be used to shorten passages containing query terms to help people quickly review many mentions of a query in context.
We apply these two ideas from text summarization in \ours~(see Section \ref{s:feed_and_viewer} and \ref{s:simplification}).

\input{figures/families2}

\subsubsection{\textbf{Time series plot}}\label{s:time_series_family}

Instead of showing text to summarize corpus contents, time series plots present the frequency of words or documents across time to offer a visual (rather than textual) corpus overview. 
This pattern is often implemented in text analysis tools \cite{voyant,twitinfo,diamonds,featurelens} and keyword search systems \cite{expedition, TimeExplorer, newspapers.com}.
Some time series visualizations \cite{Michel176, histdiv, TimeExplorer} show the frequency of a single query term across time (e.g., Figure \ref{f:time_series_plus_family}), often using a line chart. 
Others show the frequency of multiple terms (e.g., highest-count words) using a stacked area chart \cite{ThemeRiver,sotu}, and may not require a user-supplied query. 
While time series plots alone can not be used for mention gathering and analysis (because they do not show underlying text from a corpus), 
such visualizations can hint at important events or changes across documents (e.g., Michel et al.\ \cite{Michel176}).
We thus implement this design pattern in \ours~(Section \ref{s:system_ts}).

\subsection{Search design patterns}\label{s:related_work_search}

\subsubsection{\BaselongnameCap~(baseline)}\label{s:baseline}

Traditional \Baselongname~tools return relevance-ranked lists of documents on a search engine results page (SERP) in response to a free-text query \cite{irbook}.
Because historians often use such tools in practice (Section \ref{s:intro}),
we consider these systems to be baselines for mention gathering and analysis.\footnote{
One strand of humanities scholarship critically investigates how widespread adoption of \Baselongname~tools might be distorting traditional humanistic research \cite{Putnam,FrancesMaule,Underwood}.
}

Although \Baselongname~tools are widely used (Table \ref{t:baselines}), these systems have clear downsides for finding and reviewing query mentions.
First, \Baselongname~systems impose unnecessary burdens from reading and context switching. This is  described in detail in Section \ref{s:intro}. 
Additionally, \Baselongname~systems rank documents according to a computational model of relevance.
This may be undesirable for historians because relevance-ranking introduces opaque algorithmic influence over qualitative conclusions (by guiding people towards particular documents).
Section \ref{s:needs} describes the importance of neutral and comprehensive review in historical research.

\input{tables/baselines}

Ranking aside, \Baselongname~tools may also shape user perceptions of the contents of the individual documents in an archive, through displaying single-document summaries (also called query-biased snippets \cite{querybiased}) on the search engine results page.\footnote{
Google sometimes shows complex results snippets on the SERP, using proprietary techniques. 
Brin and Page briefly mention the need for such ``Result Summarization'' in their original paper \cite[Section 6.1]{pagerank}.
} 
For example, Figure \ref{f:keywordsearch_family} displays three sample single-document summaries, showing what a computer deems to be the most important information from three different search results.
Such single-document summaries may be inappropriate for historical research, as some historians may be skeptical of opaque models which select ``important'' information for their review (search engines try to include keywords in snippets, but do not try to explain summaries \cite[Section 6.3.1]{croft2010search}).
Prototypes shown in the Appendix describe our own experiences attempting to apply similar document summarization techniques for historians without success.

\subsubsection{Multi-document snippet}\label{s:snippets_family}

Where \Baselongname~systems return links to single documents in response to a user query,
other systems return collections of smaller units like paragraphs, sentences, or character spans, which are often drawn from multiple documents (see Figure \ref{f:kwic_family}).
We observe two different implementations of this multi-document snippet design pattern in interactive text analytics.

First, multi-document snippet features can be used in word clustering systems to help people investigate mentions of particular clustered words in context.
For example, TIARA \cite{tiara} allows analysts to review individual words from a cluster in underlying text.
However, because TIARA is designed for showing broad themes rather than for reviewing query mentions, 
it does not comprehensively show all mentions of a given word in its multi-document snippet.
Instead, TIARA chooses some selection of mentions for display, optimizing for diversity \cite[Section 6]{tiara}.
Such curation may introduce unwanted algorithmic bias (Section \ref{s:needs_trust}), because the system chooses some but not all query mentions for display.

Additionally, other text analysis systems which are not necessarily focused on clustering sometimes include keyword-in-context (KWIC) views \cite{voyant,oconnor-2014-mitextexplorer}, showing each mention of a query word (or a selection of such mentions) on its own line of text amid immediately surrounding tokens or characters (e.g., Figure \ref{f:kwic_family}).
While this form of multi-document snippet can be used for mention gathering and analysis, KWIC views have some limitations for historical research.
First, in many cases, historians need to investigate particular query mentions within the context of full documents (Section \ref{s:needs_context}). 
While KWIC views may include links to underlying sources, jumping from KWIC views to documents requires context switching into new windows or tabs to gather and analyze evidence.
We explain why this is undesirable in Section \ref{s:intro}.
Second, KWIC views always show some number of pixels, characters or words immediately surrounding each query mention.
This may result in awkward-sounding or choppy snippets that do not include the most salient information in source sentences;
evidence suggests that people dislike awkward-sounding snippets \cite{ryenwhitesnippets}.
Finally, KWIC views do not offer a way to keep track of which mentions have been reviewed during analysis, which may be important in historical research (Sections \ref{s:needs_comprehensive} and \ref{s:tracking}).
Noting these shortcomings, it is possible to interpret certain \ours~features (Section \ref{s:feed_and_viewer} and \ref{s:documentviewer}) as a particular form of KWIC view, addressing some of these limitations.

\subsection{Situating \ours~withing the broader literature on visual analytics for text}\label{s:related_comparison}

Researchers have proposed many approaches to interactive text analytics \cite{tiara,overview,Gorg2013JigsawReflections,serendip,HierarchicalTopics,chuangheer,termite,tiisclusterone,tiisclustertwo,tiisclusterthree,starspire,pivotpaths,eventriver,jasim2021communitypulse}. 
Within this broad literature, \ours~is unique because it is designed to help people find and analyze all occurrences of a query word across a corpus (see Section \ref{s:needs_formal_problem}).
Using terminology from Chuang et al., \ours~differs from prior work because its central ``unit of analysis'' \cite{chuangheer} is the query word in context; the system's~``visual encodings,'' ``modeling decisions'' \cite{chuangheer}, and user interactions were all designed to help people quickly and comprehensively review mentions of a query word across an archive.
For instance, \ours~includes a textual summary feature, that presents a synopsis of all occurrences of a query word in a corpus (Section \ref{s:feed_and_viewer}).

Focusing on query words in context is a departure from prior work in text analytics, which emphasizes other latent and observable textual units of analysis, such as topics \cite{tiara}, events \cite{eventriver}, document metadata \cite{pivotpaths}, interrelated entities \cite{Gorg2013JigsawReflections}, or thematic hierarchies \cite{overview}.
It is also a departure from \Baselongname~systems \cite{nytwebsite,newspapers.com}, which focus on guiding people to ranked documents  (Section \ref{s:baseline}).

Our atypical decision to design and build a text analytics system for analyzing occurrences of query words across a corpus followed from our systematic investigation into the tasks and expectations of historians and archivists (Section \ref{s:needs}), who often review queries in text.
Chuang et al.'s highly-cited guidance for text analytics \cite{chuangheer} stresses the importance of choosing units of analysis which are best ``aligned'' to the ``tasks, expectations and background knowledge'' of intended users.

Because~\ours~uses query words in context as its central unit of analysis, the system differs from prior tools in several key ways. We highlight the most important differences below.

\subsubsection{\ours~displays query words in underlying text}
Unlike \ours, some prior systems for interactive text analysis are designed for analyzing high-level textual units such as corpus themes or temporal trends.
As a result, these systems sometimes do not allow people to review occurrences of query words in underlying documents. 
For instance, structured visual summaries like the Word Tree \cite{wordtree} and Phrase Net visualizations \cite{phrasenet}, or time series displays like Theme River \cite{ThemeRiver} do not show query words in underlying text. 
Similarly, some text analytics systems such as early versions of Overview \cite{overview} experiment with alternatives to query-based paradigms.\footnote{The Overview authors describe the importance of adding query features in discussing their work \cite{overview,stray}.} 
By contrast, \ours~is designed to help people find and read query words in underlying documents; the tool's central units of analysis (i.e., query words in context) are spans of text from documents in the corpus (see Section \ref{s:needs_formal_problem}).
This design choice was informed by prior work, which emphasizes the importance of displaying underlying text in interactive text analysis (see Section \ref{s:text_seriously}).

\subsubsection{\ours~offers complete access to all query mentions in context, without extraneous mediating abstractions}
Like \ours, some text analysis tools do include features to help people navigate to query words in context.
However, because these systems are chiefly designed for analyzing other textual units (e.g., topics \cite{tiara}, events \cite{eventriver}, or document hierarchies \cite{overview}), they offer only indirect and incomplete access to query words in underlying text, via extraneous mediating abstractions.
By contrast, \ours~is designed to help people directly review all occurrences of a query in a corpus. 

For instance, EventRiver \cite{eventriver} helps people review temporal document clusters, which serve as the system's primary unit of analysis. 
In principle, a historian could use EventRiver to find and analyze query mentions in context by (1) finding document clusters containing a query word $Q$ using the tool's search-by-keyword feature, (2) clicking such clusters, and (3) using the tool's Shoebox and Storyboard features to review those occurrences of \Q~which happen to fall within documents from selected clusters.
However, such a workflow would have two downsides. 
First, the workflow would be \textit{indirect}; the historian would have to navigate through clusters to access query words in context.
Such indirect navigation would force the historian to attend to what Chuang et al.\ describe as ``extraneous information that might confuse or hamper interpretation'' \cite{chuangheer}.
Second, the workflow would be \textit{incomplete}; a historian would have no way to navigate to occurrences of query words which do not happen to fall within algorithmically-defined clusters. 
As we describe in Section \ref{s:needs}, this would likely pose a problem for historians, who need to directly and comprehensively observe all mentions of their query in context, with minimal confounding algorithmic influence.

Our focus on EventRiver merely serves as one illustrative example of a broader phenomenon. 
TIARA's mediating topic abstraction \cite{tiara}, Overview's mediating hierarchy abstraction \cite{overview}, StarSpire's mediating cluster abstraction \cite{starspire}, and Jigsaw's mediating entity abstraction \cite{Gorg2013JigsawReflections} (some queries are not entities, e.g., ``race suicide'' \cite{racesuicide}) would also force historians and archivists to navigate to query mentions in context via similarly confounding and extraneous abstractions.

\subsubsection{\ours~employs query-focused summarization to ease the burden of reading and context switching}
Like \ours, some text analytics systems include a snippet feature which shows words extracted from documents in a corpus.
Examples include TIARA \cite{tiara} Snippets, the
Overview and Footprints Document List \cite{overview,Footprints},
and results snippets from \Baselongname~systems \cite[Chp.\ 6.3]{croft2010search}.
While such snippets are visually-similar to \ours's Document Feed (Section \ref{s:feed_and_viewer}), they differ in important ways which are crucial to the work of historians and archivists.

Most importantly, many existing snippet components select and display only some occurrences of a query in a corpus.
For instance, TIARA's Snippets feature displays a selection of occurrences of a topic word in underlying documents \cite{tiara}, and the Footprints \cite{Footprints} Document List displays the first sentence in a document (regardless of whether the sentence contains a query word).
Similarly, \Baselongname~systems create and display snippets based on heterogeneous criteria \cite[Chp.\ 6]{croft2010search}, rather than display all occurrences of a query word in ranked documents.
Prior systems also do not attempt to help people understand how text is selected for a snippet.
For instance, the Overview Document List \cite{overview} selects and displays keywords based on an opaque clustering algorithm.

These properties make prior snippet features poorly suited to historians, who need to review all occurrences of a query word in a corpus with minimal algorithmic influence. 
Therefore, instead of relying on prior snippet features, a historian who wished to review all occurrences of a query word in a corpus would 
likely have to click through from snippets to underlying documents, 
which are often \cite{tiara, Gorg2013JigsawReflections, nytwebsite} shown in individual windows or tabs.
In Section \ref{s:intro} and Figure \ref{f:field_study_loop}, we explain how this workflow imposes unnecessary reading and context switching costs.
By contrast, \ours's snippet-like Document Feed employs 
novel query-focused summarization techniques in order
to allow historians and archivists to quickly scroll through and examine every single occurrence of a query term in a corpus.
We offer qualitative and quantitative evidence of the importance of this feature in Sections \ref{s:qualresults}, \ref{s:fieldstudy} and \ref{s:crowdstudy}. 

%% file: figures/families2.tex
\begin{figure}[t!]
\captionsetup[subfigure]{}

\begin{tabular}{@{}c@{}}

\begin{subfigure}{\leftwidth}
  {\vspace{\titleoffset}{\textbf{\mbox{Overview}} \\ \textbf{\mbox{patterns}} \\ { \small (Sec.\ \ref{s:related_work_overview})}} }
  \label{f:overview}
\end{subfigure}
\begin{subfigure}{\familypicwidthPlus}
  \centering
  \fbox{\includegraphics[width=\familypicwidth]{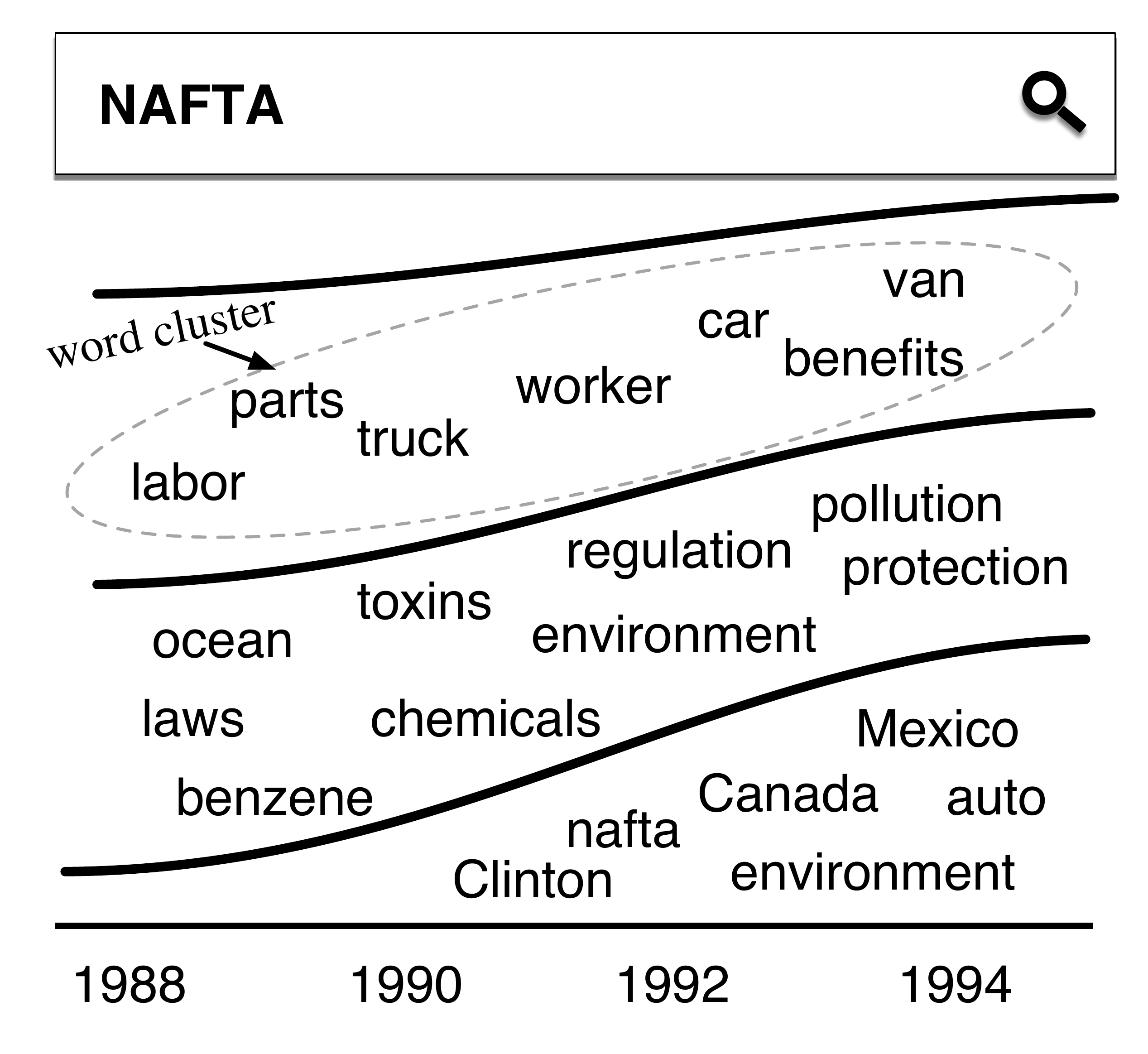}}
  \caption{\hspace{.4cm}{\small Word clustering (Section \ref{s:word_clustering_family})} \\ \centerline{ \footnotesize Examples: \cite{HierarchicalTopics,overview,termite, tiisclusterone, tiisclustertwo, tiisclusterthree}}  }
  \label{f:wordclustering_family}
\end{subfigure} 
\begin{subfigure}{\familypicwidthPlus}
  \centering
  \fbox{\includegraphics[width=\familypicwidth]{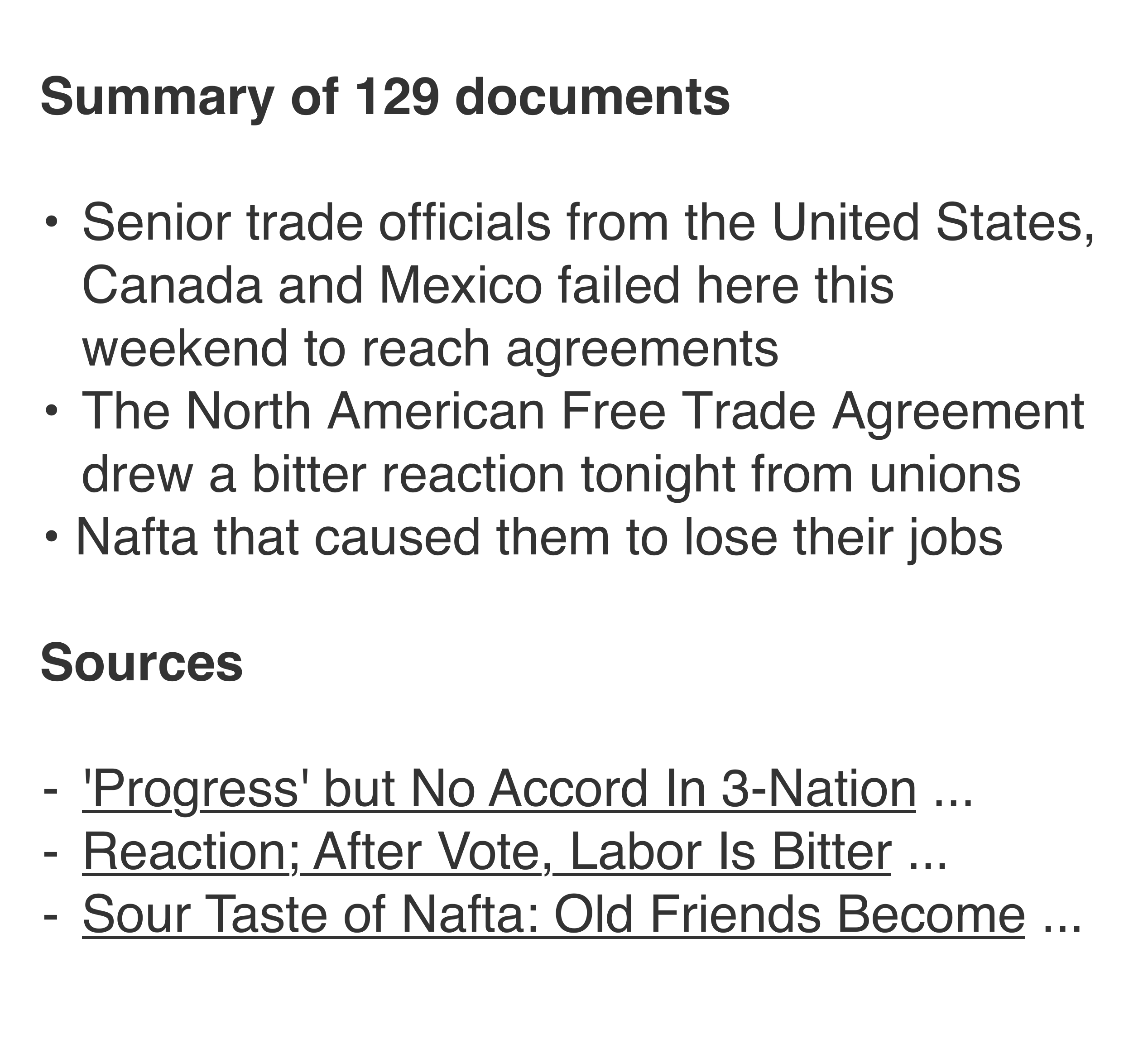}}\hspace*{-0cm}
  \hspace{.7cm}\caption{\hspace{.0cm} {\small Textual \& visual summary (Sec.\ \ref{s:textual_summary_family})} \\ { \centerline{\footnotesize  \centerline {Examples: \cite{NewsblasterMain, summons, bambrick-etal-2020-nstm} }}}}
  \label{f:textual_summary_family}
\end{subfigure} \\ \\
\begin{subfigure}{1.5cm}
  \textbf{}
\end{subfigure}
\begin{subfigure}{\familypicwidthPlus}
  \centering
  \fbox{\includegraphics[width=\familypicwidth]{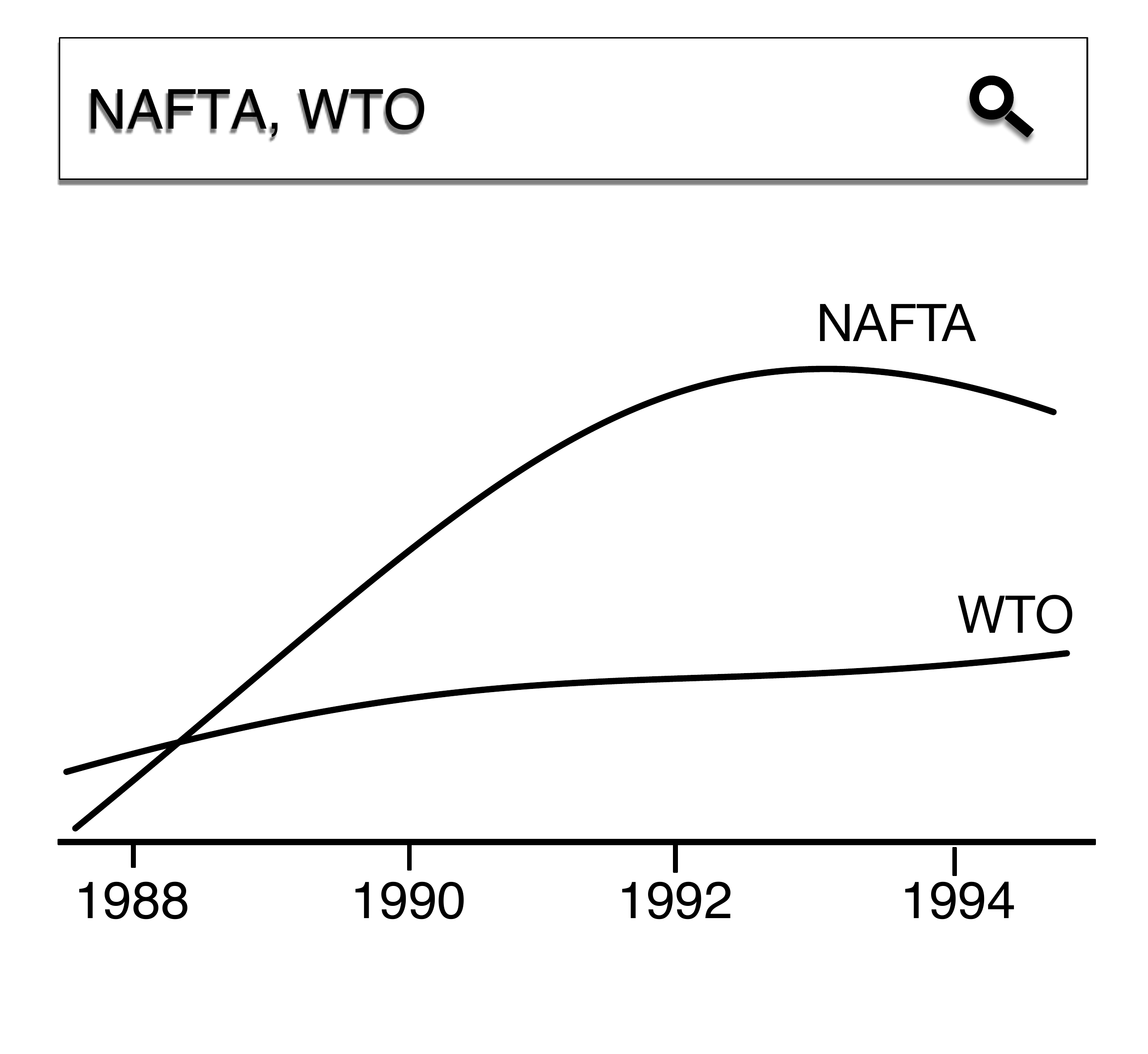}}
  \caption{\hspace{.3cm}{\small Time series plot (Sec.\  \ref{s:time_series_family})} \\ 
  {  \centerline{\footnotesize \hspace{-.4cm} Examples: \cite{ThemeRiver, Michel176, voyant, sotu}}}}
  \label{f:time_series_plus_family}
\end{subfigure} \\ \\    \hline
{} \\ 
\begin{subfigure}{\leftwidth}
  {\vspace{\titleoffset}{\textbf{Search} \\  \textbf{patterns}\\ {\small (Sec.\  \ref{s:related_work_search})}}}
\end{subfigure}
\begin{subfigure}{\familypicwidthPlus}
  \centering
  \fbox{\includegraphics[width=\familypicwidth]{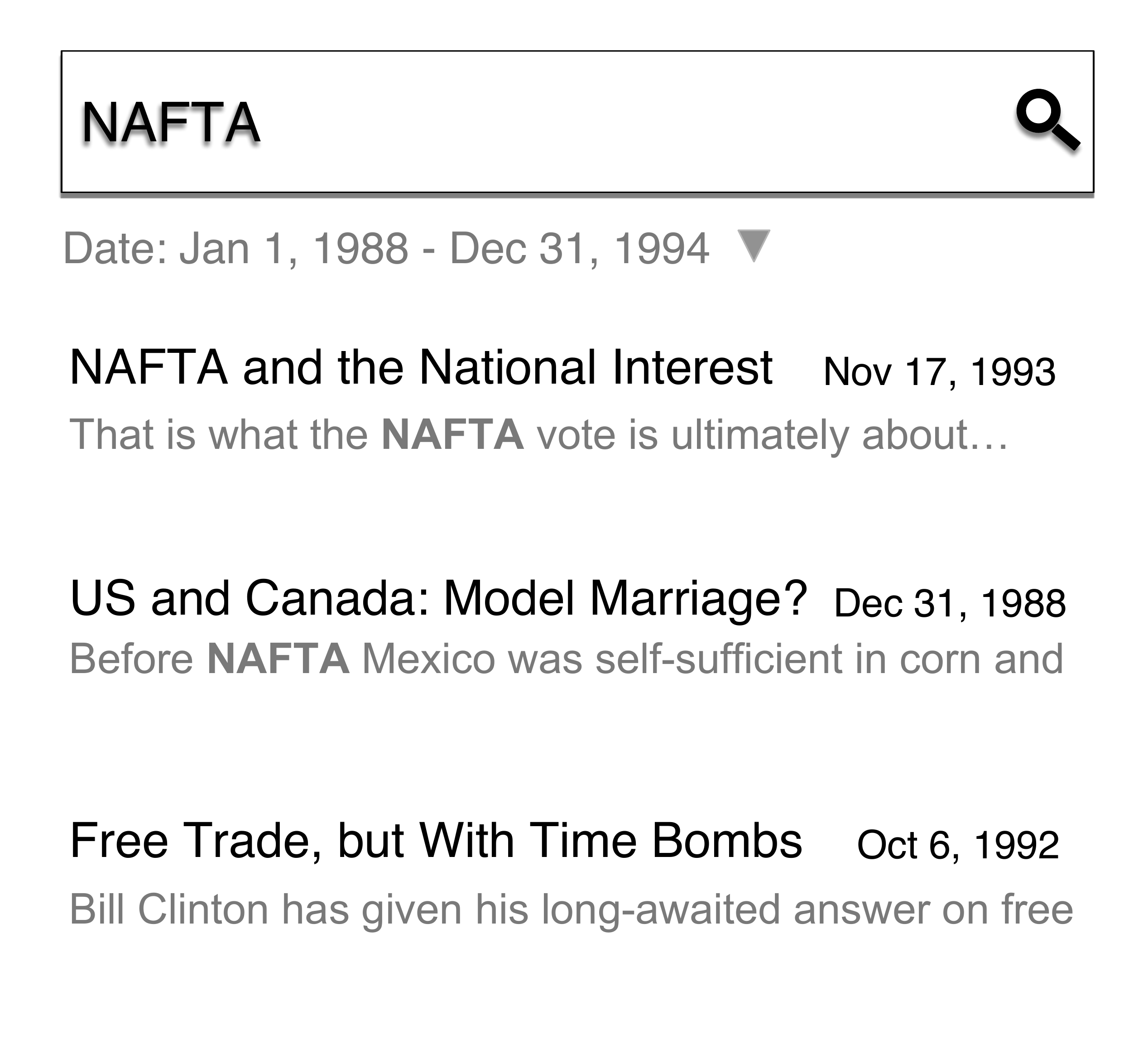}}  
  \caption{ {\hspace{.6cm}\small Keyword search (Sec.\  \ref{s:baseline}) } \\ 
   \centerline{\footnotesize Examples: \cite{nytwebsite, TileBars, hotmap, newspapers.com, lucene, voyant}} }
   \label{f:keywordsearch_family}
\end{subfigure}
\begin{subfigure}{\familypicwidthPlus}
  \centering
  \fbox{\includegraphics[width=\familypicwidth]{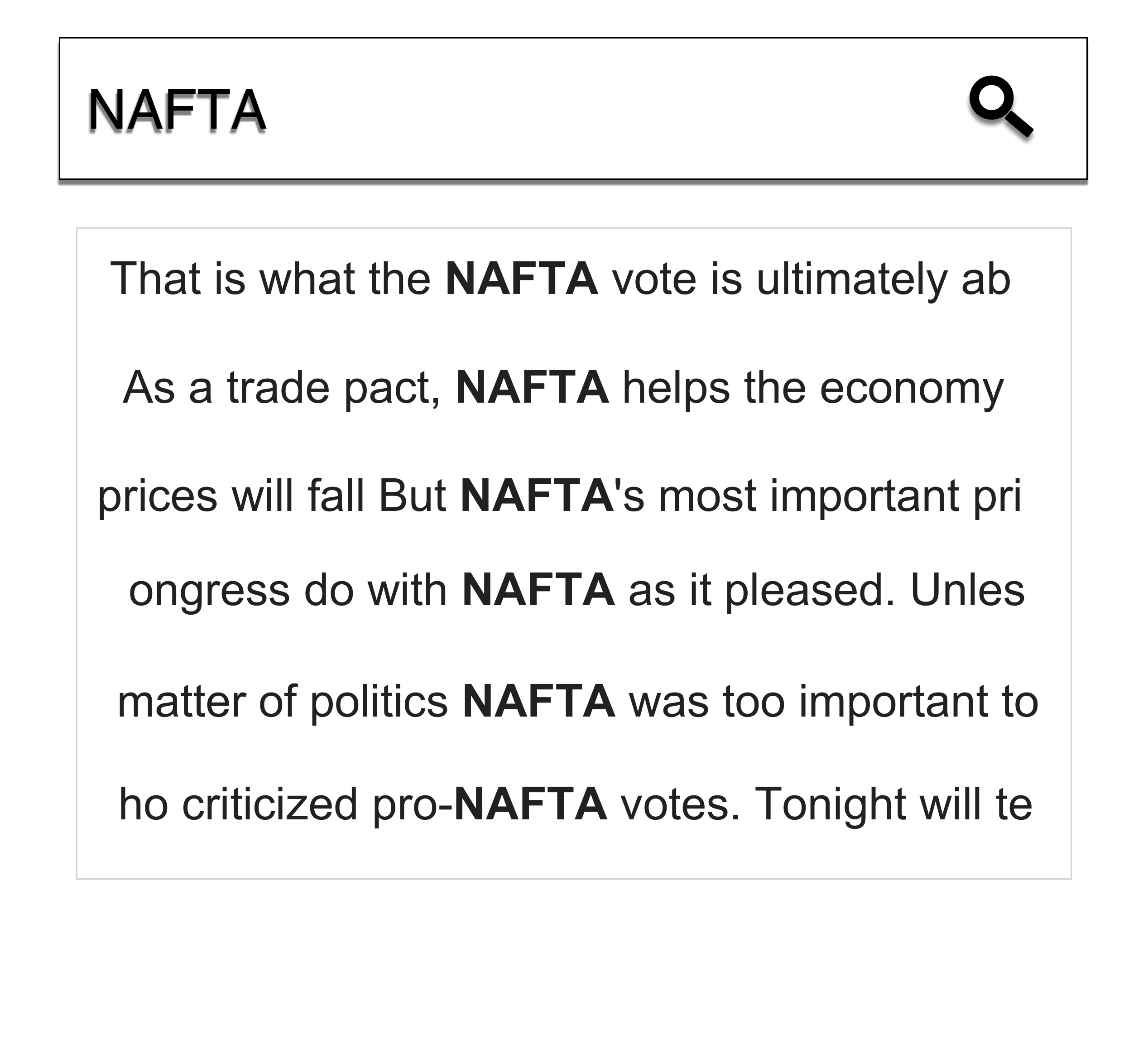}}  
  \caption{ {\hspace{.5cm}\small Multi-doc.\ snippet (Sec.\  \ref{s:snippets_family}) } \\ 
  {  \centerline{ \footnotesize Examples:  \cite{Luhn_kwic, oconnor-2014-mitextexplorer, voyant, themail}} }} 
  \label{f:kwic_family}
\end{subfigure} \\ \\ 
\end{tabular}
\caption[Five major user interface design patterns from prior work]{
\protect\input{figures/families_caption} 
}\label{f:families_all}
\end{figure}


%% file: figures/families_caption.tex
We define five major user interface design patterns from prior work devoted to helping users understand news archives and other corpora (Section \ref{s:related}).
Three of the design patterns focus on helping users gain an overview of archive contents (top two rows).
Two of the design patterns focus on helping the user to search for specific documents or passages from a corpus (bottom row).
Following Tidwell \cite{Tidwell}, this figure presents prototypical wireframes of each design pattern, created by the authors of this work. 
Each example above shows a system presenting results from 129 documents matching the query ``NAFTA'' on a corpus of \textit{New York Times} editorials published between 1988 and 1994.

%% file: tables/baselines.tex
{
\begin{table}[t!]
\centering
\begin{tabular}{rccl}
\toprule
                       & {Relevance ranking} & {Filter by Date} & {Known users}    \\ \midrule
{Chronicling America} & \checkmark                     & \checkmark               & \footnotesize{$\ithree,\ifour,\ifive$,$P4,H1$}          \\
{Newspapers.com}         & \checkmark                     & \checkmark               & \footnotesize{$\ione$,$P1,P4,H2$}     \\
{New York Times Search}       & \checkmark                     & \checkmark              &  \footnotesize{$\ithree,\ifour,\ifive$,$P1$,$P5$,$H1,H2$}  \\
{ProQuest}               & \checkmark                     & \checkmark              & \footnotesize{$\ione$ to $\ifive$, $P1,P3,P4,P5,H1,H2$}      \\  \bottomrule  
\end{tabular}
\caption[Examples of baseline keyword document search systems]{Example baseline keyword document search systems, featuring relevance-ranked search engine results pages and filtering by date. Such features are common in many news archive interfaces \cite{suveyhistorialnewsinterfaces}. Tables in the Appendix provide more details on the backgrounds of known users. 
Above, we use \ione~to \ifive~to indicate all interviewees.
}\label{t:baselines}
\end{table}
}


%% file: samples/user_needs.tex

\subsection{Formalizing historians' current practice of mention gathering and analysis}\label{s:needs_formal_problem}
In Section \ref{s:intro}, we document and informally describe historians' current practice of mention gathering and mention analysis.
We now define this work more formally. 
During mention gathering, a historian investigates a unigram query \Q~in a newspaper archive.
\Q~is a word type and each mention of the query, \specificmention, is a word token.
For instance, a historian might investigate \Q=``Falluja'' by gathering specific mentions of the word ``Falluja'' in individual documents, published on particular dates.
Using $d$ to refer to the text of a specific document and $t$ to refer to its publication date, we can formally define mention gathering as the task of finding all \mentions~in an archive \archive=$\{(d_1, t_2), (d_2, t_2) ... (d_N, t_N)\}$, which is an unordered set of $N$ timestamped documents.

The task of mention analysis consists of manually reviewing one or more query mentions in context.
We use the notation \mentionincontext~to refer to a specific passage showing a particular query mention in context, where \mentionincontext~is a token span. 
For instance, if ``Falluja'' occurs in document $d$, then \mentionincontext~might be a paragraph from $d$ that contains the string ``Falluja.'' We denote this using $\mathcal{C}_{\text{paragraph}}(i)$.
Note that different systems may define \mentionincontext~in different ways.
For example, \Baselongname~systems return whole documents.
Thus a \Baselongname~system defines \mentionincontext~as the whole document $d$ containing \specificmention.
We denote this using $\mathcal{C}_{\text{full doc.}}(i)$.

Having now formally defined historians' current practice of mention gathering and mention analysis, and explained the limitations of baseline tools for these tasks (Section \ref{s:intro}), we now describe an investigation into the needs of historians (Section \ref{s:needs_protocol}) which informs our design requirements for a text analytics system (Section \ref{s:needs_results}).

\subsection{Observing and analyzing needs from heterogeneous data}\label{s:needs_protocol}

We identified user needs by collecting and analyzing two different sources of data, described below.

\subsubsection{Observing needs from existing literature}
First, we studied historians' needs by reviewing a large literature from history, library science, and information science devoted to the systematic study of the digital and non-digital information-seeking behavior of historians.
To identify this literature, we followed citations starting from Allen and Sieczkiewicz's paper ``How Historians use Historical Newspapers'' \cite{allen}, which we first found via a search on Google Scholar. In total, we reviewed and took notes on six prior studies describing surveys and interviews with 1002 historians (shown in a table in the Appendix).
We consider our synthesis of this prior literature to be part of the contribution of our work, as we translate these prior descriptive findings (focused on how historians find information) into actionable design requirements for an interface.
The studies we review are largely unknown in computer science disciplines like NLP, IR, VIS, and HCI.

\subsubsection{Observing needs from interviews and feedback on prototypes} We additionally supplemented, contextualized, and validated existing studies by conducting five of our own one-on-one needfinding interviews with five interviewees (I1 to I5) on Zoom video chat over a period of three months.\footnote{Our needfinding interviews, expert interviews, and field study (Sections \ref{s:needs_protocol}, \ref{s:usabilitystudy}, and \ref{s:fieldstudy} respectively) were approved as exempt from review by our institution's human subjects IRB office.
All participants received a \$50 Amazon gift card for their time.} 
The Appendix describes the backgrounds of interviewees in detail.
All but one interview was 60 minutes long. (We met with \ifour~for 30 minutes, due to limited availability.) Interviews proceeded in two phases. During \textit{Phase A}, in the initial exploratory stage of our work, one researcher from our group interviewed \itwo, \ifour, and \ifive, who we recruited through convenience sampling \cite{given_sage_2008}.
The interviewer asked open-ended, exploratory questions about needs and practices, and solicited feedback on early prototypes. 
The researcher also took detailed notes.
Later, when we better understood how historians find information in archives, we began \textit{Phase B}.
During this phase, the same researcher conducted two one-on-one, video-recorded, semi-structured interviews with \ione~and\ithree, who also provided feedback on later prototypes. 
We recruited \ione~and\ithree~via email outreach.\footnote{
We emailed five PhD students in history at a nearby university. 
Each student expressed interest in media, archives or science in describing their work on their department's web page.
We also emailed all members of the editorial board at a history journal.
We do not list the name of the university or journal to ensure interviewees remain anonymous.
} 
The researcher again took detailed notes.
We include the interview script in supplemental material. 
In total, each of the five interviewees across \textit{Phase A} and \textit{Phase B} reviewed a different iterative prototype.
In the interest of space, we only present feedback on what we consider to be the two most important prototypes, shown in the Appendix.

\subsubsection{Analyzing observations of historians' needs}
Following data collection, one researcher qualitatively analyzed and organized  notes and transcripts to articulate four overall needs, and translate these needs into four corresponding design requirements (described in Section \ref{s:needs_results}).
In general, we found that feedback from needfinding interviews and feedback on early prototypes was very consistent with findings from prior work.
Nevertheless, our own needfinding interviews helped to contextualize and translate prior descriptive findings on historians' information-seeking behaviors into actionable guidelines for system design.

\subsection{Needfinding results and design requirements}\label{s:needs_results}

Following data collection and data analysis, we defined four high-level design requirements (R1-R4), based on four needs.
We describe each requirement below.

\subsubsection{\textbf{R1: A system should show a navigable overview of change over time}}\label{s:exploration}
Prior study of the information-seeking behavior of historians emphasizes the theoretical importance of \textit{``the dimension of time''} \cite{Case} in historical research, and also emphasizes historians' practical need to perform \textit{``searching and narrowing by date''} \cite{allen}. 
In our needfinding interviews, historians and archivists also stressed the theoretical and practical importance of time-based investigation. \textit{``Time is always a historian's first move},''  \ithree~explained. \textit{``It's about change over time as the fundamental thing.''} \ifive~noted: \textit{``Historians are often trying to find articles within a specific date range and about a specific topic} ... \textit{research often starts with a keyword and a date range and a source or list of sources}.'' 
Because historical research involves studying change across time, \itwo~explained how time series plots showing the frequency of query words by time period (see Figure \ref{f:time_series_plus_family}) are often useful for gaining a temporal overview of a corpus. \textit{``Bar charts [or line charts] by time are really helpful},'' \itwo~explained, \textit{``because news has these peaks where a topic becomes important and then dies down.''} Such charts \textit{``help people trace an idea or series of ideas or terminology over time}.''
Observing the centrality of temporal analysis in historical research, we assert a design requirement (R1): a system designed for historical mention gathering and analysis should show some kind of navigable, visual overview of query mentions \mentions~across the time span of a corpus.
Showing such a visual ``overview first'' \cite{HeerShneiderman, The_Eyes_Have_It} is a known best practice in visual analytics.

\subsubsection{\textbf{\rcomprehensive: A system should help people comprehensively review all query mentions in a corpus}}\label{s:needs_comprehensive}

Prior work often emphasizes the importance of gathering comprehensive evidence during historical research.  \textit{``Comprehensiveness is clearly the highest priority in searching a database,''} one study concludes \cite{DaltonCharnigo}, explaining that 70\% of 278 survey respondents would prefer to spend time filtering out irrelevant material than run the risk that relevant material \textit{``might fall through the cracks''} in a limited search. Nevertheless, some historians in prior work acknowledge that truly comprehensive search is an impossible goal. \textit{``I never think I'm going to be able to read every record,''} one reports \cite{DuffJohnson}. \textit{``I'm always creating priority orders of what I think is going to be most useful.''}

Our interviewees similarly emphasized the importance of comprehensiveness in gathering and evaluating historical evidence. \textit{``The most important thing for historical researchers is to be confident that they are being exhaustive,''} said~\ifour. \textit{``I want to know I can be confident I have been able to access everything relevant. Did my search cast a wide enough net?''} \ifour~also praised an early prototype (see Appendix) for displaying a very large number of potentially relevant passages. \textit{``The biggest fear is Type II error,''} he explained. \textit{``In doing searches, am I missing something that is crucial but I don’t know because I never looked?''}
Similarly, \ifive~explained that it is important to \textit{``be as completist as possible''} in historical research. 
\textit{``The thing about historians....they want to be as comprehensive as possible with their topics}.'' Citing the importance of comprehensiveness, \ifive~expressed deep skepticism (see Appendix) about an early prototype which omitted some information to form a summary.
However, like some interviewees in prior work, \itwo~pointed out that truly comprehensive investigation may not be possible. \textit{``Ultimately,''} she noted, \textit{``there is a limit in terms of time and money for any given project.''}
We translate the need for comprehensive archival search into a second design requirement {(\rcomprehensive): a system for mention gathering and analysis should help people comprehensively review all query mentions in a corpus.}
Expressed more formally, a system should help historians easily navigate to and review every single \specificmention~in an archive.

\subsubsection{\textbf{\rcontext: A system should present as much context as possible for any given record in an archive}}\label{s:needs_context}
Prior work emphasizes the importance of context in historical research. \textit{``Building context is the {\normalfont sine qua non [indispensable condition]} of historical research,''} Duff and Johnson write \cite{DuffJohnson}. \textit{``Without it historians are unable to understand or interpret the events or activities they are examining.''} 
In a separate study, another historian explains, \textit{``You can't have the specific facts without the context ... Where an article is in the paper, and what surrounds it, matters.''} 

During our own needfinding interviews, historians and archivists also repeatedly emphasized the importance of contextual information in archive news search.
The job of a historian is to \textit{``put facts in context,''}  \ifive~said. A historian will need to \textit{``contextualize''} facts from a periodical by examining its publishers and audience. Similarly, \ifour~noted that \textit{``as an archivist I do research to give context to collections.''}
Finally, \itwo~stressed the importance of contextualizing evidence in archive search software. \textit{``Who does the New York Times have writing this?''} \itwo~asked, while examining an early \ours~prototype. \textit{``Where does each sentence occur in the document? What section of the newspaper? You need to show more context.''}

Observing the importance of context in historical research, we assert a design requirement (\rcontext): a system for mention gathering and analysis should show each query mention amid as much surrounding context as possible.
Formally, \rcontext~implies that \mentionincontext~should be as large as possible (in token length) for each \specificmention.
However, \rcontext~must be balanced against other requirements, which impose competing demands. In particular, because screen space and human attention are limited resources, if \mentionincontext~is large (e.g., a full document) this will make it harder for a historian to comprehensively review all mentions in a corpus. Balancing a need for context and comprehensiveness is a challenge in designing for historians.

\subsubsection{\textbf{R4: A system should be as transparent, trustworthy and neutral as possible}}\label{s:needs_trust}

Prior studies of the information-seeking behavior of historians underscore the need for trustworthy tools that transparently present digital archival materials in a neutral manner. 
For instance, in one study \cite{DuffCraigCherry}, a historian reports that they prefer original sources because they can trust such sources to be \textit{``accurate, undistorted and complete.''} 
Similarly, in another study \cite{Chassanoff}, another historian explains that direct \textit{``access to the original image of the primary source rather than to a transcribed version''} is important, \textit{``especially when there is no description of what rules they used to transcribe documents.''}  
This historian reports that they do not trust and can not interpret electronic transcription, and thus must rely on direct observation of digitized images to draw conclusions.

In our interviews, historians and archivists similarly described the importance of transparently presenting digitized archives in a neutral manner. \textit{``When I see something that is trying to decide or curate for me that is a worry. That is a red flag},'' \ifour~explained. Similarly, \itwo~added, \textit{``I think the system should be as transparent as possible. I need to distinguish between what some primary source is saying versus what the computer thinks a primary source is saying.''}  \ifive~also cited the importance of transparency and trust in expressing deep skepticism about an early prototype, shown in the Appendix.\footnote{Even as some interviewees stressed the importance of unbiased, transparent and trustworthy presentation of archive evidence, \ithree~reported that, in practice, historical researchers do trust ranked results from keyword document search systems. She explained that many historians might not realize that black-box document rankings from a~\Baselongname~tool will affect conclusions from archival research.}
Because historians frequently expressed commitments to direct and neutral observation of archival evidence, we assert a design requirement {(R4): search software should show evidence in a maximally transparent and trustworthy manner.}
One consequence of R4 is that systems for mention gathering and analysis should not attempt to create a curated summary of the most ``important'' \mentions~in an archive (see Section \ref{s:discussion_NLP}).

%% file: samples/system.tex

\ours~is a visual analytics system designed to support historians in their practice of mention gathering and analysis.
The system is unique within a large literature on text analytics because it is designed to help analyze query mentions in context, which constitute the system's central ``unit of analysis'' (this terminology comes from Chuang et al.\ \cite{chuangheer}). 
By contrast, prior text analytics tools focus on the investigation of other units of analysis like topics \cite{tiara}, events \cite{eventriver}, or thematic hierarchies \cite{overview} (see Section \ref{s:related_comparison}).

\ours's unique focus on query mentions in context reflects our study into the needs and practices of historical researchers, who helped refine early iterative prototypes \cite{GouldClayton}.
Because such historians need to quickly and comprehensively review occurrences of a query word in a large corpus, \ours~uses text simplification techniques (Section \ref{s:simplification}) to create a skimmable summary of a query across an archive. 
Such text simplification techniques and skimmable summaries are unique within the large literature on text analytics, and thus constitute the chief technical contribution of our work.
We hypothesize that this query-focused summarization, along with \ours's~linked views, in-text highlighting, and history tracking features, can help experts quickly, comprehensively, and transparently gather and analyze \mentions, the comprehensive set of all mentions of a query in a news archive.

\subsection{High-level system description}\label{s:highlevell}

The \ours~web interface presents results from a Boolean search \cite[Chapter 1]{irbook}, which returns the unranked set of documents containing one or more mentions~of a unigram query term \Q~in an archive \archive. 
(This notation and terminology is defined in Section \ref{s:needs_formal_problem}; Section \ref{s:limits_and_future} discusses possible extensions to exact string matching.)
When a user enters $Q$ into the search bar at the top of the interface (Figure \ref{f:system_cc}A), \ours~identifies all documents containing $Q$ and presents the documents using three linked views \cite{BujaLinking}. 
First, \ours~includes a \textbf{Time Series View}, showing a graphical overview of the count of documents mentioning the query by year (Figure \ref{f:system_cc}D). 
Second, \ours~includes a \textbf{Document Feed} view, presenting all query mentions from across all documents in a single scrollable window (Figure \ref{f:system_cc}H).
Finally, \ours~includes a \textbf{Document Viewer}, which shows the full text of a single document from the corpus, with individual query mentions from the document highlighted in context (Figure \ref{f:system_cc}I).
\ours~also includes a \textbf{filtering system} to help users narrow the set of query mentions shown in the interface (Figure \ref{f:system_cc}B, C and F), and a \textbf{history tracking system} to automatically monitor and display reading history during comprehensive search (Figure \ref{f:system_cc}G).
All features in the interface also follow a coordinated \textbf{color coding} scheme.
For instance, the user's query word is always displayed using the purple query color {\includegraphics[scale=0.06]{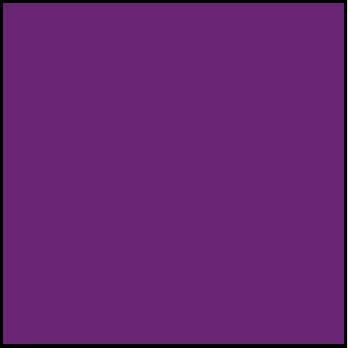}} in the Document Feed and Document Viewer, and the Time Series View also uses a {purple} line to represent query frequency (Figure \ref{f:system_cc}D).
We consider the color-coded bolding of query terms to be one form of \textbf{automatic in-text highlighting} \cite{Handler2016VisualizingTM} throughout the \ours~interface.
Automatic in-text highlighting draws user attention to some word, phrase, or passage in text by automatically setting the text's foreground color, background color or decoration (e.g., bolding).
The Appendix describes our process for selecting a colorblind safe and print-friendly palette. 
It also provides additional engineering details about our implementation of \ours.

\subsection{Overview first: a Time Series View for temporal context (\roverview)}\label{s:system_ts}

Because change across time is central to historical research (\roverview),~\ours~presents a navigable Time Series View (Figure \ref{f:system_cc}D) showing query frequency by year across a corpus.
The component's x-axis represents time (binned by year), and its y-axis represents the annual count of all documents containing the query \Q~published during a given year.
If a user also enters a subquery (Section \ref{s:dont_rank_filter}), \ours's Time Series View also shows the annual count of documents mentioning both the query and subquery.
In Figure \ref{f:system_cc}D, \ours~displays one line showing the count of documents mentioning the query term in the purple query color, and another line showing the count of documents mentioning the subquery term (as well as the query term) in the green subquery color \includegraphics[scale=0.06]{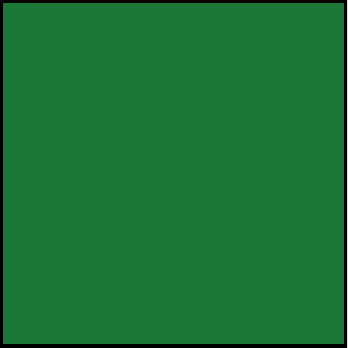}.
\ours's time series plot also shows a single rug point (small vertical line) for each document mentioning the query, just beneath the temporal x-axis (Figure \ref{f:system_cc}E).
Such rug points allow the user to easily preview and navigate to individual news stories; we describe these possible interactions in detail in the Appendix.

\subsection{A Document Feed for comprehensive search (\rcomprehensive)}\label{s:feed_and_viewer}

During needfinding, we found that experts often emphasized the importance of gathering comprehensive evidence (Section \ref{s:needs_comprehensive}), and also often search for specific query terms in news archives (Section \ref{s:intro}).
We thus designed \ours's {Document Feed} to help such users easily gather and analyze the comprehensive set of every single mention of a query term in a collection of news stories (\rcomprehensive).
We assume the user is working with a small corpus (or small set of documents from a larger corpus), where such comprehensive review is possible. 
This assumption is appropriate for our use case; 
for instance, Black reviews roughly 500 documents to analyze the racial history of ``watermelon,'' \cite{watermelon} and MacNamara reviews 605 documents to analyze ``race suicide'' \cite{racesuicide}.

After a user issues a query $Q$, \ours~populates the Document Feed to show a comprehensive, skimmable, summary which includes every single \specificmention~across the corpus (Figure \ref{f:system_cc}J). 
To create the summary, \ours~selects each sentence containing some \specificmention, and then automatically simplifies the sentence (without removing query words) so that historians can quickly read over the mention $i$ in context.
We use the notation \simplifiedsentence~to refer to a specific query mention \specificmention~shown within the context of a sentence $s$ that is simplified to $s^\prime$ for display in the Document Feed.
Section \ref{s:text_simplification_overview} provides details on how \ours~shortens sentences to create~\simplifiedsentence. To the best our knowledge, such text simplification is new to the literature on text analytics.
In presenting each~\simplifiedsentence, \ours~bolds and highlights the query word using the purple query color so that each \simplifiedsentence~is shown in a visually consistent format designed for skimming (Figure \ref{f:compressioncartoon}). 

Note that by default, \ours~displays a single~\simplifiedsentence~from each document beneath the document's headline (The Appendix describes how the sentence is chosen).
To see each \specificmention~from a document within a simplified sentence, the user can click an ``expand'' button (Figure \ref{f:field_study_loop}). 
The user can also click a star to bookmark a document in the red bookmark color \includegraphics[scale=0.06]{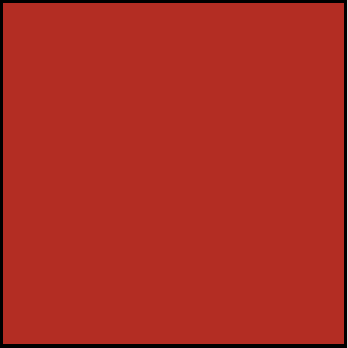}.

\input{figures/lex_summary_overview}
\input{tables/reading_volume}

\ours's Document Feed is designed to directly address two of the limitations of baseline \Baselongname~systems, described in Section \ref{s:intro}.
First, by summarizing documents mentioning $Q$, \ours~is able to fit more query mentions in limited screen space, reducing the need for context switching across individual windows or tabs.
For instance, in Figure \ref{f:system_cc}, the Document Feed saves the user from having to open 239 separate documents during comprehensive review.
Second, by selecting sentences mentioning the query from documents, and removing tokens from those sentences, \ours~reduces the user's reading burden.
For instance, in Figure \ref{f:system_cc}, the user has queried for documents mentioning Reagan and Duarte (in this example, Reagan is a subquery; subqueries are described in detail in Section \ref{s:dont_rank_filter}).
By selecting and simplifying sentences, \ours~removes 87.0\% of the tokens in all documents mentioning these two words (Table \ref{t:tokens}).
We include a detailed description of \ours's text simplification techniques in Section \ref{s:simplification}.

\subsection{A linked Document Viewer for necessary context (\rcontext, \rnoconfound)}\label{s:documentviewer}

Because historians need to evaluate evidence in context without black-box algorithmic influence (\rcontext, \rnoconfound), we anticipated that \ours~users would need to quickly review each \specificmention~from the Document Feed within the context of full underlying news articles.
Therefore \ours's Document Feed is closely linked with a corresponding \textbf{Document Viewer}, which shows the complete text of a single selected document from the corpus (Figure \ref{f:system_cc}I).
The Document Viewer satisfies \rcontext~because it shows each \specificmention~within the context of a full document, denoted $\mathcal{C}_{\text{full doc.}}(i)$.
After a user clicks a shortened sentence \simplifiedsentence~in the Document Feed, the Document Viewer updates to show the entire document containing \simplifiedsentence. 
\ours~also automatically scrolls the document so that the (just clicked) simplified sentence is visible on screen.

\ours~also makes it easy for users to locate simplified sentences, by using {automatic in-text highlighting to further link the Document Feed and Document Viewer}.
Each simplified sentence \simplifiedsentence~from the Document Feed is shown with yellow background highlighting \includegraphics[scale=0.06]{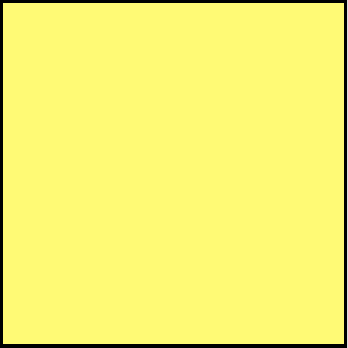} in the document shown in the Document Viewer.
Additionally, if a user hovers over a sentence in the Document Feed or Document Viewer, the sentence is highlighted in dark yellow hover color \includegraphics[scale=0.06]{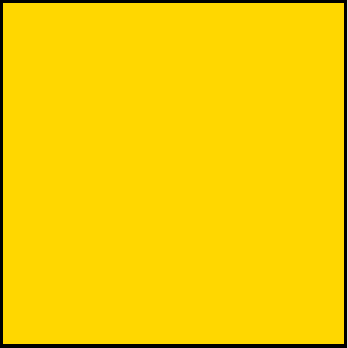} in each component (shown in Figure \ref{f:system_cc}J and \ref{f:system_cc}K).
We hypothesize that linking between shortened text and full documents helps build user trust (\rnoconfound) because it helps experts transparently see and understand how shortened mentions are drawn from underlying text.
This feature is inspired by CommunityClick \cite{CommunityClick}.

\subsection{Color-coded history tracking for systematic review of evidence (\rcomprehensive)}\label{s:tracking}
Some historical researchers emphasize the importance of comprehensively examining all available evidence during research (\rcomprehensive).
To support historians in this work, \ours~keeps track of which documents the analyst clicks in the Document Feed and opens in the Document Viewer.
\ours~also keeps track of bookmarked news stories (Figure \ref{f:system_cc}J), and displays a simple stacked horizontal bar chart (Figure \ref{f:system_cc}G) showing the proportions and counts of read, unread and bookmarked documents. 
The bar chart uses the read \includegraphics[scale=0.06]{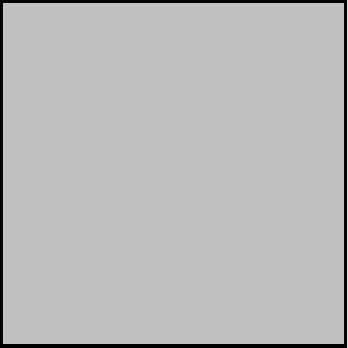}, unread \includegraphics[scale=0.06]{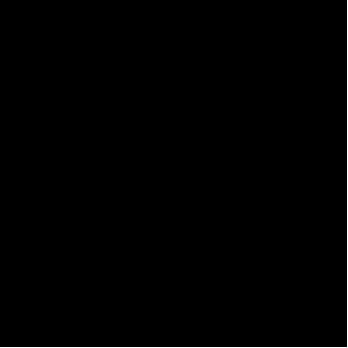}, and bookmarked \includegraphics[scale=0.06]{figures/CCRed.pdf} color scheme employed across the color-coordinated interface.
(\ours~considers all documents to be either read but not bookmarked, unread or bookmarked. We do not allow intersection between these sets.)
For instance, Figure \ref{f:system_cc}G shows 5 read, 89 unread, and 5 bookmarked documents.
The user can click check marks (Figure \ref{f:system_cc}G) to show or hide documents in each category.

\ours's Document Feed and Time Series View use the same color scheme to help users quickly identify opened and unopened documents. Stories that a user has already clicked appear with grey read text in the Document Feed, and their corresponding rug points are shown in grey in the Time Series View.
For instance, in Figure \ref{f:system_cc}, the user has read the story published on Jan. 9, 1985.
The story is greyed out in the Document Feed, and its corresponding rug point is shown in grey beneath the time series plot.
Similarly, there are five red rug points in Figure \ref{f:system_cc}E because the user has bookmarked five documents.

Note that \ours's history tracking is query-dependent; tracking resets each time a user issues a new query (unlike the history tracking mechanism in some prior work \cite[Section 6]{Footprints}).
Such query-dependent tracking is appropriate for \ours~because the system is designed to help historians review all mentions of some specific keyword in a corpus.
We hypothesize that this feature offers experts assurance they have comprehensively reviewed all \specificmention. 
We leave exploration of other forms of history tracking for future work.

\subsection{A filtering system to review many results in a neutral manner (\rnoconfound)}\label{s:dont_rank_filter}

Some prior text analysis systems designed for historians (e.g., Expedition \cite{expedition}) attempt to answer keyword queries by ranking documents to direct users towards most-relevant news articles.
Because such ranked retrieval might introduce unwanted algorithmic influence over the expert search process (\rnoconfound), \ours~responds to queries with Boolean search, which returns the unranked set of all documents containing $Q$.
(The Document Feed shows such documents in chronological order.)
\ours~then allows users to narrow down unranked search results with a filtering system, consisting of three filter controls.

The \textbf{filter-by-date} control selects documents by time period. After users select a start date and end date from date pickers at the top of the interface (Figure \ref{f:system_cc}B), \ours~updates to show only those documents mentioning the query published during the selected interval.
(Historians are often interested in specific time periods; see Section \ref{s:needs}.)
In Figure \ref{f:system_cc}B, the user has filtered to documents published in 1983--1985.

The \textbf{filter-by-subquery} control allows users to select documents that contain some additional word, called a subquery. For instance, after a user queries for the Salvadoran leader ``{Duarte}'' they might wish to further narrow results to understand the relationship between ``{Duarte}'' and his ally U.S.\ President Ronald Reagan.
To investigate, the user can enter the subquery ``{Reagan}'' to select all documents mentioning the word ``{Duarte}'' which also mention the query word ``{Reagan}'' (Figure \ref{f:system_cc}C).
We included this feature because complex Boolean queries are often popular with experts \cite[Section 1.4]{irbook}.
More complex Boolean expressions are possible in future work.

The \textbf{filter-by-count} control filters results based on the the number of times a query term is mentioned in a document.
When a user adjusts the filter-by-count slider to some value $K \in \{1,2,3,4,5\}$ all components of the interface update to show only those documents with $K$ or more \specificmention.
In cases where a user has set a subquery, the filter-by-count control allows the user to select documents which contain the subquery word at least $K$ times.
For instance, in Figure \ref{f:system_cc}F, the user selects documents which mention ``{Reagan}'' at least 3 times.

\ours~also helps users quickly see the count of query terms within documents using square-shaped, query-colored \textbf{count markers}, shown beside each document headline.
Count markers use brightness to encode the count of a query term within a document.
For instance, count markers for documents with more mentions of a query term have a darker purple color than count markers for documents with fewer mentions.
If a user enters a subquery, count markers show the count of the subquery within each document, using shades of the subquery color (as in Figure \ref{f:system_cc}F and \ref{f:system_cc}H).
This feature is inspired by TileBars \cite{TileBars}.

\subsection{Sentence simplification to help summarize a query across a corpus}\label{s:simplification}

\ours~introduces text simplification methods from NLP to the literature on text analytics. We describe these methods below.

\subsubsection{Overview of sentence simplification in \ours}\label{s:text_simplification_overview}

\ours's Document Feed displays a query-focused summary of a user's query and subquery, by first extracting and then simplifying sentences mentioning query (or subquery) words.
To simplify sentences, we turn to sentence compression techniques from the text summarization literature in NLP (introduced in Section \ref{s:textual_summary_family}).
These methods try to summarize naturally-occurring input sentences by removing words, to create shorter and well-formed output sentences which contain the most salient information from the input.
(A well-formed sentence is one that sounds natural, rather than garbled or choppy \cite{sprouseschutzeintro}.)
In particular, we turn to a specific class of sentence compression methods, which can ensure that simplified sentences both (A) fit within limited screen space in a user interface and (B) mention the user's query term or subquery term. 
Such methods are appropriate for \ours~because each line in the Document Feed has a fixed width, and must include some mention of the user's query or subquery.

More concretely, we use a \textit{query-focused clause deletion} \cite{Handler2019HumanAJ,Handler2019Query} method to shorten sentences in cases when a user has entered a query (Section \ref{s:clause_deletion}),
and also use \textit{relationship span extraction} method \cite{handler-oconnor-2018-relational} in cases when a user has entered both a query and subquery (Section \ref{s:rsum_extraction}).
We also employ a final fallback approach, \textit{character windowing},
when it is not possible to shorten a sentence using other techniques (Section \ref{s:clause_deletion}).
In the next sections, we describe each sentence shortening method in greater detail. The Appendix provides additional details on how \ours~chooses between possible sentence shortening methods.\footnote{
In Figure \ref{f:system_cc}, \ours~uses relationship span extraction to shorten and display some sentence from 31 out of 239 documents which mention ``Duarte'' and ``Reagan.'' 
It uses query-focused clause deletion to shorten and display some sentence from 85 documents, and it resorts to character windowing for the remaining 123 documents.\label{footnote_counts}}

\subsubsection{Query-focused clause deletion, and character windowing}\label{s:clause_deletion}

\ours's Document Feed requires shortened sentences that mention $Q$ and fit within available screen space.
We assume that such shortenings should also be well-formed and contain the most salient information from longer source sentences. 
Prior research in IR suggests that users prefer well-formed snippets \cite{ryenwhitesnippets}, and prior work in sentence compression \cite{Knight2000StatisticsBasedS,filippova-altun-2013-overcoming, filippova2015sentence} strives for both well-formedness and salience.
We also assume that methods for constructing shortenings must run with low latency, which is known to be important in user-facing analytics systems \cite{latencyliu}.
Different sentence shortening techniques might optimize for and manage tradeoffs between such requirements. 
But in this work we turn to a simple \textit{query-focused clause deletion} method to meet such criteria, allowing us to focus on how to apply text summarization methods in user interfaces for historical research.

Query-focused clause deletion exploits the fact that natural language sentences are sequences of words, which exhibit hierarchical and nested grammatical structure \cite{bender_linguistic_2013}.
For instance, the sequence ``She swims in the pool'' can be divided into interrelated word groups, with specific grammatical relationships; the words ``in the pool'' form a prepositional phrase that modifies the verb ``swims.''
To represent such linguistic structure, clause deletion employs a dependency parse tree \cite{Nivre2016UniversalDV} grammatical formalism.
A dependency parse is a directed tree graph with one vertex for each word in the sentence, along with a latent root vertex.\footnote{We use the UD (v1) dependency formalism \cite{Nivre2016UniversalDV}; other related formalisms allow for non-tree parses \cite{schuster-manning-2016-enhanced}. Eisenstein \cite[Chapter 11]{eisenstein2019introduction} offers a broad introduction to dependencies. We perform dependency parsing using Stanford CoreNLP \cite{corenlppipeline,chen-manning-2014-fast}.}
Each subtree in the parse corresponds to a constituent subsequence in the sentence. 
The sentence simplification literature sometimes describes such subtrees as \textit{clauses} \cite{filippova-strube-2008-dependency}.
Figure \ref{f:clause_deletion_1} shows an example dependency parse.

Sentence simplification via {clause deletion} shortens sentences by iteratively deleting clauses from a dependency parse.\footnote{Tokens from the remaining tree are then printed in left-to-right order, based on their position in the original sentence.}
Figure \ref{f:clause_deletion} shows how one sentence is shortened by iteratively deleting two clauses.
Unlike sentence compression techniques which consider individual tokens for removal (e.g., Filippova et al.\ \cite{filippova2015sentence}), deleting clauses naturally identifies and removes groups of related words.
For example, a single deletion could remove the prepositional phrase ``after the election,'' or a much longer word group with more modifiers and embedded clauses: ``after the previous election last year, which went poorly.''
Shortening sentences via clause deletion also makes it easy to ensure that output sentences must include $Q$; clauses that contain query mentions are not allowed to be removed during deletion.\footnote{
It is also possible to enforce such query constraints using integer linear programming (ILP). 
However, ILP-based sentence compression techniques (e.g., Clarke and Lapata \ \cite{clarke2008}) are NP-hard and have been shown to be orders of magnitude slower than other iterative approaches to query-focused sentence compression \cite{Handler2019Query}.
}

\input{figures/clause_deletion}

To try and create well-formed output sentences, \ours~turns to prior work on clause deletion \cite[Section 6]{Handler2019HumanAJ}, which has found that in general removing more clauses from an input sentence makes it less likely that the resulting output sentence will be well-formed.
Thus, to shorten an input sentence, \ours's clause deletion first identifies those candidate output shortenings that can be constructed by removing at most $K$ clauses from the input (without removing $Q$), and are also short enough to fit in one line of text within the Document Feed.
Because in practice it is often possible to dramatically shorten an English news sentence by removing only one or two large clauses (for example, a lengthy relative clause, such as ``Reagan met with the envoy \sout{who was sent by the ...}''), \ours~only considers shortenings which can be constructed by removing 
$0 < K \leq 2$ deletions.\footnote{
In addition to encouraging well-formed output, this strict limit ensures low latency for the user.
For a sentence $M$ words long, the worst case for performance is a tree where all words are leaf vertexes, resulting in $M+M(M-1)/2$ possible outputs of $K=1$ or $2$ deletions.
But in typical trees, there are far fewer possible deletions because: (1) the query word and all its ancestors are not allowed to be deleted, (2) after the first deletion of a clause length $C$ (i.e., the size of the deleted subtree) only $M-C$ candidates remain for the second deletion, and (3) if \ours~finds any candidate shortenings using $K$=1, it won't search for candidates using $K$=2, as shortenings which remove fewer clauses are more likely to be well-formed.
We do not consider cases where $K=0$, as most unshortened news sentences are too long to fit within the Document Viewer.}

To try and ensure that output shortenings include the most salient information from input sentences, \ours~then returns the candidate output shortening with the highest tf-idf score \cite{irbook}.
Tf-idf scores are often used in extractive sentence compression
\cite{clarke2008,filippova-strube-2008-dependency} and text summarization \cite{das2007survey} to identify salient information for inclusion in summary output;
this metric identifies words which occur with unusual frequency (relative to the overall corpus), which is an important signal of salience in summarization \cite{sumbasic}.
The Appendix includes details of how we compute td-idf in \ours~to identify words which occur frequently in documents mentioning a query.

In some cases, there is no way to shorten a sentence by removing one or two clauses while ensuring that that output sentence mentions $Q$ and will fit in the Document Feed. 
In these circumstances, \ours~resorts to shortening the sentence by extracting the span of $N$ characters to the left and right of $Q$ in the sentence, where we maximize $N$ under the constraint that the resulting character span will both fit in Document Feed and respect word boundaries.
We use this \textit{character windowing} method only as a last resort because it may cut off syntactic constituents (e.g., show only a portion of a prepositional phrase), which may create awkward-sounding output.
\Cref{footnote_counts} describes how often \ours~uses this fallback, during an example run of \ours. 

In the future, it might be possible to shorten more sentences with query-focused clause deletion by considering candidate output shortenings that are created using more than $K=2$ deletions.
(Prior work on query-focused clause deletion does not yet offer an efficient solution for considering such candidates \cite{Handler2019HumanAJ}.)
Because the number of candidates grows with $K$, developing algorithms which efficiently search over possible outputs or learn greedy deletion policies based on data (e.g., with reinforcement learning) might offer useful starting points.

\subsubsection{Relationship span extraction}\label{s:rsum_extraction}

\ours~users who search for a query term $Q$ can also filter query results by a subquery.
When a user enters both a query and a subquery term, we assume that they are broadly interested in how these two terms are related in the corpus.
For instance, a user might query for the Salvadoran leader  $Q$=``Duarte'' and apply a subquery for the then U.S.\ President ``Reagan,'' in order to understand Duarte's relationship to Reagan (Figure \ref{f:system_cc}). 

\input{figures/rsum/rsum}
To meet this information need, \ours~attempts to simplify long and complex sentences mentioning both the query and subquery terms into short sentences which concisely describe the {relationship} between the query and subquery.
We describe the process of shortening sentences in this manner as \textit{relationship span extraction} because each shortened sentence is a token span (i.e., sequence of tokens) extracted from a longer sentence. 
For instance, in Figure \ref{f:system_cc}, we extract the span ``Reagan sent congratulations to Mr.\ Duarte'' from the longer sentence ``{\color{gray}\sout{President}} Reagan sent congratulations to Mr. Duarte {\color{gray}\sout{and Ambassador Thomas R. Pickering pledged United States support for further meetings}}.''

\ours~relies on a known natural language processing technique to perform relationship span extraction, which is specified in detail in prior literature \cite[Sec.\ 4]{handler-oconnor-2018-relational}.
At a high-level, this method employs logistic regression to determine if an input sentence $s$ containing two input query words can be shortened to express a relationship between those two query words. 
To make this determination, the method first extracts a vector of linguistic features $\bm{x}$ containing information about the query words in the sentence  (e.g., is there a verb token between the query words in $s$?), and then passes the dot product of $\bm{x}$ and a learned weight vector $\bm{\theta}$ through a logistic function $\sigma$. This returns a predicted probability that the token span between the query and subquery will sound natural when removed from the sentence (Figure \ref{f:rsum}).
In our case, the input query words are the user's query and subquery;
we shorten a sentence $s$ to a relationship span if the predicted probability that the span sounds natural is greater than a threshold $T=0.5$.\footnote{We implement with Scikit-learn \cite{Pedregosa:2011:SML:1953048.2078195}. Note that setting a lower threshold $T$ might increase the total number of shortened sentences, at the cost of creating fewer well-formed extractions (and vice versa).}

In our implementation of \ours (as in prior literature \cite{handler-oconnor-2018-relational}), relationship span extraction is supervised using a benchmark corpus from \citet{filippova-altun-2013-overcoming}, consisting of single sentences paired with single-sentence summaries, which are automatically generated from news headlines.
In principle, a technically-oriented \ours~user would be able to retrain relationship span extraction on their own corpus, using the technique from Filippova and Altun to automatically generate training data from headlines in their own news archive.

%% file: figures/lex_summary_overview.tex
\begin{figure}[ht]
\centering
\includegraphics[width=.8\textwidth]{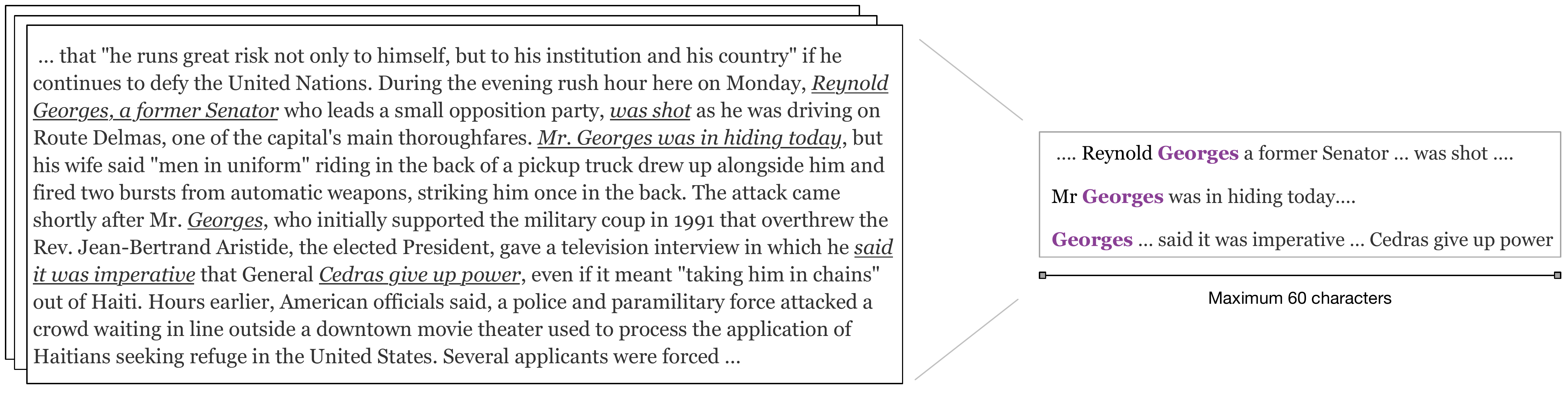}
\caption[Text simplification in \ours's Document Feed]{\ours's Document Feed uses text simplification methods from natural language processing in conjunction with color-coded, automatic, in-text highlighting, in order to summarize query mentions \mentions~in context, in a visually consistent format designed for skimming. 
Here one portion of one document (top of stack, left) containing three mentions of the query term $Q$=``Georges'' has been shortened into a summary of ``Georges'' (right). 
To create this summary, \ours~extracts each of the three sentences mentioning ``Georges'' and simplifies each sentence to render a shorter sentence \simplifiedsentence~that is fewer than 60 characters long. 
In this figure, for illustration, spans of tokens from one source document included in the summary are shown with italicized and underlined text. 
In typical use of \ours~there are often hundreds or thousands of documents containing $Q$ (shown as document stack, above left).}\label{f:compressioncartoon}
\end{figure}

%% file: tables/reading_volume.tex
{

\begin{table}[t!]
\centering
\begin{tabular}{@{}l r@{}}
\toprule
Context       & {Num.\ tokens} \\ \hline 
$\mathcal{C}_{\text{full doc.}} $      & 222,544   \\ 
$\mathcal{C}_{\text{full sent.}}$   & 49,382      \\
$\mathcal{C}_{s^\prime}$  & 28,859    \\  \bottomrule
\end{tabular}
\caption[A quantitative view of how \ours~eases reading burden]{
Total tokens presented to a historian, if each mention of the query ``Reagan'' and subquery ``Duarte'' is shown within the context of a full document ($\mathcal{C}_{\text{full doc.}}$), a full sentence  ($\mathcal{C}_{\text{full sent.}}$) or a shortened sentence ($\mathcal{C}_{s^\prime}$).
By showing shortened sentences, \ours~removes 87.0\% of tokens from all documents containing a query mention, and 41.6\% of tokens from all sentences containing a query mention.
In some cases, removing such tokens will exclude potentially relevant information from the summary; 
in these circumstances, a person would have to find and review such information using the Document Viewer.
Counts come from the example in Figure  \ref{f:system_cc}, using the Salvador corpus (Appendix) and assuming \ours~is shown on a 13-inch screen.
}\label{t:tokens}
\end{table}
}


%% file: figures/clause_deletion.tex
\begin{figure}
  \begin{subfigure}[t]{\textwidth}
    \centering
    \includegraphics[width=.65\textwidth]{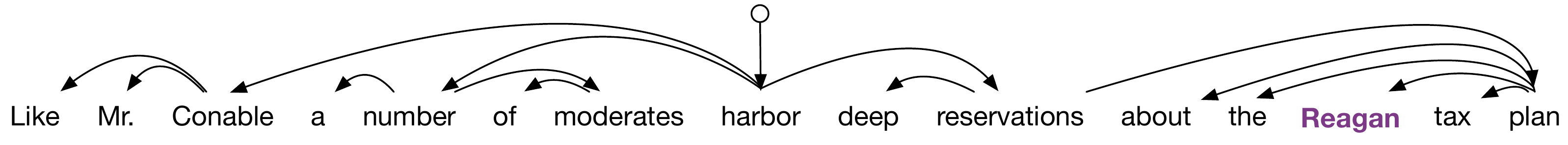}
    \caption[]{An untyped dependency parse tree 
    of an input sentence mentioning $Q$=``Reagan.''}\label{f:clause_deletion_1}
  \end{subfigure}\hfill
  \begin{subfigure}[t]{\textwidth}
    \centering
    \includegraphics[width=.65\textwidth]{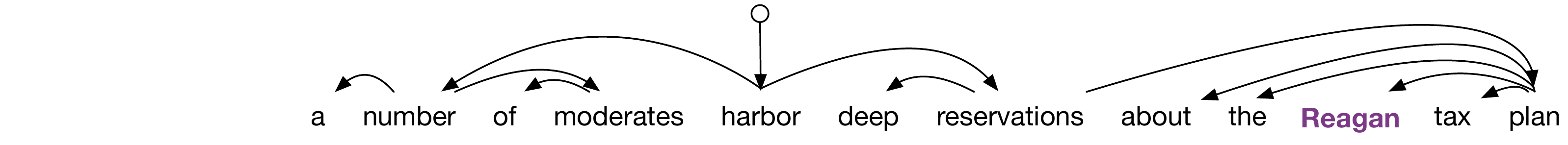}
    \caption[]{To simplify the sentence, \ours~ first removes the clause (subtree) ``Like Mr. Conable.''}\label{f:clause_deletion_2}
  \end{subfigure}
   \begin{subfigure}[t]{\textwidth}
    \centering
    \includegraphics[width=.65\textwidth]{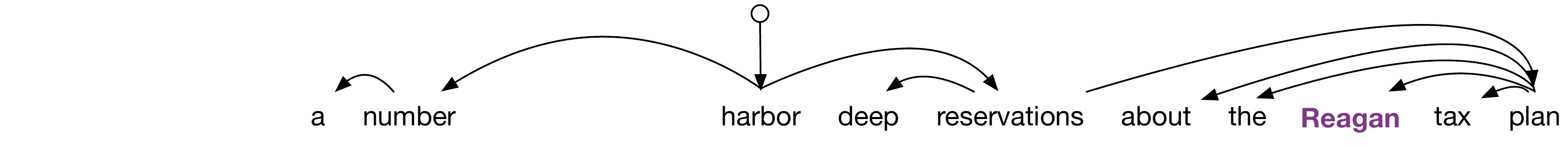}
    \caption[]{In this step, \ours~removes another clause that does not contain $Q$.\vspace{.25 cm}}\label{f:clause_deletion_3}
  \end{subfigure}
  \caption[Sentence simplification via query-focused clause deletion]{Sentence simplification via query-focused clause deletion \cite{Handler2019Query,Handler2019HumanAJ}. \ours~removes two subtrees from a dependency parse across two steps to simplify the input sentence (\ref{f:clause_deletion_1}) into the output sentence (\ref{f:clause_deletion_3}).}\label{f:clause_deletion}
\end{figure}

%% file: figures/rsum/rsum.tex
\begin{figure}[t!]
    \centering
    \includegraphics[width=.9\textwidth]{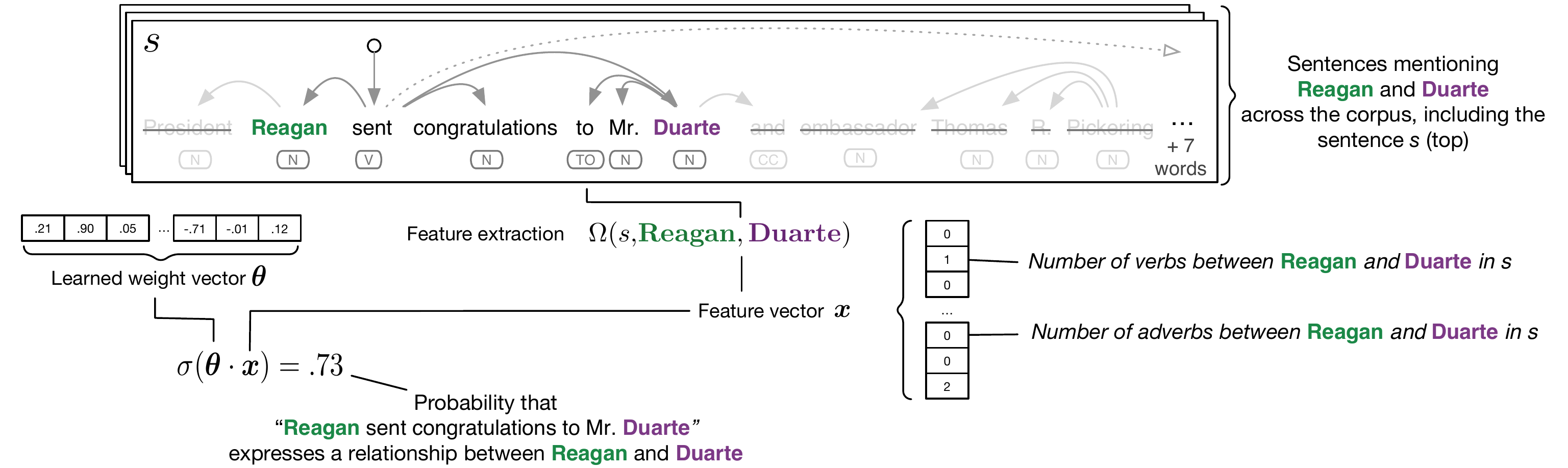}
    \caption[Sentence simplification via relationship span extraction]{Sentence simplification via relationship span extraction \cite{handler-oconnor-2018-relational}.
    This method predicts the probability that the span of tokens between the query and subquery words in a given sentence $s$ will sound natural as a shorter and standalone sentence.
    (If the predicted probability is high, the sentence can likely be shortened to the span of tokens.)
    This figure shows the predicted probability that the token span between the query \textbf{\color{CCPurple}{Duarte}} and the subquery \textbf{\color{darkgreen}{Reagan}} will sound natural as a shortened sentence when extracted from the longer sentence shown in Figure \ref{f:system_cc} (letter K).
    Seven words from $s$ are not shown in the diagram above.}\label{f:rsum}
\end{figure}


%% file: samples/expert_interview_procedure.tex

After analyzing user needs (Section \ref{s:needs}) and designing the \ours~system based on such needs (Section \ref{s:system}), we conducted an interview study over Zoom video chat to test \ours~with historians and archivists, who used the system to investigate questions from news archives.

\subsection{Recruitment, participants and corpora}

We recruited five participants (P1-P5) from two universities in the U.S., by emailing students, faculty, and staff listed on history and library department web pages. 
All participants had advanced degrees (master's or PhD) in history or library science, much like the expected users of our system. 
We provide more details on the backgrounds of participants in the Appendix. Interviewees from our needfinding study (Section \ref{s:needs}) did not participate in our expert interview study, to avoid what Sedlmair et al.\ describe as a potential form of bias \cite{Sedlmair}.
Each participant in the interview study had an established research or curatorial interest in some topic related to late 20th century or early 21st century history, which we express as a single topic word (see Appendix). 
We identified this designated topic word based on each participant's publication record and professional web presence. 
Before each interview, we then loaded \ours~with a corpus of \textit{New York Times} (NYT) editorials\footnote{Social researchers sometimes study editorials to better understand media sources \cite{gay_rights,Lule}.} published between 1987-2007 \cite{SandhausNYT} mentioning the designated topic word.

\subsection{Data collection}\label{s:datacollection}
To administer the study, one researcher from our group conducted five, one-on-one, sixty-minute interviews over Zoom video chat. 
(See supplemental materials for a detailed script.)
During each interview, the researcher asked each participant to brainstorm and then articulate a high-level research question, based on the participant's prior work (10 minutes).
They then introduced the participant to \ours~via a tutorial (7 minutes), and asked them to investigate their research question using \ours~(30 minutes). 
They concluded with a semi-structured interview (13 minutes).
Throughout, the researcher observed and recorded participant reactions and invited participants to think-aloud \cite{thinkaloud} as they used the system.
If a participant offered feedback on some portion of the interface during their investigation (e.g., offered detailed feedback on the Time Series View), the researcher did not ask about this topic again during the semi-structured interview.

\subsection{Thematic coding}\label{s:data_analysis}
The researcher who conducted the interviews analyzed automatic Zoom transcriptions for each of the video recordings, and corrected transcription errors. The researcher then extracted 183 quotes from across five interview transcripts.
Each quote consisted of a few sentences on a focused topic, along with the preceding question or comment to provide context (e.g., a quote might discuss the Document Feed). 
The researcher attempted to extract as many quotes as possible, while excluding irrelevant quotes (e.g., tutorial instructions).

The researcher then developed a codebook of six high-level codes (described in Section \ref{s:qualresults}), by grouping and re-grouping the 183 quotes to identify common themes, much like the codebook-based approach described in Miles, Huberman and Salda\~na \cite[Chp. 4]{miles_qualitative_2014}.\footnote{Miles, Huberman and Salda\~na \cite[Chp. 4]{miles_qualitative_2014} describe assigning codes in two phases; we assign codes in a single phase.}
After each assigning each quote to exactly one of the six codes, the researcher shared the codebook with an undergraduate coder with training and experience in qualitative coding (who was not involved with the development of \ours).
The second coder independently assigned codes to the same quotes, using the codebook. 
The second coder was also invited to add new codes to the codebook if needed, but reported that no new codes were necessary. 
(Thus we did not modify the codebook.)
We include a copy of the codebook in supplemental materials.

Following independent coding of each of the 183 quotes, the two coders met for 1 hour over Zoom video chat to discuss 41 disagreements, and attempted to reach consensus via discussion. In 21 cases, the two coders were able to reach agreement regarding the appropriate code. In 20 cases, the coders determined that disagreement reflected genuine ambiguity in qualitative data, and agreed to disagree.

McDonald et al.\ \cite[Section 2.2]{McDonaldCoding} use the term \emph{reliability} to describe the extent to which coders reach the same result from independent work, and use the term \emph{agreement} to describe the extent to which coders reach consensus after discussion.
Adopting this terminology, we measure the reliability of the two coders by computing a Cohen's $\kappa=0.724$ (using the R \texttt{psych} 2.0 package \cite{psych}), and we measure the agreement of the two coders by computing a $\kappa=0.855$.

%% file: samples/expert_interview_results.tex
Six themes emerged from qualitative coding, described below.

\subsection{\ours~helps with historical sensemaking}\label{s:sensemaking}

While using \ours, each of the five participants formed a question and then collected and interpreted evidence to start to answer that question.
We observed that \ours~helped historians with this investigative process, which Dalton and Charnigo \cite{DaltonCharnigo} describe as historical sensemaking. 
Our observations offered partial validation for our hypothesis that \ours~features can aid historians in their work (Section \ref{s:system}).

For instance, as part of his research, P1 studies \textit{New York Times} news coverage from journalists embedded with United States military units in the Iraqi city of Falluja during the second U.S.--Iraq war. 
From prior study, P1 understood that embedded U.S.\ journalists often published news stories reflecting the perspectives of U.S.\ military leaders. 
But while examining mentions of $Q$=``Falluja'' in \textit{New York Times} editorials using the \ours~interface, P1 expressed surprise when finding a more nuanced perspective from the opinion desk.  \textit{``I didn't see nearly as much of the sort of sensational depiction of Falluja, and the militants in Falluja [in editorials] that I expect from embedded journalists [in news stories],''} he reported. 

Similarly, P2 used~\ours~to find confirming evidence of shifting U.S.\ perspectives towards Robert Mugabe.
As P2 expected, early \textit{New York Times} editorials from the corpus praised $Q$=``Mugabe'' as a liberator, but then began to criticize \textit{``him as a bad statesman, as a tyrant and a dictator.''} 
P3 was likewise able to partially answer a research question with \ours.
She explained that while she had \textit{``a deep knowledge of [women in combat]. I don't have a deep knowledge of what the [NYT] editorial board has to say about it.''} 
Using \ours, she found evidence of editorials using \textit{``the gendered trope that women are supposed to be wives and mothers}.''
P5 also discovered an unexpected connection with musical copyright, while researching a hypothesis surrounding literary copyright. 
\textit{``The parity [with the music service] Napster that's that's really interesting ... That's not something I thought about ... I was thinking ... definitely more in literary items because that's what I deal with.''}

\subsection{\ours~features offer a corpus overview, alongside complementary context}\label{s:features_feedback}

Participants offered detailed feedback on \ours~features during interviews, which often matched our design goals for particular components of the interface.
To begin, three participants reported that \ours's \textbf{Time Series View} offered a useful overview of the entire corpus, by directing their attention to salient time periods. 
P1 said the Time Series View was an \textit{``easy way of visualizing''} corpus trends, and P5 suggested that the Time Series View might be helpful \textit{``when students are kind of in that exploratory phase ... as a way of ... coming up with research questions.''}
P4 offered similar feedback. 
\textit{``I really like this,''} she said. \textit{``This looks really functional and really useful. I like how there is quite a lot of information packed in.''} 

P1, P2 and P5 reported that the \textbf{Document Feed} was a useful feature of \ours~because it helped summarize query mentions. The Document Feed \textit{``condenses all of the essential information and sort of leaves out all the extra stuff},'' said P1. Said P2, \textit{``I found [the Document Feed] useful, especially the expand button. If I click expand I can see a rundown of the mentions right after the title without seeing the article}.'' 
P5 reported using the Document Feed to \textit{``do some ... simple kind of topic modeling in my own head ... just to see if I could pull out any ... themes there.''}
P5 added that, \textit{``having this here [i.e., Document Feed] is really helpful to kind of see what they're talking about.''}
P3 and P4 discussed the Document Feed while describing the importance of context in historical research; we include their feedback on this feature in Section \ref{s:feedback_context}. 

Several participants also reported that the \textbf{Document Viewer} helped during their research. For instance, P3 reported that automatic in-text highlighting in the feed was very helpful. \textit{``I'm a visual person. So I'm looking for the words. I like that they're in purple and green ... the words that you've given me the pop out ... and I can see if it's a pro or con article pretty quickly just from that}.'' P2 said he used the Document Viewer to \textit{``provide detail.''}

P1, P2 and P5 noted that \ours's linked \textbf{Document Feed and Document Viewer} served complementary purposes. 
They described how the Document Feed provided a summary of the query term, while the Document Viewer provided necessary and complementary details. \textit{``You need both [the Document Feed and Viewer]},'' said P2. \textit{``With just the Document Feed I won’t be able to get the full picture of the story. And with just the Document Viewer I will not be able to trace the mentions quite comprehensively and specifically}.'' P2 then added, \textit{``as a researcher, it’s important to see things in detail. If you just conclude from what you see in the Document Feed you are not going to get an objective picture of the context of the story line. But if you see the Document Feed, see the mentions, see what they imply, and then you want to understand the context of the story you are going to get to the Document Viewer}.''  
P1 said, \textit{``I like having both the Document Feed and the Document Viewer side by side. [The Document Viewer helps with] reading for more depth when I want more depth and [the Document Feed] helps with ... quick scans pretty easy.''}  
Similarly, P5 explained, \textit{``I see [the Document Feed and Viewer] working together really well ...
I start by looking at the feed to kind of pick out the articles that would want to kind of dive into deeper and then I go into the Document Viewer.''}

P4 and P5 specifically mentioned that complementary linked views from the Document Feed and Document Viewer helped with \textbf{\burdensome~and analysis}, as compared to a baseline \Baselongname~system. 
\textit{``A lot of a lot of databases that we work with do something similar to this [i.e., the Document Feed]},'' said P5, while describing a search engine results page. 
\textit{``But you often then have to click on the article to go into the article to get to that reading ... here it is nice that it was just kind of next to it and you can scroll through it.''} 
Similarly, P4 described the difficulties of context switching between documents from the Google search engine results page. 
\textit{``Obviously, it's is a time saver,''} she said, comparing \ours~to the \Baselongname~system. 
\textit{``You can tell ... just using the editorials at one newspaper.''}

Two participants relied on \ours's \textbf{filtering system} to investigate their research topics. P1 investigated the \textit{NYT} editorial board's discussion of the query term ``Falluja'' using the filter-by-subquery feature (e.g., searching for ``Falluja'' and ``resistance'' or ``Falluja'' and ``terrorist''). \textit{``It's pretty interesting to me that I get three hits with the words Falluja and resistance and only one with the word terrorist,''} he said. \textit{``That would suggest a certain orientation from the editorial board that will be unexpected}.'' P2 found the filter-by-count feature very helpful. \textit{``Oh, this is good},'' he said, while testing out the slider. \textit{``It gets us through to the most important, the most critical pieces that we want to read}.'' 

\subsection{Some disavow obligation to perform comprehensive review, noting high costs}\label{s:expert_interview_comprehensivenesss}

During needfinding, interviewees emphasized the importance of comprehensively reviewing all available evidence. 
However, to our surprise,  during the expert interview study, P4 explicitly disavowed an obligation to search comprehensively. \textit{``I don't feel like I have an obligation to look at everything,''} she said. \textit{``I have an obligation to get an overview and I think you know, with a completely unscientific measure of, oh, I think I've got enough now.''} 
Similarly, P1 commented that, 
\textit{``I don't think anyone actually does it [search comprehensively].''}
He went on \textit{``A lot of people pretend they do it ... [but] in terms of like visiting archives ... everyone's skimming ... they already know what they're looking for and they're just trying to find it.''}
P2 pointed out that comprehensive manual review was desirable but ultimately had high costs. \textit{``I am not saying we should get rid of personal scrutiny, the way you do it yourself. [But] you want to save time. If you do it [i.e., read] one-by-one it wastes too much time}.'' We discuss ambiguity surrounding comprehensive review in Section \ref{s:discussion}.

\subsection{Context is crucial in historical research, so some are wary of text summarization}\label{s:feedback_context}

Like during needfinding (Section \ref{s:needs_context}), participants often emphasized the importance of context in historical research. For instance, P3 described extensive research to prepare for oral history interviews in order to \textit{``get that context to be able to ask them the questions that I asked them}.''  P2 also reported that context is \textit{``very important''} for historians, as it \textit{``helps you understand why things are what they are.''} 

Some historians' emphasis on context informed their feedback on the Document Feed. While P1, P2 and P5 found the Document Feed useful (Section \ref{s:features_feedback}), P3 and P4 expressed reservations because they felt they needed more context to reach conclusions. P3 took the more extreme position.
\textit{``For me, I don't know if [the Document Feed] is necessary},'' she said. \textit{``As a history scholar, you can't take things out of context. You need to know the bigger context.''} 
On the other hand, P4 reported that she would need more context (i.e., longer extractions from news stories) before the feature would be useful. 
\textit{``The more context I can take in within as compact a time frame and compact a format, but sufficiently informative [the better]''} she said. \textit{``But I think these [shortened sentences in the Document Feed] might have to be longer for that to work}.''  

\subsection[Tradeoffs between neutral review and limited time]{Some users recognize a tradeoff between neutral review and limited time}\label{s:relevance_model_feedback}

During needfinding and prototyping, interviewees often stressed the importance of avoiding possible bias from software in historical research.
But during our expert interview study, P4 reported that she relied on black-box relevance models to direct her attention while searching archives.
\textit{``I do try to use the chronological sorting [when using ProQuest],''} said P4. \textit{``But it is ... too much to wade through. If your corpus is reasonably big then you have to have a relevance kind of algorithm in there.
Otherwise, it's just going to be too frustrating.''} 
P4 also recognized that reliance on ranking introduces confounds. 
\textit{``I think it would be appropriate to make people look at all of the irrelevant stuff},'' she said. 
\textit{``So they realize the algorithm is pulling the relevant stuff for you ...
but you can’t make the search s*** for people just to sort of make that point.''} 

On the other hand, P5 liked how \ours~used filtering to avoid potential bias. \textit{``I think it's better that its just showing everything,''} he explained. \textit{``I prefer having everything there to kind of whittle down ... as opposed to having certain things like cherry-picked ... I guess it's never super clear to me why certain things might be moved to the top of results ... it raises questions about how things are ordered and how they're brought to light.''}

As \ithree~predicted (Section \ref{s:needs_trust}), P1 described relying on the search function of the \textit{New York Times} website \cite{nytwebsite}, without understanding how the site was ranking search results by relevance. 
\textit{``I wasn't super aware of how they were pulling up articles for me ... They rank it in terms of views right?''} he said. 
He added, \textit{``I just don't, you know, have the knowledge of how to navigate these ... search engines well enough.''}
We discuss mixed feedback on algorithmic bias in Section \ref{s:discussion_relevance}.

\subsection{Access, integrity and integration are important to current practices}\label{s:current_practices}

Many participants commented on the importance of access, integrity and integration in describing their current practices with newspaper archives (see also Section \ref{s:limits_and_future}).
{P1} reported gathering news articles on U.S.-Iraqi relations from around the web ``for years'' by using search engines like Google or the \textit{New York Times} website \cite{nytwebsite}, saving these articles to the Internet Archive \cite{InternetArchive}, and then organizing this collection using the software program Omeka \cite{omeka}. This participant pointed out that \ours~\textit{``assumes you have found all the stuff you want to work with,''} which is not true for his current research. 
{P2} said that he had to rely on physical archives of print newspapers in Zimbabwe, which required burdensome international travel. 
{P3} said that she rarely used newspapers in her own research because many newspaper archives are often inaccessible behind paywalls, and {P4} emphasized the need for better optical character recognition technology to improve search over printed newspapers.
{P5} reported that he \textit{``used Zotero a lot''} to store and organize archival sources; he liked that Zotero is open source and integrates with Microsoft Word.

%% file: samples/field_study.tex
In their review of design study methodology, Sedlmair et al.\ emphasize the importance of deploying a designed solution ``in the wild'' to test if new software helps ``real users'' solve ``real problems'' with ``real data'' \cite{Sedlmair}. 
Thus, we deployed \ours~over the web in a field study for two historians, who used the tool to answer questions from their own research. 
Unlike in the expert interview study, during the field study, historians investigated questions over multiple meetings, and tried to reach substantive rather than preliminary conclusions.
We believe that this evaluation offers more realistic but also less uniform feedback than the one-hour expert interviews described in Section \ref{s:usabilitystudy}.

\subsection{Procedure}

We recruited two historians, $\rhone$ and $\rhtwo$ through convenience sampling \cite{given_sage_2008}. 
The Appendix includes details on their backgrounds. $\rhone$ and $\rhtwo$ did not participate in the initial design or development of \ours, to avoid what Sedlmair et al.\ \cite{Sedlmair} describe as a potential source of bias.
During the field study, one member of our research team conducted three one-on-one meetings with each historian over Zoom video chat. 
The first meeting was 30 minutes long and the subsequent meetings were 60 to 70 minutes long, with 1 to 3 weeks between each meeting.
Each meeting in the three meeting sequence had a distinct focus. 
During the first meeting, the researcher presented a tutorial of the software, described the field study process, and invited the historian to describe a question related to their research. 
After the first meeting, a member of our research team gathered the data needed to answer the historian's research question and loaded it into \ours~(the Appendix describes this data gathering). 
During the second meeting, each historian learned to use the \ours~software and performed a preliminary exploration of the data. 
Then, during the final meeting, each historian investigated some specific query by analyzing the comprehensive set of all mentions using the Document Feed and Document Viewer. 
During each meeting, the researcher observed each historian and invited the historian to think aloud \cite{thinkaloud} as they used the system. 
The researcher also asked the historian to describe their findings and explain how \ours~helped or did not help answer their research question.

\subsection{\ours~helps experts investigate by {skimming}, an advantage over baselines}

During the field study, $\rhone$ and $\rhtwo$ each used \ours~to reach substantive historical conclusions, offering additional evidence for our hypothesis (Section \ref{s:system}) that \ours~can help experts answer research questions from news archives.

${\rhone}$ used \ours~to verify a well-known claim from Herman and Chomsky, who argue that for-profit news organizations in the United States shape public opinion towards the interests of political and economic elites \cite{MC}.
To offer evidence for this theory, in their work, Herman and Chomsky assert that \textit{The New York Times} wrote five articles in February and March of 1984 describing the Salvadoran army as a protector of El Salvador's election. To verify this result, $\rhone$ searched a \textit{New York Times} corpus (see Appendix) for the query ``{election}'' and then used the filter-by-date feature to select articles from February and March of 1984. $\rhone$ then used the filter-by-subquery feature to identify those query results which contained the subquery ``{army}.'' 
$\rhone$ then systematically reviewed all 32 matching documents, through what  $\rhone$  described as \textit{``skimming highlighted parts''} in the Document Viewer. By using \ours~in this manner, $\rhone$ said that they were \textit{``able to find what might be the five articles''} Herman and Chomsky used to partially support their conclusions. $\rhone$ explained, \textit{``The tool is great for exactly this.''}

$\rhone$ found \ours's in-text highlighting helpful for their research task, drawing a comparison with a baseline \Baselongname~system (Section \ref{s:baseline}).
\textit{``I like how you have the bold highlighted and colored words in the text itself,''} they said.  \textit{``That is the advantage that this interface has over the New York Times website.''}
$\rhone$ also explained how such highlighting reduced reading burden ( compared to a \Baselongname). 
\textit{``What I need to know is the army described as a protector of the election [in an article]},'' he said.
\textit{``I don't need to read every word of the article to find that out. I can look at the paragraphs where they are describing the army and I see what they are saying in those paragraphs. That is pretty useful.''}

${\rhtwo}$ chose to use \ours~to study how the United States media represented female astronauts Svetlana Savitskaya and Sally Ride in the early 1980s.
(${\rhtwo}$ needed to answer this question to research a planned book.) 
To investigate, $\rhtwo$ used \ours's Document Feed and Document Viewer to review portrayal of Sally Ride in \textit{The New York Times}.
$\rhtwo$ queried for the word {``Ride''} and then scrolled through the Document Feed to skim over mentions of Ride in the 63 matching documents, sometimes also clicking to open individual news stories in the Document Viewer. 
\textit{``I have some hypotheses that I was able to develop very quickly through the experience of using this [system]},'' $\rhtwo$ reported. 
\textit{``One is that Ride was presented to the American public [in The New York Times] ... first as a woman and second as a scientist.''}
$\rhtwo$ asked us to continue to provide access after the study, so she could {continue researching her book using the tool}.
 
\ours's Document Feed was particularly helpful for $\rhtwo$, who found that query-focused summarization offered an advantage over a baseline \Baselongname~system.
Ride was a PhD astrophysicist turned astronaut, and $\rhtwo$ wanted to understand how the media portrayed her scientific credentials.
The Document Feed helped $\rhtwo$ quickly review this information. 
\textit{``[Here] she's called a flight engineer},'' $\rhtwo$ said, pointing to the Document Feed. \textit{``I can see this already [without opening the document].''}  
$\rhtwo$ then scrolled through the Document Feed to find shortened sentences where Sally Ride was described with her academic title (Dr. Ride), and sentences where Ride was described (or not described) as a physicist. $\rhtwo$ explained that she could identify this information \textit{``just doing the quick scan [in the Document Feed].''} 
She went on to explain how she would normally research this question with \textit{The New York Times} archive (by opening and reading individual news stories using a web browser). 
\textit{``The question is},'' she said, {\textit{``what can I do here [with \ours] that I can't do there [i.e.\ on \textit{The New York Times website]?''}}}
$\rhtwo$ continued, \textit{``It's exploring the left hand Document Feed here. This is awesome ... I am liking these short contextual pieces [i.e., shortened sentences].''}
We illustrate this comparison in Figure \ref{f:field_study_loop}; 
by using \ours, $\rhtwo$ was able to easily gather and analyze mentions of Ride across the corpus. 

%% file: samples/crowd_study.tex
\subsection{A crowdsourced historical reading comprehension task}
We designed a crowdsourced historical reading comprehension task to compare \ours~with a \Baselongname~system (IR),
 which we consider to be a baseline tool for historical research (Section \ref{s:intro} and \ref{s:related}).
Our task is designed to reflect historians' common practice of mention gathering and analysis, in which expert social researchers find and review occurrences of a query $Q$ in an archive (Section \ref{s:intro}) in order
to draw conclusions about society.
In our crowdsourced adaptation of this common historical research process, we tasked non-specialists with finding and reviewing occurrences of a query in a newspaper corpus.\footnote{We did not ask participants to take the next step of drawing substantive historical conclusions from their findings, which would have required deep historical knowledge and specialized training.}
We then used reading comprehension questions to measure how well participants performed at finding and reviewing information about the query;
many common educational assessments use similar reading comprehension questions to assess how well people learn information from documents \cite[Chp.\ 7]{reading_comp}.

To ensure we presented an ecologically-valid research prompt, we modeled our crowd task after a real historical question from P2. 
In our interview study (Section \ref{s:usabilitystudy}), P2 used \ours~to investigate if \textit{The New York Times} portrayed the controversial figure Robert Mugabe as a corrupt authoritarian, or as a hero of Zimbabwe's fight for independence. 
In our crowd study, we presented participants with one of two text analytics tools loaded with the same small corpus of 12 \textit{New York Times} editorials mentioning Robert Mugabe, published from January, 2001 to June, 2003. 
We then asked participants to ``find and remember everything the \textit{The New York Times} wrote about Robert Mugabe'' using their tool.
Because historians have only so much time for a given research project (Section \ref{s:needs_comprehensive}), we limited participants to exactly six minutes to conduct their research using their assigned interface. After six minutes, we presented eight true/false reading comprehension questions about \textit{New York Times} coverage of Mugabe, and observed the total number of correct answers for each participant.
Because scoring well on this test of reading comprehension requires finding and reviewing information about a query in a corpus, we believe our crowd task measures how well people perform mention gathering and analysis, using a particular interface.

\subsubsection{Details: reading comprehension questions and scoring}

In order to ensure that our task was as objective and neutral as possible, we created reading questions using the Wikipedia page for Robert Mugabe \cite{wikimugabe}.
Specifically, we used a semi-automated procedure based on tf-idf sentence vectors (described in detail in the Appendix) to identify Mugabe facts from Wikipedia reported in \textit{New York Times} editorials about Mugabe.
We then selected four facts from Wikipedia reported in the 12 editorials in the corpus, and four facts from Wikipedia that were not reported in the 12 editorials. These four facts were reported in some other \textit{New York Times} editorial that was not presented to participants (because the editorial was published before or after January, 2001 to June, 2003). In total, this process created a list of eight total Mugabe facts from Wikipedia.

To evaluate reading comprehension, we presented all eight facts in randomized order, and asked participants to select those facts which appeared in the articles they had reviewed during the task.
To get a perfect score of eight out of eight correct answers without guessing,\footnote{In this task, a participant would have a $.5^8 * 100 = 0.391\%$ chance of correctly guessing all 8 answers.} a participant would have to find and remember the four Mugabe facts reported in the editorials shown during the task, without selecting any of the four facts that were not reported in the editorials. 
The Appendix includes a screenshot showing the reading comprehension questions.

\subsection{Experiment design and experiment details}

We compared \ours~with a \Baselongname~system using a between-subjects experiment design with U.S.\ masters workers recruited via Amazon Mechanical Turk. (Amazon confers the master designation on crowdworkers with a record of success in crowd tasks.)
Participants were randomly assigned to complete the reading comprehension task using either \ours~or a baseline \Baselongname~interface (IR). We then measured the difference in the mean number of total correct answers from workers in each group to determine if \ours~helped people find and remember information about Robert Mugabe (as compared to the IR system).

\subsubsection{Implementation of the IR baseline}
We implemented the IR system using Whoosh, an open-source Python \Baselongname~tool which ranks results using the common BM25 metric.\footnote{Whoosh is similar to other traditional \Baselongname~tools like Lucene. \url{https://whoosh.readthedocs.io/}} The Appendix contains a screenshot of this baseline interface.

To ensure fair comparison, we tuned Whoosh to be most similar to \ours.
Specifically, Whoosh accepts a number of configuration parameters which govern how the system creates snippets on the results page (see Section \ref{s:related_work_search}). 
Because such snippets are similar to the snippets in the \ours~Document Feed, we adjusted the Whoosh snippet parameters so that that Whoosh snippets were as close as possible in length to the shortened sentences in the \ours~Document Feed.
Further details about the tuning procedure are described in the Appendix.
We also adjusted the IR system to use the same font size as \ours.

To minimize possible variation in worker behavior, we hard-coded the IR system (and the \ours~system) to use the query ``Mugabe'' during the experiment.\footnote{During the experiment, we also removed \ours~interface elements which are not relevant for the task, such as the corpus selection control, filter-by-date feature and filter-by-subquery feature.}
To rank the 12 documents in the corpus using the IR system, we loaded the IR tool with all \textit{New York Times} editorials published between 1987 and 2007 that include the word ``Zimbabwe,'' and then queried for ``Mugabe'' while applying a date filter to select only those results which were published from January, 2001 to June, 2003.

We did not implement the IR system using proprietary black-box search tools like Google or \textit{New York Times} search \cite{nytwebsite}.
This is because it is not possible to load open-source versions of such systems with a custom corpus.
Loading a custom corpus is crucial for two reasons. 
(1) Our broader goal is to design an open-source software system that can be deployed and used by historians, who are often interested in corpora that are not published on the web (Section \ref{s:limits_and_future}) and thus inaccessible to Google (or to any other web search engine product).
Comparing to an open-source search engine is thus more appropriate than comparing to a black-box system like Google. 
A historian could use an open-source search engine to index and analyze the documents they collect during their work.
(2) For a controlled experimental comparison of two user interfaces, it's necessary to fix the dataset used for both interfaces. 
In our experiment, users were evaluated based on what information they found (which would change based on the dataset).

\subsubsection{Experiment sequence, experiment pretest, and phases of data collection}

At the start of the experiment, participants in each condition watched a roughly one minute training video describing how to use their randomly assigned interface.
They also read several screens with task instructions, where they entered short phrases into text boxes to confirm they understood the task and were paying attention.
After these preliminaries, participants took an easy pretest which was very similar to the main task (but was about Iraq instead of Zimbabwe).
We describe the details of the pretest in the Appendix.
After the pretest, participants proceeded to the main Robert Mugabe task, conducted their research, and answered the eight reading comprehension questions.
The task concluded with qualitative questions, including questions about the strengths or weaknesses of the assigned interface. 
Qualitative questions are provided in the Appendix.
In total, the task took 20 minutes.

Data collection for the task proceeded in two phases. We first collected data from 18 participants in a small initial pilot. Following the initial pilot, we made adjustments to the task described in the Appendix, including fixing a bug which was favorable to the IR baseline. Following these changes, we collected data from the remaining 103 participants. We decided to include data from the pilot in our analysis because collecting data from crowdworkers was expensive, and because we had trouble recruiting participants from the limited pool of masters workers. (Pooling data is common in settings where data is sparse).
Because we struggled with recruitment, we had to increase task payment from \$2.50 to \$5.00 during data collection. 
We include details in the Appendix. 

\subsubsection{Detecting engaged and not-engaged workers}

In their highly-cited study on crowdsourcing for HCI, \citet{Kittur} emphasize the importance of detecting suspect responses from crowdworkers who may not be completing tasks in good faith.
We thus measure worker engagement in two different ways.
First, because the pretest was designed to be very easy, we assume that participants who did not score perfectly on the pretest were less engaged in the crowd task than other participants.
Second, we also assume that participants who made mistakes on task instructions were also less engaged.  
For instance, some participants made a mistake on task instructions by trying to skip ahead without watching the training video (we logged this and similar behaviors).
In subsequent analysis, we refer to participants who both completed the pretest correctly and did not make any mistakes on task instructions as \textbf{engaged participants}.
Engaged participants are a subset of \textbf{all participants}, the set of all people who completed the task.

\begin{figure}[h]
    \centering
    \subfloat{
       \includegraphics[width=8cm]{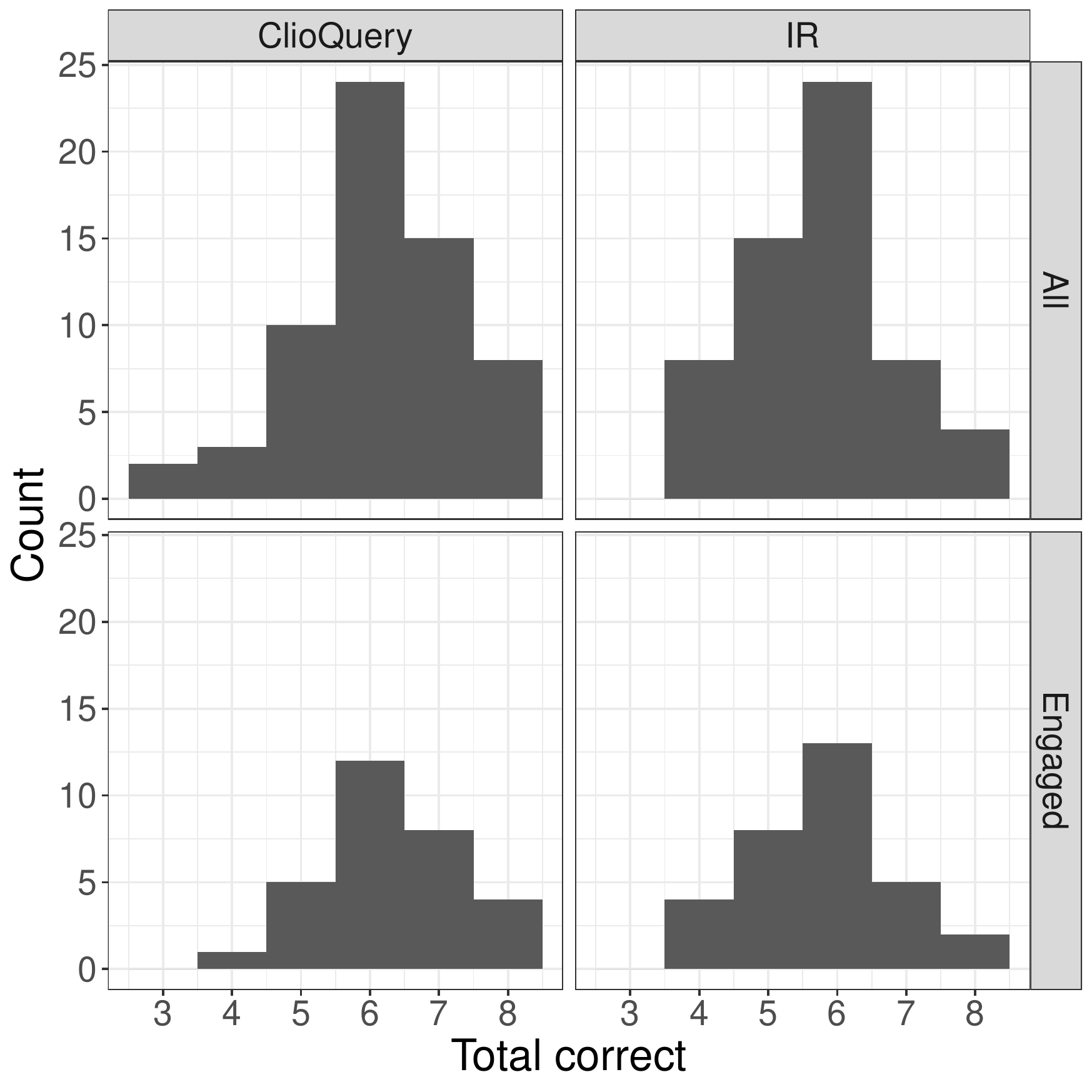}
    }\\
    \subfloat{ 
            \input{tables/crowd_table}
    }
    \caption{Total correct questions by interface, among all participants, and among engaged participants. A star* indicates a significant difference between the means of the \ours~and IR groups.}\label{f:crowdresults}
\end{figure}

\subsection{Results and analysis}

We found that participants assigned to complete the historical reading comprehension task with \ours~averaged more total correct answers than participants assigned to complete the same task with the IR system (Figure \ref{f:crowdresults}). 
Among the \nengaged~engaged participants,
workers in the \ours~group averaged \deltaCQengaged~more correct answers than workers in the IR group (Cohen's $d$=\cohensengaged).\footnote{Computed with v0.8.1 of the \texttt{effsize} package in R.}
\ours's effect was weaker among all participants, where \ours~workers averaged \deltaCQall~more correct answers than IR workers (Cohen's $d$=\cohensALL). 
We hypothesize that this weaker effect may be due to inattention among non-engaged participants, which may have introduced data collection noise.
For instance, participants who did not read task instructions carefully or who failed the pretest may have been more inclined to guess on the Mugabe task.\footnote{The number of workers in each group is not exactly equal. This is common in crowdsourced settings, where some workers may not finish a task. For instance, we used an alert to ask workers attempting to complete our task on a phone rather than a computer to not proceed with the survey.}

\begin{figure}[h]
    \centering
    \includegraphics[width=8cm]{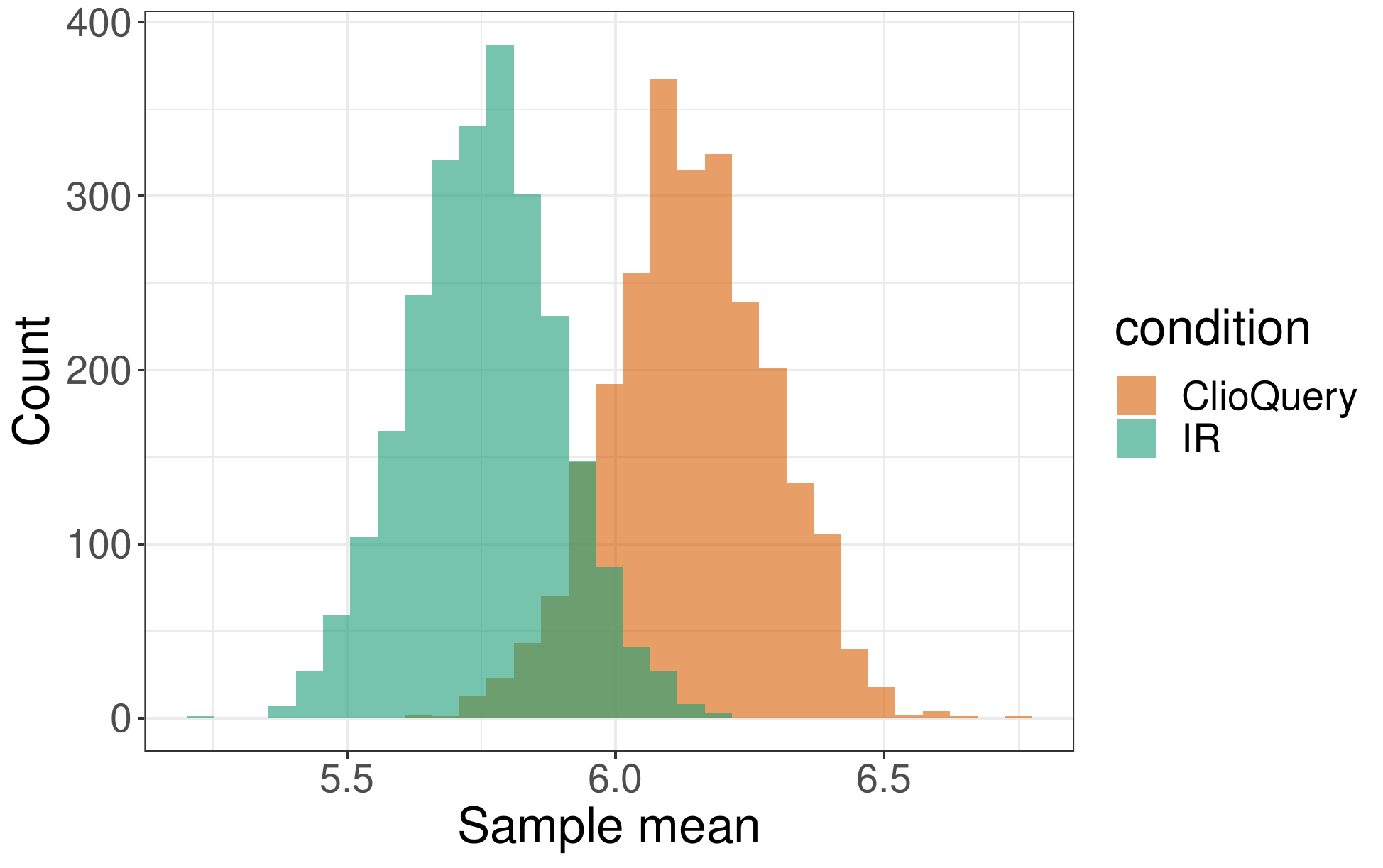}
    \caption{Distribution of sample means, across 2500 bootstrap samples of scores from the 59 IR participants, and 2500 separate bootstrap samples of scores from the 62 \ours~participants.}\label{f:samples}
\end{figure}

We tested for possible equality of means using bootstrap hypothesis testing \cite{Efron} (Algorithm 16.2).
Using 100,000 samples, we found that the difference in means among all workers in each condition was significantly different ($p=\pAll$). 
We also found that the difference in means among the subset of engaged workers in each condition was also significant ($p=\pAttentive$). 
We show separate bootstrapped distributions of sample means in Figure \ref{f:samples}.

\subsubsection{Qualitative analysis}\label{s:turk_qual}
Our experiment suggested that some properties of the \ours~interface helped participants on the historical reading comprehension task.
To try and gain a better understanding of which exact aspects of \ours~may have been helpful, we reviewed qualitative feedback from the \cqNperfectscore~\ours~participants who achieved a perfect score in reading comprehension.
Each of these participants praised one or more \ours~features in offering qualitative feedback on the system.
\textit{``I liked that I could expand the articles and filter them by the number of times the key word was mentioned,''} one top performer wrote. Said another, \textit{``I liked being able to control the number of mentions so I could determine relevance rather than trust a search engine.''} A third liked that \ours, \textit{``made it easy to see which articles I have already read and which ones I have yet to read.''} One high scorer did note that while in-text highlighting was in general helpful, \textit{``the part of this highlighting that I didn't like is that ... it was hard to gain context without reading the unhighlighted text before or after the highlighted sections.''}

On the other hand, the \irNperfectscore~IR users who got perfect scores offered scattered feedback. One liked how the results did not show \textit{``a bunch of random stuff or products to buy,''} while two others disagreed if the snippets were useful (one praised them, one said they did not help). 
Comments from the final high-scoring IR participant suggested that the IR system offered a realistic baseline. 
\textit{``There wasn't much to like or dislike,''} they said. \textit{``I really didn't find any differences how I would normally do it.''}

%% file: tables/crowd_table.tex
\begin{tabular}{llccc}
\toprule
 &  & N & Mean correct & Std. Dev.\\
\midrule
\multirow{2}{*}{All*} & ClioQuery & 62 & 6.145 & (1.185)\\
  & IR & 59 & 5.746 & (1.076)\\\midrule
\multirow{2}{*}{Engaged*} & ClioQuery & 30 & 6.300 & (1.022)\\
  & IR & 32 & 5.781 & (1.070)\\
\bottomrule
\end{tabular}

%% file: samples/discussion.tex

\subsection{\ours~suggests new features and directions for interactive text analysis}\label{s:discussion_simplification}

Much prior work in interactive text analysis focuses on helping people investigate bodies of documents by presenting textual units like topics \cite{tiara}, events \cite{eventriver}, or thematic hierarchies \cite{overview}.
But motivated by the needs of historians and archivists, \ours~instead proposes and tests a new approach to interactive corpus investigation, organized around the analysis of query mentions in context (see Section \ref{s:related_comparison}).

To help people investigate this ``unit of analysis'' \cite{chuangheer}, \ours~employs new text summarization techniques from the NLP literature to create a summary of a query term across a corpus.
The system then presents summaries alongside more traditional features from the text analytics literature, such as linked views and in-text highlighting, to help people easily and transparently review summary text in underlying documents.
During expert interview and field study evaluations, many historians said that they found such features helpful for archival research. 
They reported skimming over query mentions in the Document Feed to gain a sense of a query's use across a corpus, and then reading highlighted mentions in the Document Viewer for more context and detail. 
Several specifically mentioned that these components helped with \burdensome~and analysis.

This query-oriented approach suggests new directions for interactive text analytics in other query-oriented settings.
For instance, some marketing applications identify salient words and phrases in online forums \cite[Section 4.1]{marketingkdd};
\ours's query-focused summaries and linked views might help marketing analysts using such systems understand what people say about products online.
Features based on \ours~might also be applied in existing text analytics systems.
For instance, people might formulate a query using overview-oriented features such as word clusters, 
and then investigate this query word using a \ours-style Document Feed and Document Viewer.

\subsection{\ours~tests an idea: {``Text and its affordances should be taken seriously''}\label{s:text_seriously}}
Researchers have proposed many approaches to text visualization, which map high-dimensional text to two-dimensional graphical representations like time series plots (e.g., ThemeRiver \cite{ThemeRiver}) or bubble diagrams (e.g., EventRiver \cite{eventriver}).
By contrast, \ours's Document Feed and Document Viewer do not map text data to a graphical representation. Instead, \ours~uses text summarization methods from NLP to extract and present spans of text from a corpus for people to read, using automatic in-text highlighting to facilitate skimming.
In this sense, \ours~follows the advice of \citet{moretextplease}, who suggest that {``text and its affordances should be taken seriously''} in text analytics by making text itself ``a central piece of the visualization.''
Viewed through the lens of this recommendation, \ours~reflects one strategy for a text analytics system fundamentally organized around displaying spans from a corpus.
Other work from \citet{storifier} also explores this ``reading-centered approach.''
Some authors of prior text analytics systems have later noted the importance of showing underlying documents during interactive analysis. 
Authors of the Jigsaw system found that ``interactive visualization cannot replace the reading of reports'' \cite{Gorg2013JigsawReflections}.
Similarly, creators of both Overview \cite[Sec.\ 5]{stray} and ThemeRiver \cite[Sec.\ 7]{ThemeRiver} also describe finding that people need to read underlying text.

\subsection{User feedback on summarization has implications for natural language processing}\label{s:discussion_NLP}

\ours~applies particular ideas from query-focused text summarization for interactive text analysis.
However, building and evaluating a user-facing system forced us to reexamine several core assumptions from the text summarization literature.
In particular, early versions of \ours~applied standard optimization-based summarization methods \cite{McDonald} to select ``important'' information from a corpus.
This approach was reminiscent of prior temporally-oriented language engineering systems such as HistDiv \cite{histdiv}, TimeMine \cite{allen}, and TimeExplorer \cite{TimeExplorer}, which each attempt to automatically identify most-relevant information based on a query. 

However, during needfinding and prototyping, we found that some historians and archivists strongly disliked this approach.
Experts reported that they needed to understand why the computer was showing particular summaries, before they could actually draw conclusions from the output (see prototypes in the Appendix).
Based on this feedback, in later versions of \ours, we stopped trying to extract ``important'' mentions of a query term in search results. 
Instead, we decided to shorten and present every single sentence mentioning a user's query in the Document Feed, and allow people to easily examine such shortenings in context in the Document Viewer. 
During our expert interview and field study and evaluations, we found that this approach was more successful. 
We hypothesize that experts liked this format because they could understand {why} \ours~showed query shortenings, and thus use \ours~output in their research.

Our experiences might have implications for NLP, where research in summarization typically focuses on generating summaries which best match ``gold'' references \cite{das2007survey,nenkova2012survey} without worrying about explaining how summaries are formed.
In particular, much recent work on abstractive summarization in NLP \cite{rush-etal-2015-neural,Hermann2015TeachingMT} seeks to generate summary passages that do not occur in the input text.
Because such abstractive output can not be checked against underlying sources, and because such methods also currently suffer from frequent factual errors \cite{kryscinski-etal-2019-neural}, much more research may be required before abstractive approaches might be applied towards social research.

\subsection{Comprehensive and unbiased search costs time; transparency might help}\label{s:discussion_relevance}

During needfinding interviews, historians and archivists often emphasized the importance of directly and comprehensively examining all evidence relevant to a given research question, without allowing black-box algorithms to influence their conclusions.
We thus designed \ours~to minimize potential bias from algorithmic ranking.
Yet feedback on these aspects of \ours~was mixed (Section \ref{s:expert_interview_comprehensivenesss} and \ref{s:relevance_model_feedback}).
Some appreciated how \ours~used filters instead of ranking to narrow down search results.
But others reported that truly forgoing algorithmic curation required the researcher to spend too much time reading irrelevant documents.
For instance, some admitted that they often no have choice but to trust computer models of relevance to find evidence in archives because keyword search often turns up far more documents than they can possibly review.
While historians do sometimes work with smaller corpora (Section \ref{s:feed_and_viewer}), this issue would be particularly problematic in larger archives, where some queries will be mentioned many times. 

Why did some express deep commitment to full manual review of evidence during needfinding interviews, while others admit that they had to trust search engines to select evidence during system evaluation? 
There are at least two possibilities. 
One possibility is that historians and archivists might express commitment to comprehensive review when describing their ideal practices, but remember the limitations of this ideal when faced with a real task during system evaluation.
Some approaches to needfinding in HCI emphasize the limits of user interviews \cite{ethographic} because ``what people say and what they do can vary significantly.''
Another possibility is that there is variation in historians' commitment to comprehensiveness.
Some but not all historians may feel required to comprehensively review all evidence during research, possibly based on intellectual background or subfield.
(Other authors find similar variation among doctors \cite[Sec.\ 4.3.5]{doccurate}.)
Better understanding this apparent contradiction  between experts' stated commitments to comprehensive review and the realities of inevitable tradeoffs between recall and time \cite[Fig.\ 6]{pirolli2005sensemaking}  will require further research.

Nevertheless, future researchers might resolve the contradiction with improved user interfaces. 
Specifically, systems might transparently show which documents are selected or hidden by an algorithm, and allow people to easily override and investigate any document ranking decisions from a machine.
Such features would be particularly important for larger corpora, where historians would not be able to review all query mentions in context.
Research on tools for visually and interactively refining search results \cite{TiisVizIR} might offer a useful starting point.
Features which help groups of historians to collaborate during search could also enable teams of researchers to comprehensively review evidence from larger corpora.

%% file: samples/limitations_and_future.tex
Because it was difficult and expensive to recruit and interview highly-trained experts, this study relies on in-depth interviews with a small sample of humanists.
While such one-on-one interviews provided rich feedback, the opinions of our participants likely only approximate the true requirements of all historians and archivists. Moreover, interview studies may have limitations in unearthing design requirements (Section \ref{s:discussion_relevance}).
In the future, we thus plan to take steps to facilitate adoption in order to learn more about user needs.
In particular, we found that historians have to collect, organize, and sometimes digitize news stories before they are ready to gather and analyze query mentions (Section \ref{s:current_practices}). 
We thus plan to add features for importing news stories into \ours~from existing tools like Zotero and the Internet Archive. 
Additionally, throughout this work, we assume that query mentions are defined by exact string matches.
This simplifying assumption allows us to focus on user experience and interaction, but has clear limitations.
For instance, authors sometimes refer to ``Reagan'' using the nickname ``Dutch.''
Automatically detecting such aliases (and other deviations from exact string matching) will be important for future work.

%% file: samples/conclusion.tex
This study describes the design and evaluation of the \ours~text analytics system. 
Where prior tools focus on the analysis of textual units like topics \cite{tiara}, events \cite{eventriver}, or hierarchies \cite{overview} (Section \ref{s:related_comparison}), \ours~is uniquely organized around investigation of query words in context, which form the system's central ``unit of analysis'' \cite{chuangheer}.

\ours's unusual emphasis on the analysis of query words in context emerged from our study into the needs and practices of historians and archivists, who find and analyze occurrences of queries in their research.
Working with and studying historians revealed that analyzing change across time, undertaking comprehensive review of evidence, evaluating contextual information, and conducting neutral observation were each central to the practice of historical research.
Based on these insights, we designed the \ours~system, which applied query-focused text summarization methods from NLP to create skimmable summaries of a query term across an archive.
\ours~then used more traditional analytics features like linked views and automatic in-text highlighting to show summary text within the context of underlying news stories, in order to build expert trust in automatic summaries.

We tested \ours~in two separate user studies with historians, where we found that \ours's approach to organizing and presenting query mentions could help experts answer real questions from news archives.
Many historians reported that \ours's~Document Feed facilitated rapid analysis of query mentions, and that \ours's linked Document Viewer offered complementary context and detail.
In a separate quantitative comparison study, we found that \ours~helped crowd participants answer significantly more questions than a \Baselongname~tool.

Together, our work on \ours~suggests possible new directions for interactive text analysis.
In particular, \ours's~combination of text summarization and linked in-text highlighting could be applied in other query-oriented settings, where people also need to investigate query words in context.
For instance, some marketing applications suggest notable keywords from comments in online forums \cite[Section 4.1]{marketingkdd}.
\ours~methods might be applied to help marketers gather and analyze keyword mentions, or to help others investigate queries in other domains.

%% file: appendix_content.tex
\section{Appendix}
\input{samples/appendix/system_details}
\input{samples/appendix/corpus_collection}
\input{samples/appendix/crowd_experiment}

\clearpage
\subsection{Additional Figures}

\input{figures/crowd_appendix}

\input{figures/irscreen1}

\input{figures/irscreen2}

\clearpage

\input{figures/prototype_1}

\input{figures/prototype_2}

\clearpage
\subsection{Additional Tables}
\input{tables/participants} 
\input{tables/field_study_historians} 
\input{tables/interviewees}
\input{tables/libraryscience} 

%% file: samples/appendix/system_details.tex
\subsection*{The \ours~system: additional details}

\subsubsection*{Implementation details}
\ours~is a web application written in Python 3, using the Flask and React libraries.\footnote{\url{https://flask.palletsprojects.com/en/1.1.x/} and \url{https://reactjs.org/}} The text simplification methods in the paper use Stanford CoreNLP \cite{corenlppipeline} for tokenization, dependency parsing, and part-of-speech tagging.\footnote{Eisenstein \cite{eisenstein2019introduction} offers a detailed introduction to these NLP techniques.}  \ours's relationship span extraction method also employs logistic regression; we use the implementation from Scikit-learn \cite{Pedregosa:2011:SML:1953048.2078195}.
In the future, rewriting our Python-based prototype~in a faster language like Java or C would reduce our system's latency, helping \ours~scale to larger corpora.
It might also be possible to further improve performance by employing time and space efficient IR methods for efficiently indexing and retrieving the locations of query words in documents \cite{irbook}.

\subsubsection*{Time Series View: additional details}
\ours's {Time Series View} shows a single rug point (small vertical line) for each document mentioning the query.
These markings both help explain aggregated count statistics encoded in the time series plot (more rug points mean an higher annual count), and help link the Time Series View with the Document Feed. 
If a user hovers over a rug point, \ours~displays the headline of the corresponding news story using a tooltip; if the user clicks a rug point, \ours~updates so that the story is displayed in the Document Feed and in the Document Viewer.
When a user hovers over some year in the Time Series View, \ours~displays a tooltip showing the total count of documents containing the query for that year.

\subsubsection*{Default system behaviors}
If a user has not yet entered a query, \ours's time series plot simply shows the overall counts of documents by year across the entire corpus, shown with a neutral black line.
In this case, the Document Feed also shows all documents in the corpus. Moreover, when filter-by-date is not used, \ours~shows documents from the time span of the corpus.

\subsubsection*{Choosing colors}
We {chose colors for \ours} using Colorbrewer \cite{colorbrewer}, a common resource, which offers colorblind safe and print-friendly palettes.
Hall and Hanna \cite{HallAndHanna} test how foreground and background color affects how people read, retain, and experience text on screen. Our study focuses on testing the utility of in-text highlighting and text simplification for expert social researchers; future work might test the effect of varying the foreground or background color.

\subsubsection*{Handling token gaps during clause deletion}
In some cases, there may be gaps between tokens in simplified mentions, where tokens have been removed from the middle of a sentence. 
(These are shown with ellipses in the Document Feed).
In these cases, in performing {automatic in-text highlighting to link the Document Feed and Document Viewer}, we highlight the span in the Document Viewer which begins with and ends with the first and last token of the corresponding simplified mention, shown in the Document Feed.

\subsubsection*{Computing tf-idf scores of iterative clause deletion}
To compute tf-idf scores during iterative clause deletion, we assign each word in each possible output candidate shortening a word-level tf-idf score, and average the word-level tf-idf scores of all words in each possible candidate shortening to compute an overall, sentence-level tf-idf score. 
We assign each word a tf score equal to the total occurrences of the word among all documents that contain $Q$, and an idf score equal to 1 divided by the count of documents containing the word across the corpus.
We then multiply each word's tf score by its idf score to get a word-level td-idf score.
We then select the candidate shortening with the highest overall tf-idf score for display in the Document Feed.

\subsubsection*{Choosing among possible sentence shortening methods}
In the System section, we describe three different sentence shortening techniques, which are applied in the \ours~interface. 
Below, we describe how \ours~chooses to apply the three different methods.

After a user enters a query $Q$, for each document mentioning $Q$, \ours's Document Feed displays the first sentence within the document mentioning $Q$ that can be shortened via query-focused clause deletion.
If no such sentence exists, \ours~resorts to shortening the first sentence mentioning $Q$ via {character windowing}. 
(Character windowing is only used as a last resort because it does not attempt to create well-formed output containing salient words from the input.)

In cases when a user has entered both a query and subquery, for each document mentioning the query or subquery, \ours~will attempt to display the first sentence in the document that can be shorted via relationship span extraction.
This is because we assume the user is interested in the relationship between the query and subquery.
If there is no sentence that can be shortened via relationship span extraction, \ours~will display the first sentence that can be shortened via query-focused clause deletion.
If no sentence can be shortened via clause deletion, it will resort to shortening the first sentence mentioning the query or subquery via character windowing.

\ours~also allows the user to click ``expand'' to see all sentences mentioning the query within the document, as described in the System section. 
In this case, \ours~will first attempt to shorten each sentence mentioning $Q$ via query-focused clause deletion, before resorting to shortening the sentence with character windowing. 
If the user has also set a subquery (in addition to $Q$), \ours~will first try to shorten each sentence mentioning the query and subquery using relationship span extraction (and then attempt clause deletion, and character windowing).

%% file: samples/appendix/corpus_collection.tex
\subsection*{Field study: additional details}

To help $H1$ answer their question using \ours, we gathered a custom corpus of articles from \textit{The New York Times} (NYT). To gather the corpus, we searched for ``El Salvador'' on \textit{The New York Times} website \cite{nytwebsite}, and then automatically downloaded all query-matching articles published between 1980 and 1985 in the World News and Week in Review sections of the newspaper. 
We filtered downloaded articles to create a corpus of NYT articles containing the word ``Salvador,'' and we loaded this corpus into \ours~for $H1$. 

To help $H2$ answer their research question, we similarly gathered a second custom corpus of articles by searching for ``astronaut'' on the \textit{New York Times} website \cite{nytwebsite}, and then automatically downloading all query-matching articles published between 1980 and 1985. We then similarly filtered the documents to ensure that all query-matching mentioned ``astronaut'' and loaded the corpus into \ours~for $H2$.

%% file: samples/appendix/crowd_experiment.tex
\subsection*{Quantitative comparison study: additional details}

\subsubsection*{Additional details regarding creation of reading comprehension questions}
We used a semi-automated procedure to create reading comprehension questions for our quantitative crowd study.
Specifically, we first collected all editorials from The New York Times Annotated Corpus \cite{SandhausNYT} which included the words ``Zimbabwe'' and ``Mugabe''.
We then used the TfidfVectorizer class from scikit-learn \cite{Pedregosa:2011:SML:1953048.2078195} with default settings to construct tf-idf vectors for all 1,689 sentences in the editorials. 
We also similarly constructed tf-idf vectors for all 597 sentences from the Wikipedia page on Robert Mugabe \cite{wikimugabe}. 
We then computed the cosine similarity of each sentence pair in the Cartesian product of Wikipedia and \textit{New York Times} sentences. We manually reviewed the 200 sentence pairs with the highest cosine similarities, and manually labeled 37 total sentences from \textit{New York Times} editorials which reported a fact described in some sentence from Wikipedia.
This process identified 37 facts about Mugabe from Wikipedia reported in editorials in \textit{The New York Times}. 
We selected 8 of these facts to create reading comprehension questions for our task.



\subsubsection*{Additional details regarding tuning of IR baseline}
We implemented the IR baseline using Whoosh, an open-source Python search engine. Like many search engines, Whoosh shows small document snippets from ranked documents on the search engine results page (Figure \ref{f:appendix_ir_serp}). To encourage fair comparison between Whoosh and \ours, we tuned Whoosh so that document snippets contained roughly as much text as the shortened sentences in the \ours~Document Feed. 
Specifically, Whoosh allows snippet customization by setting the \texttt{maxchars} and \texttt{surround} parameters in its \texttt{Highlighter} module.
We set these parameters by performing a grid search over all possible values from 10 to 100 (for each parameter), in order to maximize the average number of characters per Whoosh document snippet, under the constraint that the average was less than or equal to 90 characters (the length of the longest-possible shortened sentence in the \ours~Document Feed).
The final setting for the \texttt{surround} parameter was 27 characters and the final setting for the \texttt{maxchars} parameter was of 10 characters. Using these settings, we observe a mean snippet length of exactly 90 characters using the IR system on the crowd task.
Beyond tuning these parameters, we use default settings for the Whoosh search engine.

\subsubsection*{Additional details regarding the crowd study pretest}
Before beginning the main task in our crowd study, participants in each condition used their interface to complete a three minute pretest using a small corpus of six \textit{New York Times} editorials mentioning ``Iraq''. 
The pretest was very similar to the main task; each interface was hard-coded to use the query ``Falluja'' and 
participants were instructed to ``find and remember everything the \textit{New York Times} wrote about Falluja'' using their tool.
After participants typed this exact phase into a text box to confirm they understood the instructions, they conducted research using their assigned interface.
After 3 minutes, participants were then presented with a screen with four facts about U.S.\ involvement in Falluja (included in supplemental materials),
and asked to identify which facts were reported in the six articles.
Because only one fact from the list was reported in the articles, 
to get a perfect score of 4 out of 4 on the pretest, workers had to both correctly identify the reported fact, and refrain from guessing any of the other three facts. 
The pretest was designed to be very easy for attentive workers.

\subsubsection*{Additional details regarding data collection phases for the crowd task}
Data collection for the crowd task proceeded in two phases: an initial pilot phase and a main data collection phase.
After the small pilot, we added two training screens for \ours~participants (shown in supplemental materials) to help \ours~users gain practice using unfamiliar features. 
We also fixed a bug in the pilot in which \ours~users were shown an extra two editorials. 
We emphasize that these two editorials did not contain any facts about Mugabe which could be used to answer the reading comprehension questions, and also note that the 
two extra editorials would have made the task harder for \ours~participants (because they would have had to read extra text during the task, which was not relevant to the reading comprehension questions).
Finally, after the pilot, we adjusted the random assignment mechanism so that participants were assigned to conditions in an alternating fashion following an initial random draw (i.e.\ first \ours, then IR, then \ours...). In the pilot, participants were assigned to conditions at random when they loaded the first screen in the task.

\subsubsection*{Additional details regarding task payment}
Because we had trouble recruiting qualified masters workers for our lengthy and complex task we increased payment during data collection. The first 18 participants were paid \$2.50 to complete the pilot. After the pilot, we increased payment to \$3.00 and collected data from 75 more participants. Because data collection was still very slow (e.g.\ 10 workers over a 24 hour period) we further increased payment to \$4.00 for the task and collected data from 26 more workers. Finally, we increased payment to \$5.00 for the task. When only 2 workers signed up over a half-day period at the \$5.00 rate we ended data collection.

%% file: figures/crowd_appendix.tex
\begin{figure}[h]
\fbox{\includegraphics[width=14cm]{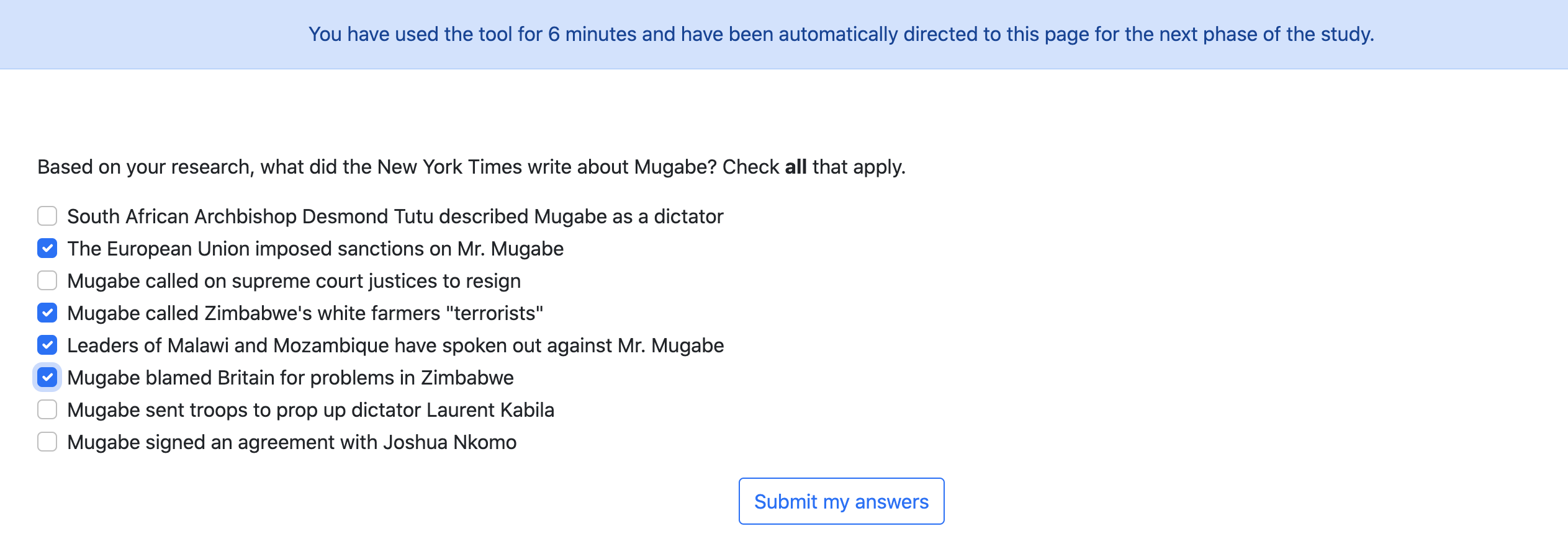}}
\caption{Participants answered eight true/false questions about what \textit{The New York Times} wrote about Robert Mugabe, using the form shown above. The four facts shown with checkboxes were described in editorials available to participants during the study. 
The four false facts shown without checkboxes were described in other editorials, not available to participants during the study. 
Participants who found and remembered the four facts from the corpus and who also did not incorrectly guess any of the four facts not described in the corpus scored 8 out of 8 on the reading comprehension task.
The order of questions was randomized.}
\end{figure}

\begin{figure}[h]
\fbox{\includegraphics[width=14cm]{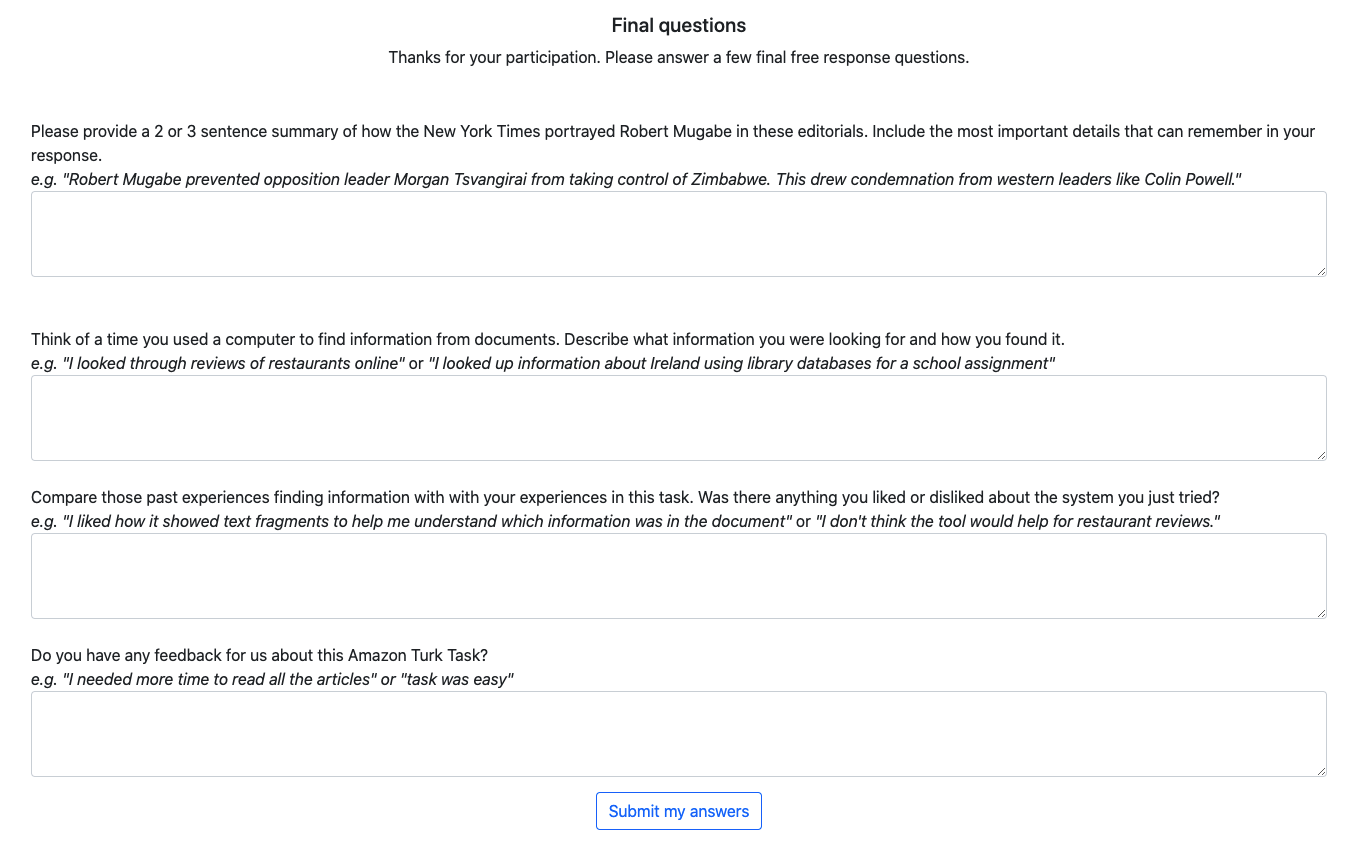}}
\caption{Qualitative questions for participants at the end of the crowd task}
\end{figure}

%% file: figures/irscreen1.tex
\begin{figure}[t!]
\centering
\fbox{\includegraphics[width=1.0\linewidth]{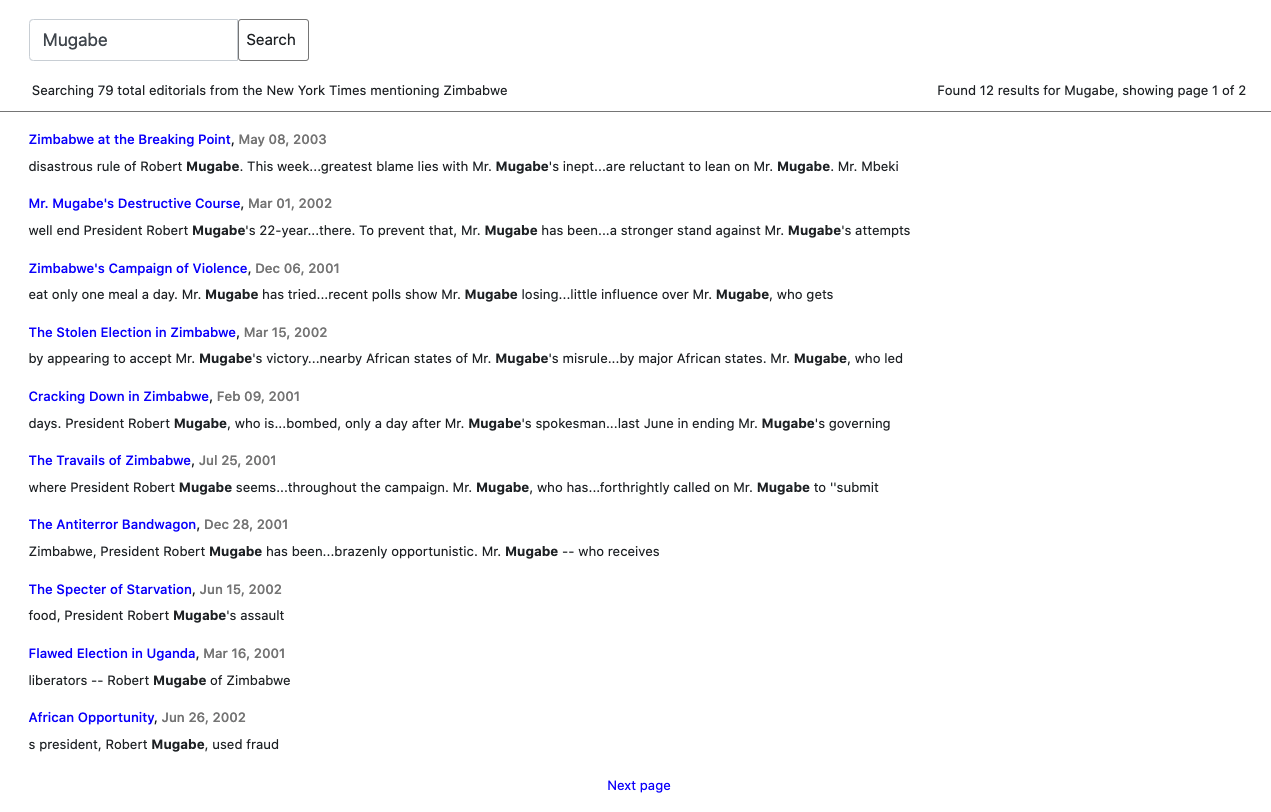}}
\caption{The IR baseline interface in our crowd study}
\label{f:appendix_ir_serp}
\end{figure}

%% file: figures/irscreen2.tex
\begin{figure}[t!]
\centering
\fbox{\includegraphics[width=1.0\linewidth]{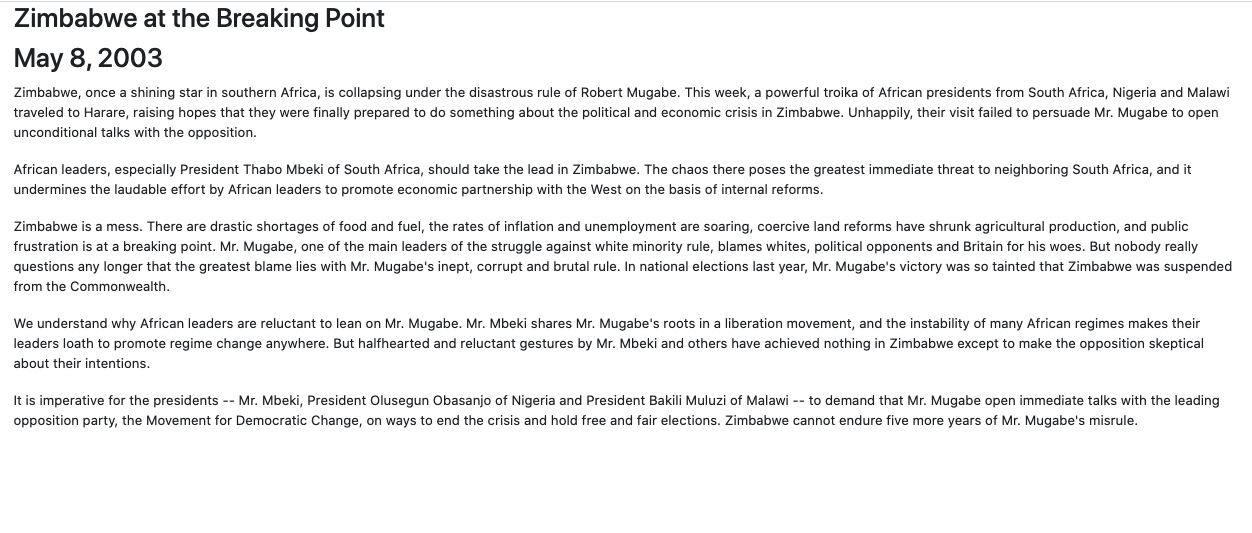}}
\caption{A single search result from the IR interface in our crowd study}
\label{f:appendix_ir}
\end{figure}

%% file: figures/prototype_1.tex
\begin{figure}[t!]
\centering
\fbox{\includegraphics[width=.55\linewidth]{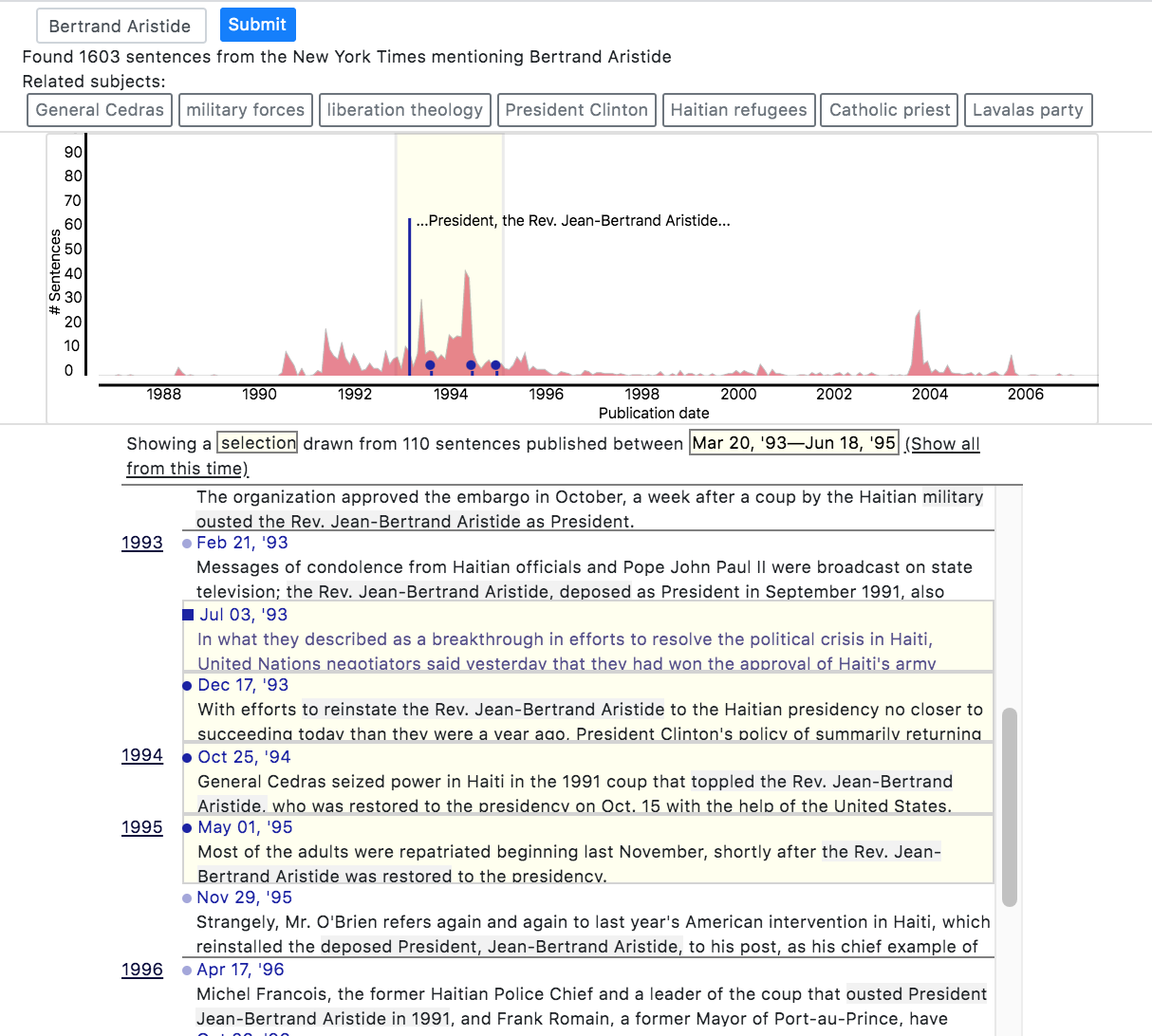}}
  \caption[Another early prototype of \ours]{
An early prototype of \ours, which used traditional, optimization-based text summarization methods from natural language processing \cite{McDonald} to try and select the most salient information from a given time period for display. This prototype selects four most ``important'' articles (shaded in light yellow in the feed below), to summarize the hundreds of articles mentioning the query ``Bertrand Aristide'' in \textit{The New York Times} from May 20, '93 to June 18, '95 (shown shaded in light yellow in the time series above).
\ifive~strongly disliked this approach, prompting a shift towards interfaces emphasizing transparency and trustworthiness. 
\textit{``I need to know what is included and why,''} \ifive~ explained. \textit{``I need to know why it is showing this limited view.''} \ifive~continued, \textit{``I am wary of algorithms that choose for me what the important facts are. I am a PhD historian. Leaving stuff out. We are taught to be critical of that.''} Ultimately, \ifive~noted, \textit{``History is written by the victors. What actually matters is what people choose to put in the timeline.''} We theorize that \ifive~could not trust the prototype because it seemed to lack the capacity to select important facts or the integrity to adhere to historical research principles; prior work in HCI (e.g.\ SMILY \cite{smiley}) assumes that in order to earn user trust, a system must both have the capacity to help the user and the integrity to adhere to principles which are important in a given domain.
}\label{f:prototype1}
\end{figure}

%% file: figures/prototype_2.tex
\begin{figure}[t!]
\centering
\fbox{\includegraphics[width=.55\linewidth]{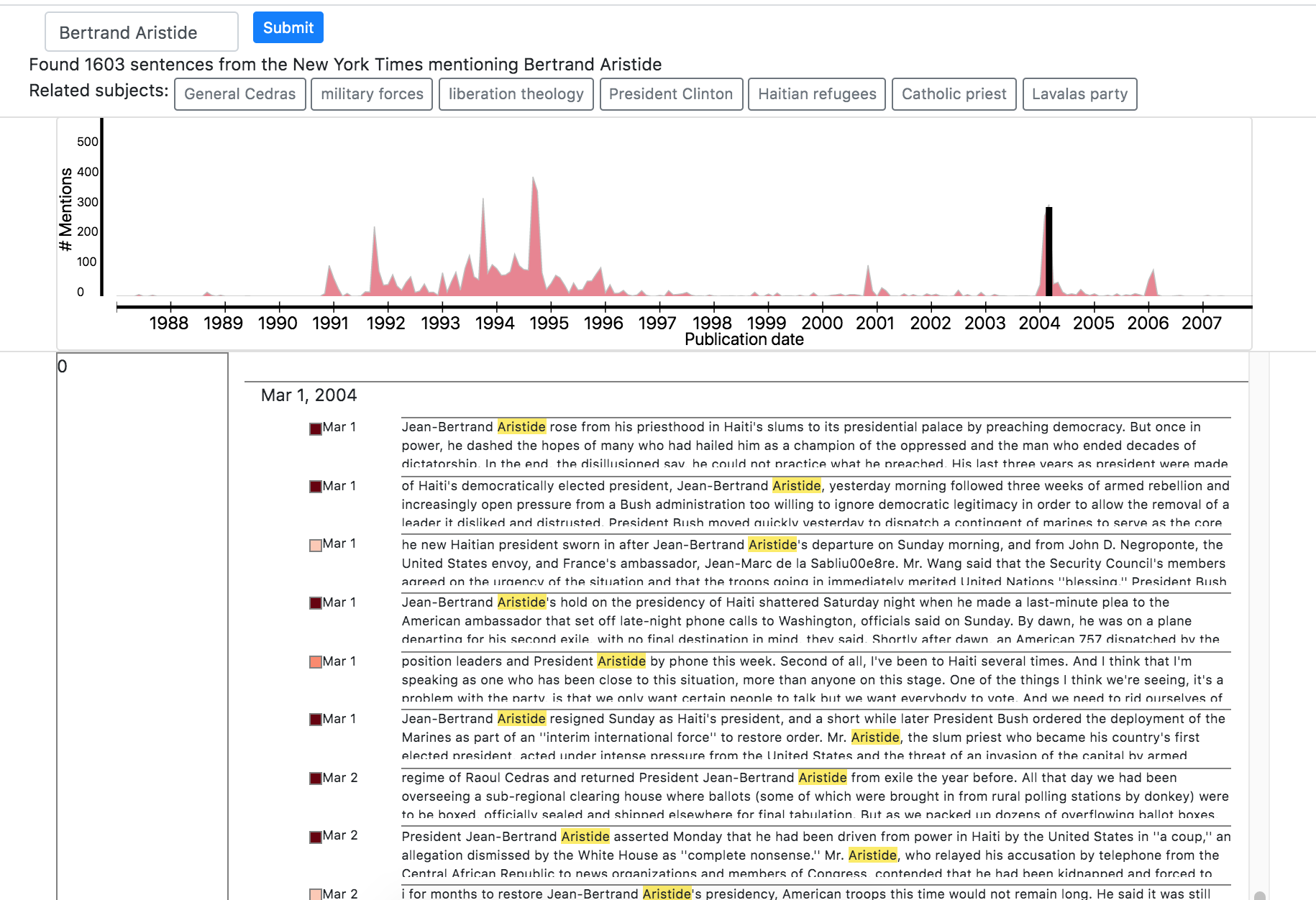}}
  \caption[An early prototype of \ours]{  
  An early prototype of \ours, displaying and highlighting every single mention of the query term ``Aristide'' in \textit{New York Times} articles mentioning ``Haiti''. ``Are we showing too much information in this interface?'' one researcher from our group asked \ifour, when presenting the prototype. ``This is literally every mention of your query term.''
\textit{``No this is good},''  \ifour~explained, \textit{``because of what I was calling the type II error concern [i.e.\ the fear of missing relevant material]. When I see something that is trying to decide or curate for me that is a worry. That is a red flag.''} 
However, \ifour~went on to explain how the interface needed to provide more context and transparency surrounding highlighted snippets. 
\textit{``With this design you have to click or read each snippet to see if it is relevant,''} he said.  ``\textit{The snippets are valuable and good but very small and you have to look at the contents of the article. Sometimes you can eliminate that by just quickly scanning the article title ... there needs to be a way to provide the information in a more transparent way.}'' 
(The search bar shown at the top is non-functioning mockup; the ``0'' on the left hand side is a placeholder.)
}\label{f:prototype2}
\end{figure}

%% file: tables/participants.tex
{
\begin{table}[h]
\begin{tabular}{lccccc}
\toprule
ID     & Research experience & Library experience  & University role                & \multicolumn{1}{c}{Gender} & Topic     \\  \midrule
{P1} & 6                   & 0                  & PhD candidate                & Male                       & Iraq      \\
P2 & 5                   & 0                  & PhD candidate                & Male                       & Zimbabwe  \\
P3 & 4                   & 0                  & PhD candidate                & Female                     & combat    \\
P4 & 20                  & 0                  & Instructor/researcher        & Female                     & wages     \\
P5 & 0                   & 3                  & History librarian & Male                       & copyright \\ \bottomrule 
\end{tabular}
\caption[Participants in the interview study]{Interview study participants. We report history and library experience in years.}\label{t:particpants}
\end{table}
}

%% file: tables/field_study_historians.tex
{
\begin{table}[h]
\begin{tabular}{@{}lccccc@{}}
\toprule
ID     & Research experience & Library experience & Academic role   & \multicolumn{1}{c}{Gender} & Research area     \\ \midrule
\rhone     & 5                   & 0                  & PhD student     & Male                       & Media and society \\
\rhtwo & 25                  & 0                  & Tenured faculty & Female                     & Space exploration \\ \bottomrule

\end{tabular}
\caption[Historians in the field study]{Historians in the field study. History and library experience are listed in years.}\label{t:field_study_particpants}
\end{table}
}

%% file: tables/interviewees.tex
{
\begin{table}[h]
\begin{tabular}{rccccc}
\toprule
    ID        & {Research experience} & {Library experience }   & {University role}   & {Gender}    &  {Field}   \\  \midrule
{I1} & 5-10  &  1-5    & PhD Candidate & Male & History \\
{I2} & 0 &  20-30  & Librarian & Female & Lib.\ Science \\ 
{I3} & 10-20   &  0      &  Junior Faculty  &   Female & Am.\ Studies \\
{I4} & 10-20 & 10-20 & Archivist & Male & Lib.\ Science  \\
{I5} & 10-20 & 10-20 & Librarian & Non-binary   & History  \\  \bottomrule
\end{tabular}
\caption[Interviewees in the needfinding study]{Interviewees in our needfinding study. We list research experience and library experience as a range of years. 
We abbreviate American Studies as Am.\ Studies and Library Science as Lib.\ Science.
}\label{t:interviewees}
\end{table}
}

%% file: tables/libraryscience.tex

\begin{table}[h]
\begin{tabular}{@{}llcc@{}}
\toprule
Author(s) & Venue &  Study type & Participants  \\ \midrule
Allen and Sieczkiewicz \cite{allen}     &   \textit{Proc. ASIS\&T} (Info. Science)  &        Interview      &       8     \\
Case \cite{Case}     &  \textit{The Library Quarterly }   &     Interview         &     20       \\
Duff and Johnson     \cite{DuffJohnson}   &   \textit{The Library Quarterly}    &     Interview         &  10  \\     \midrule
Chassanoff \cite{Chassanoff}     &   \textit{The American Archivist}    &  Survey  &  86      \\
Dalton and Charnigo \cite{DaltonCharnigo}      &   \textit{College \& Research Libraries}    &   Survey           &     278       \\ 
Duff, Craig, and Cherry  \cite{DuffCraigCherry}      &   \textit{The Public Historian}    &    Survey         &    600         \\ \bottomrule
\end{tabular}\caption[A selection from prior work in library science and information science]{A selection from prior work in library science and information science, focused on the information-seeking behavior of historians.
These papers describe studies of $N=1002$ historians (in total).}\label{t:libraryscience}
\end{table}
